\documentclass[USenglish]{llncs} \usepackage[latin1]{inputenc} \usepackage[english]{babel}

\usepackage{algorithm}
\usepackage{algorithmic}
\usepackage{amsmath}
\usepackage{amssymb}
\usepackage{amsthmWithoutProof}
\usepackage{array}
\usepackage{balance}
\usepackage{color}
\usepackage{listings}
\usepackage{paralist}
\usepackage{tabularx}
\usepackage{times}
\usepackage{url}

\definecolor{darkgreen}{rgb}{0,.5,.2}
\definecolor{gray}{rgb}{.5,.5,.5}

{\vspace{5mm}\begin{sloppypar}
  \noindent\hrulefill\, BEGIN \,\hrulefill
  \end{sloppypar}
}%
{\begin{sloppypar}
  \noindent\hrulefill\, END   \,\hrulefill
  \end{sloppypar}\vspace{5mm}
}

\newcommand{\todo}[1]{}

\newcommand{\Todo}[2]{#1}

\newcounter{question}

\newcommand{\removable}[1]{#1}

\lstdefinelanguage{PseudoCode}{%
  morekeywords={FUNCTION,CALL,LET,RETURN,IF,THEN,ELSE,AND,OR,NOT,WHILE,DO,FOR,EACH},%
  sensitive=true,%
  morecomment=[l]--,%
  morecomment=[s]{/*}{*/},%
  morestring=[d]',%
  morestring=[d]"%
}[keywords,comments,strings]

\lstdefinelanguage{SPARQL}{%
  morekeywords={BASE,PREFIX,SELECT,DISTINCT,DESCRIBE,CONSTRUCT,%
        ASK,FROM,NAMED,WHERE,ORDER,BY,ASC,DESC,LIMIT,OFFSET,OPTIONAL,%
        GRAPH,UNION,FILTER,STR,LANG,DATATYPE,REGEX,BOUND,ISURI,ISBLANK,
        ISLITERAL,TRUE,FALSE},%
  sensitive=false,%
  morecomment=[l]--,%
  morecomment=[s]{/*}{*/},%
  morestring=[d]',%
  morestring=[d]"%
}[keywords,comments,strings]

\lstdefinelanguage{N3}{%
  morekeywords={@prefix},%
  sensitive=true,%
}[keywords,comments,strings]
\newcommand{\code}[1]{\texttt{#1}}

\newcommand{\rdfTermInFigure}[1]{\begin{scriptsize}#1\end{scriptsize}}

\theoremstyle{plain}
\newtheorem{fact}{Fact}

\theoremstyle{definition}
\newcommand{\definedTerm}[1]{\textbf{#1}}

\newcommand{\webproblem}[4]{%
	\noindent
	\begin{tabularx}{\columnwidth}{|p{22.5mm} >{\normalsize}X|} \hline
		\textbf{Problem:} &
			\problemName{#1}
		\\ Web Input: &
			#2
		\\ Ordinary Input: &
			#3
		\\ Question: &
			#4
		\\ \hline
	\end{tabularx}%
}

\newcommand{\problemName}[1]{\textsc{#1}}

\newenvironment{myproof}[1]{%

	\noindent \textbf{Proof #1.}%
}{%
	\qed \vspace{2ex}

}

\newenvironment{proofdeclitems}{%
	\noindent
	Let:
	\begin{itemize} \itemsep1mm%
}{%
	\end{itemize}%
}

\newcommand{\IFF}{if and only if } 
\newcommand{\LDdoc}{LD document}
\newcommand{\ID}{URI} 
\newcommand{\triple}{RDF triple} 
\newcommand{\TP}{triple pattern}

\newcommand{\true}{\mathrm{true}} 
\newcommand{\false}{\mathrm{false}} 
\newcommand{\fctDomName}{\mathrm{dom}}
\newcommand{\fctDom}[1]{\fctDomName(#1)}

\newcommand{\symAllURIs}{\mathcal{U}} 
\newcommand{\symAllLiterals}{\mathcal{L}} 
\newcommand{\symAllBNodes}{\mathcal{B}} 
\newcommand{\symAllVariables}{\mathcal{V}} 
\newcommand{\symURI}{u} 

\newcommand{\fctTermsName}{\mathrm{terms}}
\newcommand{\fctTerms}[1]{\fctTermsName(#1)} 
\newcommand{\fctIDsName}{\mathrm{uris}} 
\newcommand{\fctIDs}[1]{\fctIDsName(#1)} 
\newcommand{\fctVarsName}{\mathrm{vars}}
\newcommand{\fctVars}[1]{\fctVarsName(#1)} 

\newcommand{\WoD}{W} 
\newcommand{\dataFct}{data} 
\newcommand{\adocFct}{adoc} 
\newcommand{\fctAllDataName}{\mathrm{AllData}} 
\newcommand{\fctEncName}{\mathrm{enc}}
\newcommand{\fctEnc}[1]{\fctEncName(#1)}

\newcommand{\LDSPARQLFull}{SPARQL$_\text{\textsf{LD}}$}

\newcommand{\LDSPARQLReach}{SPARQL$_\text{\textsf{LD}(R)}$}

\newcommand{\ReachPartScPW}[4]{#4^{(#1,#3)}_{#2}} 
\newcommand{\Reach}[1]{#1_\mathfrak{R}} 

\newcommand{\cNone}{c_\mathsf{None}}
\newcommand{\cAll}{c_\mathsf{All}}
\newcommand{\cMatch}{c_\mathsf{Match}}

\newcommand{\evalOrigPG}[2]{[\![#1]\!]_{#2}} 
\newcommand{\evalFullFctP}[1]{\mathcal{Q}^{#1}}
\newcommand{\evalFullPW}[2]{\evalFullFctP{#1}\!\bigl(#2\bigr)} 
\newcommand{\evalReachFctPSc}[3]{\mathcal{Q}^{#1,#2}_{#3}}
\newcommand{\evalReachPScW}[4]{\evalReachFctPSc{#1}{#2}{#3}\!\bigl(#4\bigr)} 

\newcommand{\OpAND}{\text{ \normalfont\scriptsize\textsf{AND} }}
\newcommand{\OpUNION}{\text{ \normalfont\scriptsize\textsf{UNION} }}
\newcommand{\OpOPT}{\text{ \normalfont\scriptsize\textsf{OPT} }}
\newcommand{\OpFILTER}{\text{ \normalfont\scriptsize\textsf{FILTER} }}

\newcommand{\fctBoundName}{\mathrm{bound}}
\newcommand{\fctBound}[1]{\fctBoundName(#1)}

\renewcommand{\emptyset}{\varnothing}

\definecolor{darkgray}{rgb}{.3,.3,.3}

\def\ojoin{\setbox0=\hbox{$\Join$}\rule[.01ex]{.25em}{.45pt}\llap{\rule[1.05ex]{.25em}{.4pt}}}
\def\LJoin{\mathbin{\ojoin\mkern-7.5mu\Join}}

\begin{document}

\title{SPARQL for a Web of Linked Data:\\Semantics and Computability}
\subtitle{(Extended Version)\footnote{This report presents an extended version of a paper published in ESWC 2012~\cite{Hartig12:SPARQL4LinkedDataPaper}. The extended version contains proofs for all
	technical results
in the paper (cf.~\appendixname~\ref{Appendix:Proofs}).}
\vspace{-5mm} 
}

\author{Olaf Hartig} 
\authorrunning{O. Hartig} 
\tocauthor{           
    Olaf Hartig (Humboldt-Universit\"at zu Berlin)
}
\institute{
Humboldt-Universit\"at zu Berlin\\
\email{hartig@informatik.hu-berlin.de} 
}

\maketitle

\vspace{-5mm} 
\begin{abstract}
	The World Wide Web currently evolves into a Web of Linked Data where content providers publish and link data as they have done with hypertext for the last 20 years. While the declarative query language SPARQL is the de facto for querying a-priory defined sets of data from the Web, no language exists for querying the Web of Linked Data itself. However, it seems natural to ask whether SPARQL is also suitable for such a purpose.

In this paper we formally investigate the applicability of SPARQL as a query language for
Linked Data on the Web.
	In particular, we study two query models: 1) a \emph{full-Web semantics} where the scope of a query is the complete set of Linked Data on the Web and 2) a family of \emph{reachability-based semantics} which restrict the scope to data that is reachable by traversing certain data links.
		For both models we discuss properties such as monotonicity and computability as well as the implications of querying a Web that is infinitely large due to data generating servers.

\end{abstract}

\enlargethispage{\baselineskip} 
\vspace{-10mm} 

\section{Introduction} \label{Section:Introduction}

%

\noindent
The emergence of vast amounts of RDF data on the WWW has spawned research on storing and querying large collections of such data efficiently. The prevalent query language in this context is SPARQL~\cite{Perez09:SemanticsAndComplexityOfSPARQL} 
	which defines queries as functions over
		an RDF dataset, that is,
	a fixed, a-priory defined collection of sets of RDF triples. This definition
naturally fits the use case of querying a repository of RDF data copied from the Web.
%

However, most RDF data on the Web is published following the Linked Data principles%
	~\cite{BernersLee07:LinkedData}%
, contributing to the emerging Web of Linked Data%
	~\cite{Bizer09:TheStorySoFar}.
This practice allows for query approaches that access the most recent version of remote data on demand. More importantly, query execution systems may automatically discover new data by traversing data links. As a result, such a system answers queries based on data that is not only up-to-date but may also include initially unknown data. These features
	are the foundation for true serendipity, which we regard as the most distinguishing advantage of querying the Web itself, instead of a predefined, bounded collection of data.

While several research groups work on systems that
	evaluate SPARQL basic graph patterns
over the Web of Linked
	Data (cf. \cite{Harth10:DataSummariesForLDQueryProcessing}, \cite{
Hartig11:HeuristicForQueryPlanSelection,Hartig09:QueryingTheWebOfLD} and~\cite{Ladwig10:LinkedDataQueryProcessingStrategies,Ladwig11:SIHJoin}),
we notice a shortage of work on theoretical foundations and properties of such queries. Furthermore, there is a need to support
	queries that are more expressive than conjunctive (basic graph pattern based) queries%
~\cite{
Picalausa11:RealSPARQLqueries}. However,
	it seems natural to assume that SPARQL could be used in this context
because the Web of Linked Data is based on the RDF data model and SPARQL is a query language for RDF data.
	In this paper we challenge this assumption.

\vspace{1ex} \noindent \textbf{Contributions} 
In this paper we understand queries as functions over the Web of Linked Data as a whole. 
To analyze the suitability of SPARQL as a language for such queries, we have to adjust the semantics of SPARQL. More precisely, we have to redefine the scope for evaluating SPARQL algebra expressions. In this paper we discuss two approaches for such an adjustment. The first approach uses a semantics where the scope of a query is
	the complete set of
Linked Data on the Web.
	We
call this semantics \emph{full-Web semantics}. The second approach introduces a family of \emph{reachability-based semantics} which restrict the scope to data that is reachable by traversing certain data links. We emphasize that both approaches allow for query results that are based on data from initially unknown sources and, thus, enable applications to tap the full potential of the Web. Nevertheless, both approaches precisely define the (expected) result for any
query.

\todo{add a few words about what exactly we are investigating (computability), why (copy from the corresponding section in the fullWeb sect.), what is needed and why (sat and mono b/c we assume a dependency -- which we also shall be able to show), and what the main result is (limited computability in both cases)}

	As a prerequisite for defining the aforementioned semantics and for studying theoretical properties of queries under these semantics, we introduce a theoretical framework.
%
	The basis of this framework is a data model that
captures the idea of a Web of Linked Data. We model such a Web as an infinite structure of documents that contain RDF data and that are interlinked via this data.
Our model allows for infiniteness because the number of entities described in a Web of Linked Data may be infinite; so may the number of documents. The following example illustrates such a case:
\begin{example} \label{Example:LinkedOpenNumbers}
	Let $u_i$ denote an HTTP
		\removable{scheme based}
	URI that identifies the natural number $i$. There is a countably infinite number of such URIs. The WWW server which is responsible for these URIs may be set up to provide a document for each natural number. These documents may be generated upon request and may contain RDF data including the RDF triple $( u_i , \text{\rdfTermInFigure{http://.../next}} , u_{i+1} )$. This
	triple associates
		\removable{the natural number}
	$i$ with its successor $i$\textnormal{+1} and, thus, links to the data about $i$\textnormal{+1}~\textnormal{\cite{Vrandecic09:LinkedGeoData}}. An example for such a server is provided by the Linked Open Numbers project\footnote{http://km.aifb.kit.edu/projects/numbers/}.
\end{example}

\noindent
	In addition to the data model our theoretical framework comprises a computation model. This model is based on a particular type of Turing machine which
%
	formally
captures the limited data access capabilities of computations over the Web.

We summarize the main contributions of this paper as follows:

\begin{itemize}
	\vspace{-2mm} 
	\item We present a data model and a computation model that provide a theoretical framework to define and to study query languages for the Web of Linked Data.
	\item We introduce a full-Web semantics and a family of reachability-based semantics
	for a (hypothetical) use of SPARQL as a language for queries over Linked Data.
	\item We systematically analyze SPARQL queries under the semantics that we introduce. This analysis includes a discussion of satisfiability, monotonicity, and computability of queries under the different semantics, a comparison of the semantics, and
		a study
	of the implications of querying a Web of Linked Data that is infinite.
\end{itemize}

%
\noindent \textbf{Related Work} \,
Since its emergence the WWW has attracted research on declarative query languages for the Web.
For
	an overview
on
	early work in this area we refer to%
~\cite{Florescu98:DBTechniquesForWWW}.
Most of
	this work
understands the WWW as a hypertext Web.
Nonetheless, some of the foundational work can be adopted for
	research on
Linked Data. The computation model that we use in this paper is an adaptation of the ideas
	presented in~\cite{Abiteboul00:QueriesAndComputationOnTheWebArticle} and~\cite{Mendelzon98:FormalModelsOfWebQueries}.

In addition to the early work on Web queries, query execution over Linked Data on the WWW has attracted much attention recently~\cite{Harth10:DataSummariesForLDQueryProcessing,%
Hartig11:HeuristicForQueryPlanSelection,Hartig09:QueryingTheWebOfLD,Ladwig10:LinkedDataQueryProcessingStrategies,Ladwig11:SIHJoin}.
However, existing work primarily focuses on various aspects of (query-local) data management, query execution, and optimization. The only work we are aware of that aims to formally capture the concept of Linked Data and to provide a well-defined semantics for queries in this context is Bouquet et al.'s~\cite{Bouquet09:QueryingWebOfData}. They define three types of query methods for conjunctive queries: a bounded method which only uses
	RDF data
referred to in queries, a direct access method which assumes an oracle that provides all RDF graphs which are ``relevant'' for a given query, and a navigational method which corresponds to
	a particular
reachability-based semantics. For
	the latter
%
	Bouquet et al.~%
define a notion of reachability that allows a query execution system to follow \emph{all} data links. As a consequence, the semantics of queries using this navigational method is equivalent
	to, what we call, $\cAll$-semantics~(cf.~Section~\ref{Subsection:Reach:Definitions}); it is the most general of our reachability-based semantics.
Bouquet et al.'s navigational query model does not support other, more restrictive notions of reachability, as is possible with our model. Furthermore, Bouquet et al.~do not discuss full SPARQL, theoretical properties of queries, or the infiniteness of the WWW.

While we focus on the
	query language
SPARQL in the context of Linked Data on the Web, the theoretical properties of SPARQL as a query language for a fixed, predefined collection of RDF data are well understood today~\cite{%
Angles08:ExpressivePowerOfSPARQL,%
Arenas11:SPARQLTutorialAtPODS,%
Perez09:SemanticsAndComplexityOfSPARQL,%
Schmidt10:FoundationsOfSPARQLOptimization}.
	Particularly interesting in our context are semantical equivalences between SPARQL expressions~\cite{Schmidt10:FoundationsOfSPARQLOptimization} because
these equivalences may also be used for optimizing
	SPARQL queries over Linked Data.


\vspace{1ex} \noindent \textbf{Structure of the paper}\,
The remainder of this paper is organized as follows. Section~\ref{Section:Preliminaries} introduces the preliminaries for our work. In Section~\ref{Section:WebOfLinkedData} we present the data model and the computation model. Sections~\ref{Section:FullWeb} and~\ref{Section:Reach} discuss the full-Web semantics and the reachability-based semantics for SPARQL, respectively.
	We
conclude the paper in Section~\ref{Section:Conclusion}.
For full technical proofs of all results in this paper we refer to \appendixname~\ref{Appendix:Proofs}.

\section{Preliminaries} \label{Section:Preliminaries}
This section provides a brief introduction of
RDF and the query language SPARQL.

We assume
pairwise disjoint, countably infinite sets $\symAllURIs$ (all HTTP scheme based URIs\footnote{For the sake of simplicity we assume in this paper that URIs are HTTP scheme based URIs. However, our models and result may be extended easily for all possible types of URIs.}), $\symAllBNodes$ (blank nodes), $\symAllLiterals$ (literals), and $\symAllVariables$ (variables, denoted by a leading
	'?'
symbol). An {\triple} $t$ is a tuple $(s,p,o) \in (\symAllURIs \cup \symAllBNodes) \times \symAllURIs \times (\symAllURIs \cup \symAllBNodes \cup \symAllLiterals)$. For any {\triple} $t = (s,p,o)$ we define $\fctTerms{t} = \lbrace s,p,o \rbrace$ and $\fctIDs{t} = \fctTerms{t} \cap \symAllURIs$. Overloading function $\fctTermsName$, we write
	$\fctTerms{G} = \bigcup_{t \in G} \fctTerms{t}$ for any (potentially infinite) set $G$ of {\triple}s.
In contrast to the usual formalization of RDF we
	allow for
infinite sets of {\triple}s which we require to study infinite Webs of Linked Data.

In this paper we focus on the core fragment of SPARQL discussed by P\'{e}rez et al.~\cite{Perez09:SemanticsAndComplexityOfSPARQL} and we adopt their formalization approach, that is, we use the algebraic syntax and the compositional set semantics introduced in~\cite{Perez09:SemanticsAndComplexityOfSPARQL}.
%
	\emph{SPARQL expression}s are defined recursively:
\begin{inparaenum}[i)]
	\item A \emph{triple pattern} $(s,p,o) \in (\symAllVariables \cup \symAllURIs) \times (\symAllVariables \cup \symAllURIs) \times (\symAllVariables \cup \symAllURIs \cup \symAllLiterals)$ is a SPARQL expression\footnote{For the sake of a more straightforward formalization we do not permit blank nodes in triple patterns. In practice, each blank node in a SPARQL query can be replaced by a new variable.}.
	\item If $P_1$ and $P_2$ are SPARQL expressions, then $(P_1 \OpAND P_2)$, $(P_1 \OpUNION P_2)$, $(P_1 \OpOPT P_2)$, and $(P_1 \OpFILTER R)$ are SPARQL expressions where $R$ is a filter condition.
\end{inparaenum}
For a
	formal
definition of
filter conditions we refer to~\cite{Perez09:SemanticsAndComplexityOfSPARQL}.
To denote the set of all variables in all triple patterns of a SPARQL expression $P$ we write $\fctVarsName(P)$.

To define the semantics of SPARQL we 
	introduce \emph{valuations}, that are, partial mappings $\mu : \symAllVariables \rightarrow \symAllURIs \cup \symAllBNodes \cup \symAllLiterals$.
%
The \emph{evaluation} of a SPARQL expression $P$ over a potentially infinite set $G$ of {\triple}s, denoted by $\evalOrigPG{P}{G}$, is a set of valuations. In contrast to the usual case, this set may be infinite in our scenario.
The evaluation function $\evalOrigPG{\cdot}{\cdot}$ is defined recursively over the structure of SPARQL expressions. Due to space limitations, we do not
	reproduce
the full formal definition of $\evalOrigPG{\cdot}{\cdot}$ here. Instead, we refer the reader to the definitions given by P\'{e}rez et al.~\cite{Perez09:SemanticsAndComplexityOfSPARQL}; even if P\'{e}rez et al.~define
	$\evalOrigPG{\cdot}{\cdot}$
for finite sets of {\triple}s, it is trivial to extend their formalism for infiniteness (cf.~%
	Appendix~\ref{Appendix:DefinitionSPARQL}).


A SPARQL expression $P$ is \emph{monotonic} if for any pair $G_1,G_2$ of (potentially infinite) sets of {\triple}s such that $G_1 \subseteq G_2$, it holds that $\evalOrigPG{P}{G_1} \subseteq \evalOrigPG{P}{G_2}$.
A SPARQL expression $P$ is \emph{satisfiable} if there exists a (potentially infinite) set $G$ of {\triple}s such that
$\evalOrigPG{P}{G} \neq \emptyset$.
It is trivial to show that any non-sat\-is\-fi\-a\-ble
expression is monotonic.

In addition to the traditional notion of satisfiability we shall need a more restrictive notion for the discussion in this paper%
: A SPARQL expression $P$ is \emph{nontrivially satisfiable} if there exists a (potentially infinite) set $G$ of {\triple}s and a valuation $\mu$ such that i)~$\mu \in \evalOrigPG{P}{G}$ and ii)~%
	$\mu$ provides a binding for at least one variable; i.e.~%
$\fctDom{\mu} \neq \emptyset$.

\begin{example} \label{Example:Preliminaries:NonTrivialSatisfiability}
	Let $P_{\mathsf{Ex}\ref{Example:Preliminaries:NonTrivialSatisfiability}} = tp$ be a SPARQL expression that consists of a single triple pattern $tp=(\symURI_1,\symURI_2,\symURI_3)$ where $\symURI_1,\symURI_2,\symURI_3 \in \symAllURIs$; hence, $tp$ actually is an {\triple}. 
	For any set $G$ of {\triple}s for which $(\symURI_1,\symURI_2,\symURI_3) \in G$ it is easy to see that the evaluation of $P_{\mathsf{Ex}\ref{Example:Preliminaries:NonTrivialSatisfiability}}$ over $G$ contains a single, empty
		valuation \removable{$\mu_\emptyset$},
	that is, $\evalOrigPG{P_{\mathsf{Ex}\ref{Example:Preliminaries:NonTrivialSatisfiability}}}{G} = \lbrace \mu_\emptyset \rbrace$ where $\fctDom{\mu_\emptyset}=\emptyset$. In contrast, for any other set $G$ of {\triple}s it holds $\evalOrigPG{P_{\mathsf{Ex}\ref{Example:Preliminaries:NonTrivialSatisfiability}}}{G} = \emptyset$. Hence, $P_{\mathsf{Ex}\ref{Example:Preliminaries:NonTrivialSatisfiability}}$ is \emph{not} nontrivially satisfiable (although it is satisfiable).
\end{example}

\section{Modeling a Web of Linked Data} \label{Section:WebOfLinkedData}
\noindent
In this section we introduce theoretical foundations which shall allow us to define and to analyze query models for Linked Data.
%
In particular, we propose a data model and introduce a computation model. For these models we assume a \emph{static view} of the Web; that is, no changes are made to the data on the Web during the execution of a query.

\subsection{Data Model} \label{Subsection:DataModel}
\noindent
%
%
We model
	the
Web of Linked Data as a potentially infinite structure of interlinked documents. Such documents, which we call Linked Data documents, or \emph{{\LDdoc}}s for short, are accessed via {\ID}s
and contain data that is represented as a set of {\triple}s.

\begin{definition} \label{Definition:WebOfData}
	Let
		$\mathcal{T}= (\symAllURIs \cup \symAllBNodes) \times \symAllURIs \times (\symAllURIs \cup \symAllBNodes \cup \symAllLiterals)$
	be the infinite set of all possible {\triple}s.
	A \definedTerm{Web of Linked Data} is a tuple $\WoD = ( D,\dataFct,\adocFct )$ where:
	\begin{itemize}
		\vspace{-2mm} 
		\item $D$
			is a (finite or countably infinite) set of
				symbols that represent {\LDdoc}s.
		\item $\dataFct$
			is a total mapping
				\,$\dataFct \!:\! D \rightarrow 2^\mathcal{T}$\,
			such that $\forall \, d \in D : \dataFct(d) \text{ is finite}$ and $\forall d_1,d_2 \in D : d_1 \neq d_2 \Rightarrow \fctTermsName\bigl(\dataFct(d_1)\bigr) \cap \symAllBNodes \neq \fctTermsName\bigl(\dataFct(d_2)\bigr) \cap \symAllBNodes$.
		\item $\adocFct$
			is a partial, surjective mapping \,$\adocFct \!: \symAllURIs \rightarrow D$.
	\end{itemize}
\end{definition}

\noindent
	While the three elements $D$, $\dataFct$, and $\adocFct$ completely define a Web of Linked Data in our model, we point out that these elements
		are abstract concepts and, thus,
	are not available to a query execution system. However, by retrieving {\LDdoc}s, such a system may gradually obtain information about the Web. Based on this information the system may (partially) materialize these three elements. In the following we discuss the three elements
and introduce additional concepts that we need to define
	queries.

We say a Web of Linked Data $\WoD = ( D,\dataFct,\adocFct )$ is \emph{finite} \IFF $D$ is finite; otherwise,
$\WoD$ is \emph{infinite}.
Our model allows for infiniteness to cover cases where Linked Data about an infinite number of identifiable entities is generated on the fly. The Linked Open Numbers project (cf.~Example~\ref{Example:LinkedOpenNumbers}) illustrates that such cases are possible in practice. Another example is the LinkedGeoData project\footnote{http://linkedgeodata.org} which provides Linked Data about any circular and rectangular area on Earth~\cite{Auer09:LinkedGeoData}. Covering these cases
enables us to model queries over such data and analyze the effects of executing such queries.

Even if a Web of Linked Data
	$\WoD = \left( D,\dataFct,\adocFct \right)$
is infinite, Definition~\ref{Definition:WebOfData} requires countability for
	$D$. We emphasize that this requirement
does not restrict us in modeling the WWW as a Web of Linked Data: In the WWW we use URIs to locate documents that contain Linked Data. Even if URIs are not limited in length, they are words over a finite alphabet. Thus, the infinite set of all possible URIs is countable, as is the set of all documents that may be retrieved
	using URIs.

The mapping $\dataFct$ associates each
	\removable{{\LDdoc}}
$d \in D$ in a Web of Linked Data $\WoD = \left( D,\dataFct,\adocFct \right)$ with a finite set of {\triple}s. In practice, these
	triples
are obtained by parsing $d$ after $d$ has been retrieved from the Web. The actual retrieval mechanism
	is not relevant for our model.
However, as prescribed by the RDF data model, Definition~\ref{Definition:WebOfData} requires that the data of each $d \in D$ uses a unique set of
	blank~nodes.

To denote the (potentially infinite but countable) set of \emph{all {\triple}s} in
	$\WoD$
we write $\fctAllDataName( W )$; i.e.~it holds:
$
	\fctAllDataName( W ) = \big\lbrace \dataFct(d) \,|\, d \in D \big\rbrace
$.

Since we use
	{\ID}s
as identifiers for entities, we say that an {\LDdoc} $d \in D$ \emph{describes} the entity identified by
{\ID} $\symURI \in \symAllURIs$ if
	there exists $(s,p,o) \in \dataFct(d)$ such that $s=\symURI$ or $o=\symURI$.
Notice,
there might be multiple {\LDdoc}s
that describe an entity identified by $\symURI$%
. However, according to the Linked Data principles, each $\symURI \in \symAllURIs$ may also serve as a reference to a specific {\LDdoc} which is considered as an authoritative source of data about the entity identified by $\symURI$. We model the relationship between {\ID}s and authoritative {\LDdoc}s by mapping $\adocFct$. Since some {\LDdoc}s may be authoritative for multiple entities, we do not require injectivity for $\adocFct$. The ``real world'' mechanism for dereferencing {\ID}s (i.e.~learning about the location of the
authoritative {\LDdoc})
is not relevant for our model.
	For each $\symURI \in \symAllURIs$ that cannot be dereferenced (i.e.~``broken links'') or that is not used in $W$ it holds $\symURI \notin \fctDom{\adocFct}$.

	A
{\ID} $\symURI \!\in\! \symAllURIs$ with $\symURI \!\in\! \fctDom{\adocFct}$ that is used in the data of an {\LDdoc} $d_1 \!\in\! D$ constitutes a \emph{data link} to the {\LDdoc} $d_2 = \adocFct(\symURI) \in D$.
	These data links form a graph structure which we call \emph{link graph}.
The vertices in such a graph represent the {\LDdoc}s of the corresponding Web of Linked Data; edges represent
	data links.

	To study the monotonicity of queries over a Web of Linked Data we require a concept of containment for such Webs. For this purpose, we introduce
the notion of an induced subweb which resembles the concept of induced subgraphs in graph theory%
	.
\begin{definition} \label{Definition:InducedSubWeb}
	Let $\WoD \!=\! ( D,\dataFct,\adocFct )$ and $\WoD' \!=\! ( D' \!, \dataFct' \!, \adocFct' )$ be Webs of Lin\-ked Data.
	$\WoD'$ is an \definedTerm{induced subweb of $\WoD$} if
	\begin{inparaenum}[i)]
		\item $D' \subseteq D$, \label{DefinitionRequirement:InducedSubWeb:D} 
		\item $\forall \, d \in D' : \dataFct'(d) = \dataFct(d)$, and \label{DefinitionRequirement:InducedSubWeb:data}
		\item
			$\forall \, \symURI \in \symAllURIs_{D'} : \adocFct'(\symURI) = \adocFct(\symURI)$ where $\symAllURIs_{D'} = \lbrace \symURI \in \symAllURIs \,|\, \adocFct(\symURI) \in D' \rbrace$.
		 \label{DefinitionRequirement:InducedSubWeb:adoc}
	\end{inparaenum}
\end{definition}

\noindent
It can be easily seen from Definition~\ref{Definition:InducedSubWeb} that specifying $D'\!$ is sufficient to unambiguously define an induced subweb $\left( D' \!, \dataFct' \!, \adocFct' \right)$ of a given Web of Linked Data.
Furthermore, it is easy to verify that
for an induced subweb $\WoD'$ of a Web of Linked Data $\WoD$ it holds $\fctAllDataName(\WoD') \subseteq \fctAllDataName(\WoD)$%
	.

In addition to the structural
	part,
our data model
	introduces
a general understanding of queries over a Web of Linked Data:
	\begin{definition} \label{Definition:LDQuery}
		Let $\mathcal{W}$ be the infinite set of all possible Webs of Linked Data
			(i.e.~all 3-tuples that correspond to Definition~\ref{Definition:WebOfData})
		and let $\Omega$ be the infinite set of all possible valuations. A \definedTerm{Linked Data query} $q$ is a total
		function \,$q \!: \mathcal{W} \rightarrow 2^\Omega$.
	\end{definition} \noindent
%
%
The notions of
	satisfiability and monotonicity
carry over naturally to Linked Data queries:
	A Linked Data query $q$ is
		\emph{satisfiable}
	if there exists a Web of Linked Data $\WoD$ such that
		$q( \WoD )$ is not empty.
	A Linked Data query $q$ is \emph{nontrivially satisfiable} if there exists a Web of Linked Data $\WoD$ and a valuation $\mu$ such that i)~$\mu \in q( \WoD )$ and ii)~$\fctDom{\mu} \neq \emptyset$.
	A Linked Data query $q$ is
		\emph{monotonic}
	if for every pair $\WoD_1$, $\WoD_2$ of Webs of Linked Data it holds: If $\WoD_1$ is an induced subweb of $\WoD_2$, then $q( \WoD_1 ) \subseteq q( \WoD_2 )$.

\subsection{Computation Model} \label{Subsection:ComputationModel}
\noindent
Usually, functions are computed over structures that are assumed to be fully (and directly) accessible. In contrast,
	we focus
on Webs of Linked Data in which accessibility is limited: To discover {\LDdoc}s and access their data we have to dereference {\ID}s, but the full set of those {\ID}s for which we may retrieve documents is unknown.
Hence, to properly analyze a query model for Webs of Linked Data we must define a model for computing functions on such a Web. This section introduces such a model.

	In the context of queries over a hypertext-centric view of
the WWW, Abiteboul and Vianu
	introduce
a specific Turing machine
called Web machine~\cite{Abiteboul00:QueriesAndComputationOnTheWebArticle}. Mendelzon and Milo
	propose
a similar machine model~\cite{Mendelzon98:FormalModelsOfWebQueries}. These machines formally capture the limited data access capabilities
	on the WWW and thus present an adequate abstraction for computations over a structure such as the WWW.
Based on these machines the authors introduce particular notions of computability for queries over the WWW. These notions are: \emph{(finitely) computable queries}, which correspond to the traditional notion of computability; and \emph{eventually computable queries} whose computation may not terminate but each element of the query result will eventually be reported during the computation.
We adopt the ideas of Abiteboul and Vianu and of Mendelzon and Milo for our work. More precisely, we adapt the idea of a Web machine to our
	scenario of a Web of Linked Data.
We call our machine a \emph{Linked Data machine} (or LD machine, for short). Based on this machine we shall define finite
and eventual computability for Linked Data queries.

Encoding (fragments of) a Web of Linked Data
	$\WoD = ( D,\dataFct,$ $\adocFct )$
on the tapes of such an LD machine is straightforward because all relevant structures, such as the sets $D$ or $\symAllURIs$, are countably infinite. In the remainder of this paper we write $\fctEnc{x}$ to denote the encoding of some element $x$ (e.g.~a single {\triple}, a set of triples, a full Web of Linked Data, a valuation, etc.). For a detailed definition of the encodings we use in this paper, we refer to
	Appendix~\ref{Appendix:Encoding}.
We
	now define
	LD machine:

\begin{definition} \label{Definition:LDMachine}
	An \definedTerm{LD machine} is a multi-tape
		Turing machine
	with five tapes and a finite set of states, including a special state called \emph{expand}.
	%
	%
		The five tapes include two, read-only input tapes:
		i)~an ordinary input tape and
		ii)~a right-infinite \emph{Web tape} which can only be accessed in the expand state;
		two work tapes:
		iii)~an ordinary, two-way infinite work tape and
		iv)~a right-infinite \emph{link traversal tape};
	and v)~a right-infinite, append-only output tape.
	Initially, the work tapes and the output tape are empty, the Web tape contains a (potentially infinite) word that encodes a Web of Linked
		Data,
	and the ordinary input tape contains an encoding of further input (if any).
	Any LD machine operates like an ordinary
		multi-tape
		Turing machine
	except when it reaches the expand state. In this case
		LD machines perform the following \emph{expand procedure}: The machine inspects
	the word currently stored on the link traversal tape. If the suffix of this word is the encoding $\fctEncName(\symURI)$ of some {\ID} $\symURI \in \symAllURIs$ and the word on the Web tape contains \,$\sharp \, \fctEncName(\symURI) \, \fctEncName( \adocFct(\symURI) ) \, \sharp$\,, then the machine appends \,$\fctEncName( \adocFct(\symURI) ) \, \sharp$\, to the (right) end of the word on the link traversal tape by copying from the Web tape; otherwise, the machine appends \,$\sharp$\, to the word on the link traversal tape.
\end{definition}

\noindent
Notice how any LD machine $M$ is limited in the way it may access a Web of Linked Data $\WoD \!=\! ( D,\dataFct,\adocFct )$ that is encoded on its Web
tape: $M$ may use the data of any particular $d \!\in\! D$ only after it performed the expand procedure using a {\ID} $\symURI \!\in\! \symAllURIs$ for which $\adocFct(\symURI) \!=\! d$. Hence, the expand procedure simulates a URI based lookup which conforms to the (typical) data access method on the WWW.
%
%
We now use
	LD machines
to adapt
	the notion of
finite
and eventual computability~\cite{Abiteboul00:QueriesAndComputationOnTheWebArticle} for Linked Data queries:
\begin{definition} \label{Definition:FinitelyComputable}
	A Linked Data query $q$  is \definedTerm{finitely computable} if there exists an LD machine which, for any Web of Linked Data $\WoD$ encoded on the Web tape, halts after a finite number of
	steps and produces a possible encoding of $q(W)$ on its output tape.
\end{definition}
\begin{definition} \label{Definition:EventuallyComputable}
	A Linked Data $q$ query is \definedTerm{eventually computable} if there exists an LD machine whose computation on any Web of Linked Data $\WoD$ encoded on the Web tape has the following two properties:
	\begin{inparaenum}[1.)]
		\item the word on the output tape at each step of the computation is a prefix of a possible encoding of $q(W)$ and \label{DefinitionRequirement:EventuallyComputable:Prefix}
		\item the encoding $\fctEncName(\mu')$ of any $\mu' \in q(\WoD)$ becomes part of the word on the output tape after a finite number of computation steps. \label{DefinitionRequirement:EventuallyComputable:All}
	\end{inparaenum}
\end{definition}

\noindent
	Any machine for a non-satisfiable query may immediately report the empty result. Thus:
\begin{fact} \label{Fact:NonSat:Computability}
	Non-satisfiable Linked Data queries are finitely computable.
\end{fact}

\enlargethispage{\baselineskip} 

\noindent
	In our analysis of SPARQL-based Linked Data queries
we shall discuss decision problems that have a
Web of Linked Data $\WoD$ as input. For
	such
problems we assume
the computation may only be performed by an LD machine with $\fctEnc{\WoD}$ on its Web~tape:
\begin{definition}
	Let $\mathcal{W}'$ be a (potentially infinite) set of Webs of Lin\-ked Data (each of which may be infinite itself);
	let $\mathcal{X}$ be an arbitrary (potentially infinite) set of finite structures;
	and let $DP \!\subseteq\! \mathcal{W}' \!\times\! \mathcal{X}$. 
	The decision problem for $DP$, that is,
	decide for any $(\WoD,X) \in \mathcal{W}' \!\times\! \mathcal{X}$ whether $(\WoD,X) \!\in\! DP$, is \definedTerm{LD machine decidable} if there exist an LD machine whose computation on any $\WoD \!\in\! \mathcal{W}'$ encoded on the Web tape and any $X \!\in\! \mathcal{X}$ encoded on the ordinary input tape, has the following property: The machine halts in an accepting state if $(\WoD,X) \!\in\! DP$; otherwise the machine halts in a rejecting state.
\end{definition}

\noindent
Obviously, any (Turing) decidable problem that does not have a Web of Linked Data as input, is also LD machine decidable because LD machines are Turing machines%
	\removable{; for these problems the corresponding set $\mathcal{W}'$ is empty}
.
\section{Full-Web Semantics} \label{Section:FullWeb}

\noindent
Based on the concepts introduced in the previous section we now define and study approaches that adapt SPARQL as a language for expressing Linked Data queries.


The first approach that we discuss is full-Web semantics where the scope of each query is
	the complete set of
Linked Data on the Web. Hereafter, we refer to SPARQL queries under this full-Web semantics as {\LDSPARQLFull} queries. The definition of these queries is straightforward and makes use of SPARQL expressions and their semantics:

\begin{definition} \label{Definition:FullWebSemantics}
	Let $P$ be a SPARQL expression.
	The \definedTerm{{\LDSPARQLFull} query that uses $P$}, denoted by $\evalFullFctP{P}$\!, is a Linked Data query that, for any Web of Linked Data $\WoD$, is defined as:
		$\evalFullPW{P}{\WoD} = \evalOrigPG{P}{\fctAllDataName(\WoD)}$.
	Each valuation $\mu \in \evalFullPW{P}{\WoD}$ is a \definedTerm{solution} for $\evalFullFctP{P}$ in $\WoD$.
\end{definition}

\noindent
	In the following we study satisfiability, monotonicity, and computability of {\LDSPARQLFull} queries and
we discuss implications of querying Webs of Linked Data that are infinite.


\subsection{Satisfiability, Nontrivial Satisfiability, Monotonicity, and Computability}
\noindent
For
	satisfiability
and monotonicity we may show
	the following dependencies.
\begin{proposition} \label{Proposition:FullWeb:SatAndMonoAsInSPARQL}
	Let $\evalFullFctP{P}$ be a {\LDSPARQLFull} query \removable{that uses SPARQL expression $P$}.
	\begin{enumerate}
		\vspace{-2mm} 
		\item $\evalFullFctP{P}$ is satisfiable \IFF $P$ is satisfiable. \label{Proposition:FullWeb:SatAndMonoAsInSPARQL:CaseSat}
		\item $\evalFullFctP{P}$ is nontrivially satisfiable \IFF $P$ is nontrivially satisfiable. \label{Proposition:FullWeb:SatAndMonoAsInSPARQL:CaseNonTrivSat}
		\item $\evalFullFctP{P}$ is monotonic \IFF $P$ is monotonic. \label{Proposition:FullWeb:SatAndMonoAsInSPARQL:CaseMono}
	\end{enumerate}
\end{proposition}


\todo{emphasize that proving Proposition~\ref{Proposition:FullWeb:SatAndMonoAsInSPARQL} is not as trivial as one might think - maybe start this by outlining the proof idea}

%

\noindent
We now discuss computability. Since all non-satisfiable {\LDSPARQLFull} queries are
finitely computable (recall Fact~\ref{Fact:NonSat:Computability}), we focus on satisfiable {\LDSPARQLFull} queries. Our first main result shows that the computability of such queries depends on their monotonicity:

\begin{theorem} \label{Theorem:FullWeb:ComputabilityDependsOnMonotonicity}
	If a satisfiable {\LDSPARQLFull} query
	is monotonic, then
		it
	is eventually computable (but not finitely computable); otherwise,
		it
	is
		not even eventually computable.
\end{theorem}

\noindent
In addition to a direct dependency between monotonicity and computability, Theorem~\ref{Theorem:FullWeb:ComputabilityDependsOnMonotonicity} shows that not any satisfiable {\LDSPARQLFull} query is finitely computable\removable{; instead, such queries are at best eventually computable}. The reason for
	this limitation
is the infiniteness of $\symAllURIs$: To (fully) compute a satisfiable {\LDSPARQLFull}
	query,
an LD machine requires access to the data of \emph{all} {\LDdoc}s in
	the queried Web of Linked Data.
Recall that, initially, the machine has no information about what {\ID} to use for performing an expand procedure with which it may access any particular
	document.
Hence, to ensure that all
	documents
have been accessed, the machine must expand all $\symURI \in \symAllURIs$. This process never terminates because $\symAllURIs$ is
infinite.
Notice, a real query system for the WWW would have a similar problem: To guarantee that such a system sees all documents, it must enumerate and lookup all (HTTP scheme) URIs.

	The computability of any Linked Data query is a general, input independent property which covers the worst case (recall, the requirements given in Definitions~\ref{Definition:FinitelyComputable} and~\ref{Definition:EventuallyComputable} must hold for any Web of Linked Data). As a consequence, in certain cases the computation of some (eventually computable) {\LDSPARQLFull} queries may still terminate:

\begin{example} \label{Example:FullWeb:DifferentComputability}
	Let $\evalFullFctP{P_{\mathsf{Ex}\ref{Example:Preliminaries:NonTrivialSatisfiability}}}$ be a monotonic {\LDSPARQLFull} query which uses the SPARQL expression $P_{\mathsf{Ex}\ref{Example:Preliminaries:NonTrivialSatisfiability}} = (\symURI_1,\symURI_2,\symURI_3)$ that we introduce in Example~\ref{Example:Preliminaries:NonTrivialSatisfiability}. Recall, $P_{\mathsf{Ex}\ref{Example:Preliminaries:NonTrivialSatisfiability}}$ is satisfiable but \emph{not} nontrivially satisfiable. The same holds for $\evalFullFctP{P_{\mathsf{Ex}\ref{Example:Preliminaries:NonTrivialSatisfiability}}}$ (cf.~Proposition~\ref{Proposition:FullWeb:SatAndMonoAsInSPARQL}).
	An LD machine for $\evalFullFctP{P_{\mathsf{Ex}\ref{Example:Preliminaries:NonTrivialSatisfiability}}}$ may take advantage of this fact: As soon as
		the machine
	discovers an {\LDdoc} which contains
	{\triple} $(\symURI_1,\symURI_2,\symURI_3)$, the machine may
	halt (after reporting $\lbrace \mu_\emptyset \rbrace$ with $\fctDom{\mu_\emptyset}=\emptyset$ as the complete query result). In this particular case the machine would satisfy the requirements for finite computability. However,
	$\evalFullFctP{P_{\mathsf{Ex}\ref{Example:Preliminaries:NonTrivialSatisfiability}}}$ is still only eventually computable because there exist Webs of Linked Data that do not contain any {\LDdoc} with
	{\triple} $(\symURI_1,\symURI_2,\symURI_3)$; any (complete) LD machine based computation of $\evalFullFctP{P_{\mathsf{Ex}\ref{Example:Preliminaries:NonTrivialSatisfiability}}}$ over such a Web cannot halt (cf.~proof of Theorem~\ref{Theorem:FullWeb:ComputabilityDependsOnMonotonicity}).
\end{example}

\noindent
The example illustrates that the computation of an eventually computable query over a particular Web of Linked Data may terminate.
%
%
%
			This observation leads us to a decision problem which we denote as \problemName{Termination(\LDSPARQLFull)}. This problem takes a Web of Linked Data $\WoD$ and a satisfiable {\LDSPARQLFull} query $\evalFullFctP{P}$ as input and asks whether an LD machine exists that computes $\evalFullPW{P}{\WoD}$ and halts.
%
For discussing this problem we note that the query in Example~\ref{Example:FullWeb:DifferentComputability} represents a special case, that is,
	\removable{{\LDSPARQLFull}} queries which are satisfiable but
not nontrivially satisfiable.
	\removable{The reason why an LD machine for such a query may halt, is the implicit knowledge that the query result is complete once the machine identified the empty valuation $\mu_\emptyset$ as a solution. Such a completeness criterion does not exist for any nontrivially satisfiable {\LDSPARQLFull} query:}

\todo{introduce the following lemma before Theorem~\ref{Theorem:FullWeb:ComputabilityDependsOnMonotonicity} - it immediately shows that (most) {\LDSPARQLFull} queries are not finitely computable - the corresponding part of the proof Theorem~\ref{Theorem:FullWeb:ComputabilityDependsOnMonotonicity} must then only discuss the special case}
\begin{lemma} \label{Lemma:FullWeb:NoTerminationWithVars}
	There is not any nontrivially satisfiable {\LDSPARQLFull} query $\evalFullFctP{P}$ for which exists an LD machine that, for any Web of Linked Data $\WoD$ encoded on the Web tape, halts after a finite number of computation steps
		and outputs an encoding of $\evalFullPW{P}{\WoD}$.
\end{lemma}


\noindent
	Lemma~\ref{Lemma:FullWeb:NoTerminationWithVars} shows that the answer to \problemName{Termination(\LDSPARQLFull)} is negative
%
	in most cases%
%
	. However,
the problem in general is undecidable (for LD machines)
	since the~in\-put
for the problem includes queries that correspond to the aforementioned special case.

\begin{theorem} \label{Theorem:FullWeb:Problem:Termination}
	\problemName{Termination(\LDSPARQLFull)} is not LD machine decidable.
\end{theorem}

\subsection{Querying an Infinite Web of Linked Data} \label{Subsection:FullWeb:Infiniteness}
\noindent
The limited computability of {\LDSPARQLFull} queries that our results in the previous section show, is a consequence of the infiniteness of $\symAllURIs$ and not of a possible infiniteness of the queried Web. We now focus on the implications of potentially infinite Webs of Linked Data for {\LDSPARQLFull} queries.
	However, we assume a \emph{finite} Web first:

\begin{proposition} \label{Proposition:FullWeb:FiniteWeb}
	{\LDSPARQLFull} queries over a finite Web of Linked Data have a finite result.
\end{proposition}

\noindent
The following example illustrates that a similarly general statement does not exist when the queried Web is infinite such as the WWW.

\begin{example} \label{Example:FullWeb:DifferentResultFiniteness}
 	Let $\WoD_\mathsf{inf} = \left( D_\mathsf{inf},\dataFct_\mathsf{inf},\adocFct_\mathsf{inf} \right)$ be an \emph{infinite} Web of Linked Data that contains {\LDdoc}s for all natural numbers (similar to the documents in Example~\ref{Example:LinkedOpenNumbers}).
	Hence, for each natural number%
		\footnote{In this paper we write $\mathbb{N}^+$ to denote the set of all natural numbers without zero.} 
	$k \in \mathbb{N}^+$, identified by $\symURI_k \in \symAllURIs$, exists an {\LDdoc} $\adocFct_\mathsf{inf}(\symURI_k) = d_k \in D_\mathsf{inf}$ such that
	$
		\dataFct_\mathsf{inf}( d_k ) =
		\big\lbrace ( \symURI_k, \mathsf{succ} , \symURI_{k+1} ) \big\rbrace
	$ 
	where $\mathsf{succ} \in \symAllURIs$ identifies the successor relation
		for $\mathbb{N}^+$.
	Furthermore, let $P_1 = ( \symURI_1, \mathsf{succ} , ?v )$ and $P_2 = ( ?x, \mathsf{succ} , ?y )$ be SPARQL expressions.
	\Todo{It can be seen easily}{use a more precise text here for a more comprehensive text} that the result of {\LDSPARQLFull} query $\evalFullFctP{P_1}$ over $\WoD_\mathsf{inf}$ is finite, whereas, $\evalFullPW{P_2}{\WoD_\mathsf{inf}}$ is infinite.
\end{example}

\noindent
The example demonstrates that some {\LDSPARQLFull} queries have a finite result over some infinite Web of Linked Data and some queries have an infinite result. Consequently, we are interested in
%
%
a decision problem \problemName{Finiteness(\LDSPARQLFull)} which asks, given a (potentially infinite) Web of Linked Data $\WoD$ and a satisfiable SPARQL expression $P$, whether
	$\evalFullPW{P}{W}$
is finite.
	Unfortunately, we cannot answer the problem in general:


\begin{theorem} \label{Theorem:FullWeb:Problem:Finiteness}
	\problemName{Finiteness(\LDSPARQLFull)} is not LD machine decidable.
\end{theorem}
\section{Reachability-Based Semantics} \label{Section:Reach}
\noindent
Our results in the previous section show that SPARQL queries under full-Web semantics have a very limited computability. As a consequence, any
	\removable{SPARQL-based}
query approach for Linked Data that uses full-Web semantics requires some ad hoc mechanism to abort query executions and, thus, has to accept incomplete query results. Depending on the abort mechanism the
	query execution
may even be nondeterministic.
	If we take these issues as an obstacle, we are interested in an alternative, well-defined semantics for SPARQL over Linked Data.
In this section we discuss a family of such semantics which we call \emph{reachability-based semantics}.
These semantics restrict the scope of queries to data that is reachable by traversing certain data links using a given set of {\ID}s as starting points.
%
%
	Hereafter, we
refer to queries under any reachability-based semantics as \emph{{\LDSPARQLReach} queries}.
In the
	remainder of this section
we
	formally introduce reachability-based semantics,
discuss theoretical properties of
	{\LDSPARQLReach} queries,
and compare
	{\LDSPARQLReach} to {\LDSPARQLFull}.

\subsection{Definition} \label{Subsection:Reach:Definitions}
\noindent
The basis of any reachability-based semantics is a notion of reachability of {\LDdoc}s. Informally, an {\LDdoc} is reachable if there exists a (specific) path in the link graph of a Web of Linked Data to the document in question; the potential starting points for such a path are {\LDdoc}s that are authoritative for a given set of entities. However, allowing for arbitrary paths might be questionable in practice because this approach would require following \emph{all} data links (recursively) for answering a query completely.
	Consequently, we introduce the notion of a reachability criterion that supports an explicit specification of what data links should be followed.

\begin{definition} \label{Definition:ReachabilityCriterion}
	Let
		$\mathcal{T}$
	be the infinite set of all possible {\triple}s
	and let $\mathcal{P}$ be the~infinite set of all possible SPARQL expressions.
	A \definedTerm{reachability criterion} $c$ is a
	(Turing) computable function
	$
		c : \mathcal{T} \times \symAllURIs \times \mathcal{P} \rightarrow \lbrace \true, \false \rbrace
	$. 
\end{definition}

\noindent
An example for a reachability criterion is $c_{\mathsf{All}}$ which corresponds to the aforementioned approach of allowing for arbitrary paths to reach {\LDdoc}s; hence, for each tuple $(t,\symURI,Q)\in \mathcal{T} \times \symAllURIs \times \mathcal{Q}$ it holds $\cAll(t,\symURI,Q) = \true$.
The complement of $c_{\mathsf{All}}$ is $c_\mathsf{None}$ which \emph{always} returns $\false$%
.
	Another example is
$c_{\mathsf{Match}}$
	which specifies the notion of reachability that we use for link traversal based query execution~\cite{Hartig11:HeuristicForQueryPlanSelection,Hartig09:QueryingTheWebOfLD}.
\begin{equation*} \label{Equation:CMatch}
	c_{\mathsf{Match}}\Bigl( t,\symURI, P \Bigr) =
	\begin{cases}
		\true & \text{if there exists a triple pattern $tp$ in $P$ and $t$ matches $tp$}, \\
		\false & \text{else}.
	\end{cases}
\end{equation*}
\noindent
where an {\triple} $t = (x_1,x_2,x_3)$ \emph{matches} a {\TP} $tp = (\tilde{x_1},\tilde{x_2},\tilde{x_3})$ if
	for all $i \in \lbrace 1,2,3 \rbrace$ holds: If $\tilde{x_i} \notin \symAllVariables$, then $\tilde{x_i} = x_i$.

We call a reachability criterion $c_1$ \emph{less restrictive than} another criterion $c_2$
	if
i)~for each
$(t,\symURI,P)\in \mathcal{T} \times \symAllURIs \times \mathcal{P}$ for which $c_2(t,\symURI,P)=\true$, also holds $c_1(t,\symURI,P)=\true$ and ii)~there exist a $(t',\symURI',P')\in \mathcal{T} \times \symAllURIs \times \mathcal{P}$ such that $c_1(t',\symURI',P')=\true$ but $c_2(t',\symURI',P')=\false$.
It can be
seen that $c_{\mathsf{All}}$ is the least restrictive
criterion, whereas $c_{\mathsf{None}}$ is the most restrictive criterion.
%
	We now
define reachability of {\LDdoc}s:

\begin{definition} \label{Definition:QualifiedReachability}
	Let $S \subset \symAllURIs$ be a finite set of seed {\ID}s;
	let $c$ be a reachability criterion;
	let $P$ be a SPARQL expression; and
	let $\WoD=(D,\dataFct,\adocFct)$ be a Web of Linked Data.
	An {\LDdoc} $d \in D$ \definedTerm{is $(c,P)$-reachable from $S$ in $\WoD$} if either
	\begin{enumerate}
		\vspace{-1mm} 
		\item
			there exists a {\ID} $\symURI \in S$ such that $\adocFct(\symURI) = d$; or%
			\label{DefinitionCase:QualifiedReachability:IndBegin}
		\item
			there exist
			$d' \in D$,
			$t \in \dataFct(d')$, and
			$\symURI \in \fctIDsName(t)$ such that i)~$d'$ is $(c,P)$-reachable from $S$ in $\WoD$,
				ii)~$\adocFct(\symURI)=d$, and iii)~$c(t,\symURI,P)=\true$.%
			 \label{DefinitionCase:QualifiedReachability:IndStep}
	\end{enumerate}
\end{definition}

\noindent
%
%
Based on reachability of {\LDdoc}s we define reachable parts of a Web of Linked Data. Such a part is an induced subweb covering all reachable {\LDdoc}s.
Formally:

\begin{definition} \label{Definition:ReachablePart}
	Let $S \subset \symAllURIs$ be a finite set of
	{\ID}s;
	let $c$ be a reachability criterion;
	let $P$ be a SPARQL expression; and
	let $\WoD = (D,\dataFct,\adocFct)$ be a Web of Linked Data.
	The \definedTerm{$(S,c,P)$-reachable part of $\WoD$}, denoted by $\ReachPartScPW{S}{c}{P}{\WoD}$, is an induced subweb
		$( \Reach{D} ,$ $\Reach{\dataFct} ,\Reach{\adocFct} )$
	of $\WoD$ such that
	$
		\Reach{D}=\big\lbrace d \in D \, | \, d \text{ is $(c,P)$-reachable from $S$ in $\WoD$} \big\rbrace
	$.
\end{definition}

\noindent
We now use the concept of reachable parts
%
	to define
%
	{\LDSPARQLReach} queries.

\begin{definition} \label{Definition:ReachSemantics}
	Let $S \subset \symAllURIs$ be a finite set of
	{\ID}s;
	let $c$ be a reachability criterion; and
	let $P$ be a SPARQL expression.
	The \definedTerm{{\LDSPARQLReach} query that uses $P$, $S$, and $c$}, denoted by $\evalReachFctPSc{P}{S}{c}$, is a Linked Data query that, for any Web of Linked Data $\WoD$, is defined as $\evalReachFctPSc{P}{S}{c}(\WoD) = \evalOrigPG{P}{\fctAllDataName(\ReachPartScPW{S}{c}{P}{\WoD})}$ (where $\ReachPartScPW{S}{c}{P}{\WoD}$ is the $(S,c,P)$-reach\-able part of $\WoD$).
\end{definition}

\noindent
As can be seen from Definition~\ref{Definition:ReachSemantics}, our notion of {\LDSPARQLReach} consists of a family of (reachability-based) query semantics, each of which is characterized by a certain reachability criterion. Therefore, we refer to {\LDSPARQLReach} queries for which we use a particular reachability criterion $c$ as {\LDSPARQLReach} queries \emph{under $c$-semantics}.

Definition~\ref{Definition:ReachSemantics} also shows that query results depend on the given set $S \subset \symAllURIs$ of seed {\ID}s.
	It is easy to see that any {\LDSPARQLReach} query which uses an empty set of seed {\ID}s is not satisfiable and, thus, monotonic and finitely computable.
We therefore consider only
	nonempty
sets of seed {\ID}s in the remainder of this paper.

\enlargethispage{\baselineskip} 

\subsection{Completeness and Infiniteness}
\noindent
	Definition~\ref{Definition:ReachSemantics} defines precisely what the sound and complete result of any {\LDSPARQLReach} query $\evalReachFctPSc{P}{S}{c}$ over any Web of Linked Data $\WoD$ is. However, in contrast to {\LDSPARQLFull}, it is not guaranteed that such a (complete) {\LDSPARQLReach} result is complete w.r.t.~all data on $\WoD$. This difference 
can be attributed to the fact that the corresponding $(S,c,P)$-reachable part of $\WoD$ may not cover $\WoD$ as a whole.
	\removable{We emphasize that such an incomplete coverage is even possible for the reachability criterion $\cAll$ because the link graph of $\WoD$ may not be connected; therefore, $\cAll$-semantics differs from full-Web semantics.}
The following
	result
relates {\LDSPARQLReach} queries to their {\LDSPARQLFull} counterparts.

\begin{proposition} \label{Proposition:Reach:ComparisonToFullWeb}
	Let $\evalReachFctPSc{P}{S}{c}$ be a {\LDSPARQLReach} query; let $\evalFullFctP{P}$ be the {\LDSPARQLFull} query that uses the same SPARQL expression as $\evalReachFctPSc{P}{S}{c}$; let $\WoD$ be a Web of Linked Data.
	It holds:
	\begin{enumerate}
		\vspace{-1mm} 
		\item If $\evalFullFctP{P}$ is monotonic, then $\evalReachPScW{P}{S}{c}{\WoD} \subseteq \evalFullPW{P}{\WoD}$. \label{Proposition:Reach:ComparisonToFullWeb:Case1}
		\item $\evalReachPScW{P}{S}{c}{\WoD} = \evalFullPW{P}{ \ReachPartScPW{S}{c}{P}{\WoD} }$. ~ (recall, $\ReachPartScPW{S}{c}{P}{\WoD}$ is the $(S,c,P)$-reachable part of $\WoD$) \label{Proposition:Reach:ComparisonToFullWeb:Case2}
	\end{enumerate}
\end{proposition}

\noindent
Since
any {\LDSPARQLFull} query over a finite Web of Linked Data
	has a finite result
(cf. Proposition~\ref{Proposition:FullWeb:FiniteWeb}),
we use Proposition~\ref{Proposition:Reach:ComparisonToFullWeb}, case~\ref{Proposition:Reach:ComparisonToFullWeb:Case2}, to show the same for
	{\LDSPARQLReach}:

\begin{proposition} \label{Proposition:Reach:FiniteWeb}
	The result of any {\LDSPARQLReach} query $\evalReachFctPSc{P}{S}{c}$ over a finite Web of Linked Data $\WoD$ is finite; so is the $(S,c,P)$-reach\-able part of $\WoD$.
\end{proposition}

\noindent
For the case of an \emph{infinite} Web of Linked Data the results of {\LDSPARQLReach} queries may be either finite or infinite.
		\removable{In Example~\ref{Example:FullWeb:DifferentResultFiniteness} we found the same heterogeneity for {\LDSPARQLFull}.} However, for {\LDSPARQLReach} we may identify
	the following dependencies.

\begin{proposition} \label{Proposition:Reach:InfiniteneWebFindings}
	Let $S \subset \symAllURIs$ be a
		finite,
	nonempty set of
	{\ID}s;
	let $c$ and $c'$ be reachability criteria; and
	let $P$ be a SPARQL expression. 
		Let
	$\WoD$ be an \emph{infinite} Web of Linked Data.
	\begin{enumerate}
		\vspace{-2mm} 
		\item $\ReachPartScPW{S}{\cNone}{P}{\WoD}$ is always finite; so is $\evalReachPScW{P}{S}{\cNone}{\WoD}$. \label{Proposition:Reach:InfiniteneWebFindings:Case1}
		\item If $\ReachPartScPW{S}{c}{P}{\WoD}$ is finite, then $\evalReachPScW{P}{S}{c}{\WoD}$ is finite. \label{Proposition:Reach:InfiniteneWebFindings:Case3}
		\item If $\evalReachPScW{P}{S}{c}{\WoD}$ is infinite, then $\ReachPartScPW{S}{c}{P}{\WoD}$ is infinite. \label{Proposition:Reach:InfiniteneWebFindings:Case4}
		\item If $c$ is less restrictive than $c'$ and $\ReachPartScPW{S}{c}{P}{\WoD}$ is finite, then $\ReachPartScPW{S}{c'}{P}{\WoD}$ is finite. \label{Proposition:Reach:InfiniteneWebFindings:Case5}
		\item If $c'$ is less restrictive than $c$ and $\ReachPartScPW{S}{c}{P}{\WoD}$ is infinite, then $\ReachPartScPW{S}{c'}{P}{\WoD}$ is infinite. \label{Proposition:Reach:InfiniteneWebFindings:Case6}
	\end{enumerate}
\end{proposition}

\noindent
Proposition~\ref{Proposition:Reach:InfiniteneWebFindings} provides valuable insight into the dependencies between reachability criteria, the (in)finiteness of reachable parts of an infinite Web, and the (in)finiteness of query results. In practice, however, we are primarily interested in answering two decision problems: \problemName{FinitenessReachablePart} and \problemName{Finiteness(\LDSPARQLReach)}. While the latter
	\removable{problem}
is the {\LDSPARQLReach} equivalent to \problemName{Finiteness(\LDSPARQLFull)} (cf.~Section~\ref{Subsection:FullWeb:Infiniteness}), the former has the same input as \problemName{Finiteness(\LDSPARQLReach)} (that is, a Web of Linked Data and a {\LDSPARQLReach} query) and asks whether the corresponding reachable part of the given Web is finite.
	Both
problems are undecidable in our context:

%
%

\begin{theorem} \label{Theorem:Reach:Problem:Finiteness}
	\problemName{FinitenessReachablePart} and \problemName{Finiteness(\LDSPARQLReach)} are
\\
	not LD machine decidable.
\end{theorem}

\subsection{Satisfiability, Nontrivial Satisfiability, Monotonicity, and Com\-put\-abil\-i\-ty}
\noindent
We now investigate satisfiability, nontrivial satisfiability, monotonicity, and computability of {\LDSPARQLReach} queries. First, we identify the following dependencies.


\begin{proposition} \label{Proposition:Reach:SatAndMonoAsInSPARQL}
	Let $\evalReachFctPSc{P}{S}{c}$ be a {\LDSPARQLReach} query that uses a nonempty
		$S \subset \symAllURIs$.
	\begin{enumerate}
		\vspace{-2mm} 
		\item $\evalReachFctPSc{P}{S}{c}$ is satisfiable \IFF $P$ is satisfiable. \label{Proposition:Reach:SatAndMonoAsInSPARQL:CaseSat}
		\item $\evalReachFctPSc{P}{S}{c}$ is nontrivially satisfiable \IFF $P$ is nontrivially satisfiable. \label{Proposition:Reach:SatAndMonoAsInSPARQL:CaseNonTrivSat}
		\item $\evalReachFctPSc{P}{S}{c}$ is monotonic if $P$ is monotonic. \label{Proposition:Reach:SatAndMonoAsInSPARQL:CaseMono}
	\end{enumerate}
\end{proposition}

\noindent
	Proposition~\ref{Proposition:Reach:SatAndMonoAsInSPARQL}
reveals a first major difference between {\LDSPARQLReach} and {\LDSPARQLFull}: The statement about monotonicity in
	that proposition
%
is only a material
	conditional, whereas it is a biconditional
in the case of
	{\LDSPARQLFull} (cf.~Proposition~\ref{Proposition:FullWeb:SatAndMonoAsInSPARQL}).
The reason for this disparity
	are
{\LDSPARQLReach} queries for which monotonicity is independent of
	the corresponding SPARQL expression.
The following proposition identifies such a case.

\begin{proposition} \label{Proposition:Reach:MonotonicityCNone}
	Any {\LDSPARQLReach} query $\evalReachFctPSc{P}{S}{\cNone}$
	is monotonic if $\left|S\right|=1$.
\end{proposition}

\noindent
Before we may come back to the aforementioned disparity, we focus on the computability of {\LDSPARQLReach} queries. We first show the following,
	noteworthy
result.

\begin{lemma} \label{Lemma:Reach:TerminationCriterion}
	Let $\evalReachFctPSc{P}{S}{c}$ be a {\LDSPARQLReach} query that is nontrivially satisfiable.
	There exists an LD machine that computes $\evalReachFctPSc{P}{S}{c}$ over any (potentially infinite) Web of Linked Data $\WoD$ and that halts
		after a finite number of computation steps with an encoding of $\evalReachPScW{P}{S}{c}{\WoD}$ on its output tape
	\IFF the $(S,c,P)$-reachable part of $\WoD$ is finite.
\end{lemma}

\noindent
The importance of Lemma~\ref{Lemma:Reach:TerminationCriterion} lies in showing that some computations of nontrivially satisfiable {\LDSPARQLReach} queries may terminate. This possibility presents another major difference between {\LDSPARQLReach} and {\LDSPARQLFull} (recall Lemma~\ref{Lemma:FullWeb:NoTerminationWithVars} which shows that any possible computation of nontrivially satisfiable {\LDSPARQLFull} queries never terminates).
Based on Lemma~\ref{Lemma:Reach:TerminationCriterion} we may even show that a particular class of satisfiable {\LDSPARQLReach} queries are finitely
	computable.
This class comprises all queries that use a reachability criterion which ensures the finiteness of reachable parts of any queried Web of Linked Data. We define this property
	\removable{of reachability criteria}
as follows:

\begin{definition} \label{Definition:Reach:EnsuresFiniteness}
	A reachability criterion $c$ \definedTerm{ensures finiteness} if for any Web of Linked Data $\WoD$, any (finite) set $S \subset \symAllURIs$ of seed {\ID}s, and any SPARQL expression $P$, the $(S,c,P)$-reachable part of $\WoD$ is finite.
\end{definition}

\noindent
We may now show the aforementioned result:
\begin{proposition} \label{Proposition:Reach:ComputabilityFiniteness}
	Let $c$ be a reachability criterion that ensures finiteness.
	{\LDSPARQLReach} queries under $c$-semantics are finitely computable.
\end{proposition}

\noindent
While it remains an open question whether the property to ensure finiteness is decidable for all reachability criteria, it is easy to verify the property for
	criteria which always only accept a given, constant set of data links. For a formal discussion of such criteria, which we call \emph{constant reachability criteria}, we refer to
		Appendix~\ref{Appendix:ConstReachCriteria}.
	$\cNone$ is a special case of these criteria; Proposition~\ref{Proposition:Reach:InfiniteneWebFindings}, case~\ref{Proposition:Reach:InfiniteneWebFindings:Case1}, verifies that $\cNone$ ensures finiteness.

Notice, for any reachability criterion $c$ that ensures finiteness,
the computability of {\LDSPARQLReach} queries under $c$-semantics does not depend on the monotonicity of these queries.
This independence is another difference to {\LDSPARQLFull} queries (recall Theorem~\ref{Theorem:FullWeb:ComputabilityDependsOnMonotonicity}). However,
	for any other reachability
criterion (\Todo{including $\cMatch$ and $\cAll$}{This claim must be proved, which should be easy using examples}), we have
	a similar
dependency between monotonicity and computability of (satisfiable) {\LDSPARQLReach} queries, that we have for {\LDSPARQLFull} queries (recall Theorem~\ref{Theorem:FullWeb:ComputabilityDependsOnMonotonicity}):


\begin{theorem} \label{Theorem:Reach:ComputabilityDependsOnMonotonicity}
	Let $c_{n\!f}$ be a reachability criterion that does not ensure finiteness. If a satisfiable {\LDSPARQLReach} query $\evalReachFctPSc{P}{S}{c_{n\!f}}$ (under $c_{n\!f}$-semantics) is monotonic, then $\evalReachFctPSc{P}{S}{c_{n\!f}}$ is either finitely computable or eventually computable;
	otherwise, $\evalReachFctPSc{P}{S}{c_{n\!f}}$
		may not even be eventually computable.
\end{theorem}

\noindent
By comparing Theorems~\ref{Theorem:FullWeb:ComputabilityDependsOnMonotonicity} and~\ref{Theorem:Reach:ComputabilityDependsOnMonotonicity} we notice that {\LDSPARQLFull} queries and {\LDSPARQLReach} queries (that use a reachability criterion which does not ensure finiteness) feature
	a similarly
limited computability. However, the reasons for
	both of these
results differ significantly:
	\removable{In the case of}
{\LDSPARQLFull} the limitation follows from the infiniteness of $\symAllURIs$, whereas, for {\LDSPARQLReach} the limitation is a consequence of the possibility to query an infinitely large Web of Linked Data.

However, even if the computability of many {\LDSPARQLReach} queries is as limited as that of their {\LDSPARQLFull} counterparts, there is another major difference: Lemma~\ref{Lemma:Reach:TerminationCriterion} shows
	that
for (nontrivially satisfiable) {\LDSPARQLReach} queries which are not finitely computable, the computation over some Webs of Linked Data may still terminate; this includes all finite Webs (cf.~Proposition~\ref{Proposition:Reach:FiniteWeb}) but also some infinite Webs (cf.~proof of Lemma~\ref{Lemma:Reach:TerminationCriterion}). Such a possibility does not exist for nontrivially satisfiable {\LDSPARQLFull} queries (cf.~Lemma~\ref{Lemma:FullWeb:NoTerminationWithVars}).
Nonetheless, the termination problem for {\LDSPARQLReach} is undecidable in our context.


\begin{theorem} \label{Theorem:Reach:Problem:Termination}
	\problemName{Termination(\LDSPARQLReach)} is not LD machine decidable.
\end{theorem}

\noindent
We now come back to the impossibility for showing that {\LDSPARQLReach} queries (with a nonempty set of seed {\ID}s) are monotonic \emph{only if} their SPARQL expression is monotonic. Recall, for some {\LDSPARQLReach} queries monotonicity is irrelevant for identifying the computability (cf.~Proposition~\ref{Proposition:Reach:ComputabilityFiniteness}). We are primarily interested in the monotonicity of all other (satisfiable) {\LDSPARQLReach} queries because for those queries computability depends on monotonicity
	as we show in Theorem~\ref{Theorem:Reach:ComputabilityDependsOnMonotonicity}.
Remarkably,
	for those queries it is possible to show the required dependency that was missing from Proposition~\ref{Proposition:Reach:SatAndMonoAsInSPARQL}:

\begin{proposition} \label{Proposition:Reach:Mono2AsInSPARQL}
	Let $\evalReachFctPSc{P}{S}{c_{n\!f}}$ be a {\LDSPARQLReach} query that uses a finite, nonempty
		$S \subset \symAllURIs$
	and a reachability criterion $c_{n\!f}$ which does not ensure finiteness.
	$\evalReachFctPSc{P}{S}{c_{n\!f}}$ is monotonic only if $P$ is monotonic.
\end{proposition}

\todo{add a few words about the proof}

\vspace{-3mm} 

\section{Conclusions} \label{Section:Conclusion}
Our investigation
of SPARQL as a
language for Linked Data queries reveals the following main results.
	Some special cases aside,
the computability of queries under any of the studied semantics is limited and no guarantee for termination can be given. For reachability-based semantics it is at least possible that some of the (non-special case) query computations terminate; although, in general it is undecidable which.
As a
	consequence,
any SPARQL-based query system for Linked Data
	\removable{on the Web}
must be prepared for query executions that discover an infinite amount of data and that do not terminate.

	Our results
also show that --for reachability-based semantics-- the aforementioned issues must be attributed to the possibility for infiniteness in the queried Web (which is a result of data generating servers). Therefore, it seems worthwhile to study approaches for detecting whether the execution of a {\LDSPARQLReach} query traverses an infinite path in the queried Web.
However, the mentioned issues may also be addressed by
	another, \removable{alternative}
well-defined semantics
that restricts the scope of queries even further (or differently) than our reachability-based semantics. It remains an open question how such an alternative may still allow for queries that tap the full potential of the Web.

	We
also show that computability depends on satisfiability and monotonicity and that for (almost all) SPARQL-based Linked Data queries, these two properties directly correspond to the same property for the used SPARQL expression. While
	Arenas and P{\'e}rez show
that the core fragment of SPARQL without \!$\OpOPT$\! is monotonic~\cite{Arenas11:SPARQLTutorialAtPODS}, it requires further work to identify (non-)satisfiable and (non-)monotonic fragments
	and, thus,
enable an explicit classification of SPARQL-based Linked Data queries w.r.t.~computability.


\bibliographystyle{abbrv}
\bibliography{main}

\newpage
\appendix
\section*{\appendixname}
\noindent
The Appendix is organized as follows:
\begin{itemize}
  \item Appendix~\ref{Appendix:Encoding} describes how we encode relevant structures (such as a Web of Linked Data and a valuation) on the tapes of Turing machines.
  \item Appendix~\ref{Appendix:DefinitionSPARQL} provides a formal definition of SPARQL.
  \item Appendix~\ref{Appendix:Proofs} contains the full technical proofs for all results in the paper.
  \item Appendix~\ref{Appendix:ConstReachCriteria} provides a formal discussion of constant reachability criteria.
\end{itemize}

\section{Encoding} \label{Appendix:Encoding}

To encode Webs of Linked Data and query results on the tapes of a Turing machine we assume the existence of a total order $\prec_\symAllURIs$, $\prec_\symAllBNodes$, $\prec_\symAllLiterals$, and $\prec_\symAllVariables$ for the {\ID}s in $\symAllURIs$, the blank nodes in $\symAllBNodes$, the constants in $\symAllLiterals$, and the variables in $\symAllVariables$, respectively; in all three cases $\prec_x$ could simply be the lexicographic order of corresponding string representations. Furthermore, we assume a total order $\prec_t$ for {\triple}s that is based on the aforementioned orders.

For each $\symURI \in \symAllURIs$, $c \in \symAllLiterals$, and $v \in \symAllVariables$ let $\fctEncName(\symURI)$, $\fctEncName(c)$, and $\fctEncName(v)$ be the binary representation of $\symURI$, $c$, and $v$, respectively. The encoding of a {\triple} $t=(s,p,o)$, denoted by $\fctEncName(t)$, is a word
\,$
	\langle \, \fctEncName(s) \, , \, \fctEncName(p) \, , \, \fctEncName(o) \, \rangle
$. 

The encoding of a finite set of {\triple}s $T = \lbrace t_1, ... \, , t_n \rbrace$, denoted by $\fctEncName(T)$, is a word
\,$
	\langle\!\langle \, \fctEncName(t_1) \, , \, \fctEncName(t_2) \, , \, ... \, , \, \fctEncName(t_n) \, \rangle\!\rangle
$\, 
where the $\fctEncName(t_i)$ are ordered as follows: For each two {\triple}s $t_x,t_y \in T$, $\fctEncName(t_x)$ occurs before $\fctEncName(t_y)$ in $\fctEncName(T)$ if $t_x \prec_t t_y$.

For a Web of Linked Data $\WoD = (D,\dataFct,\adocFct)$, the encoding of {\LDdoc} $d \in D$, denoted by $\fctEncName(d)$, is the word $\fctEncName( \dataFct(d) )$. The encoding of $\WoD$ itself, denoted by $\fctEncName(\WoD)$, is a word
\begin{equation*}
	\sharp \, \fctEncName(\symURI_1) \, \fctEncName( \adocFct(\symURI_1) ) \, \sharp \, ... \, \sharp \, \fctEncName(\symURI_i) \, \fctEncName( \adocFct(\symURI_i) ) \, \sharp \, ...
\end{equation*}
where $\symURI_1,...,\symURI_i,...$ is the (potentially infinite but countable) list of {\ID}s in $\fctDom{\adocFct}$, ordered according to $\prec_\symAllURIs$.

The encoding of a valuation $\mu$ with $dom(\mu) = \lbrace v_1 ,...\,, v_n \rbrace$, denoted by $\fctEncName(\mu)$, is a word
\begin{equation*}
	\langle\!\langle \, \fctEncName(v_1) \rightarrow \fctEncName\bigl( \mu(v_1) \bigr) \, , \, ... \, , \, \fctEncName(v_n) \rightarrow \fctEncName\bigl( \mu(v_n) \bigr) \, \rangle\!\rangle
\end{equation*}
where the $\fctEncName(\mu(v_i))$ are ordered as follows: For each two variables $v_x,v_y \in dom(\mu)$, $\fctEncName(\mu(v_x))$ occurs before $\fctEncName(\mu(v_y))$ in $\fctEncName(\mu)$ if $v_x \!\prec_V\! v_y$.

Finally, the encoding of a (potentially infinite) set of valuations $\Omega = \lbrace \mu_1 , \mu_2 , ... \rbrace$, denoted by $\fctEncName(\Omega)$, is a word
\,$
	\fctEncName( \mu_1 )  \, \fctEncName( \mu_2 ) \, ...
$\, 
where the $\fctEncName( \mu_i )$ \Todo{may occur in any order}{that might be a problem}.
\section{Formal Definition of SPARQL} \label{Appendix:DefinitionSPARQL}

\todo{introduce: $\fctTerms{P}$}

\noindent
A SPARQL \emph{filter condition} is defined recursively as follows:
\begin{inparaenum}[i)]
	\item If $?x,?y \in \symAllVariables$ and $c \in (\symAllURIs \cup \symAllLiterals)$ then $?x=c$, $?x = \, ?y$, and $\fctBound{?x}$ are filter conditions;
	\item If $R_1$ and $R_2$ are filter conditions then $(\neg R_1)$, $(R_1 \land R_2)$, and $(R_1 \lor R_2)$ are filter conditions.
\end{inparaenum}

\begin{definition} \label{Definition:ExpressionSPARQL}
	A \definedTerm{SPARQL expression} is defined recursively as follows:
	\begin{enumerate}
		\item A tuple $(s,p,o) \in (\symAllVariables \cup \symAllURIs) \times (\symAllVariables \cup \symAllURIs) \times (\symAllVariables \cup \symAllURIs \cup \symAllLiterals)$ is a SPARQL expression. We call such a tuple a \definedTerm{triple pattern}.
		\item If $P_1$ and $P_2$ are SPARQL expressions, then $(P_1 \OpAND P_2)$, $(P_1 \OpUNION P_2)$, and
\par
			$(P_1 \OpOPT P_2)$ are SPARQL expressions.
		\item If $P'$ is a SPARQL expression and $R$ is a filter condition, then $(P' \OpFILTER R)$ is a SPARQL expression.
	\end{enumerate}
\end{definition}

Let $\mu$ be a valuation and let $R$ be a filter condition. We say $\mu$ satisfies $R$ iff either
\begin{inparaenum} [i)]
	\item $R$ is $?x=c$, $?x \in \fctDom{\mu}$ and $\mu(?x)=c$;
	\item $R$ is $?x = \, ?y$, $?x,?y \in \fctDom{\mu}$ and $\mu(?x)=\mu(?y)$;
	\item $R$ is $\fctBound{?x}$ and $?x \in \fctDom{\mu}$;
	\item $R$ is $(\neg R_1)$ and $\mu$ does not satisfy $R_1$;
	\item $R$ is $(R_1 \land R_2)$ and $\mu$ satisfies $R_1$ and $R_2$; or
	\item $R$ is $(R_1 \lor R_2)$ and $\mu$ satisfies $R_1$ or $R_2$.
\end{inparaenum}


Let $\Omega_l$, $\Omega_r$ and $\Omega$ be (potentially infinite but countable) sets of valuations; let $R$ be a filter condition. The binary operations \emph{join}, \emph{union}, \emph{difference}, and \emph{left outer-join} between $\Omega_l$ and $\Omega_r$ are defined as follows:
\begin{align*}
	\Omega_l \Join \Omega_r & = \lbrace \mu_l \cup \mu_r \,|\, \mu_l \in \Omega_l \text{ and } \mu_r \in \Omega_r \text{ and } \mu_l \sim \mu_r \rbrace \\
	\Omega_l \cup \Omega_r & = \lbrace \mu \,|\, \mu \in \Omega_l \text{ or } \mu \in \Omega_r \rbrace \\
	\Omega_l \setminus \Omega_r & = \lbrace \mu_l \in \Omega_l \,|\, \forall \mu_r \in \Omega_r : \mu_l \not\sim \mu_r \rbrace \\
	\Omega_l \LJoin \Omega_r &= \left( \Omega_l \Join \Omega_r \right) \cup \left( \Omega_l \setminus \Omega_r \right) \\
	\sigma_R ( \Omega ) &= \lbrace \mu \in \Omega \,|\, \mu \text{ satisfies } R \rbrace
\end{align*}
%

\begin{definition} \label{Definition:EvaluationSPARQL}
	Let $P$ be a SPARQL expression and let $G$ be a (potentially infinite but countable)
		set of RDF triples.
%
	The \definedTerm{evaluation of $P$ over $G$}, denoted by $\evalOrigPG{P}{G}$, is defined recursively as follows:
	\begin{enumerate}
		\item If $P$ is a triple pattern $tp$, then
		\begin{align*}
			\evalOrigPG{P}{G} = \lbrace \mu \,|\, & \mu \text{ is a valuation with } \fctDom{\mu} = \fctVarsName(tp) \\ & \text{ and } \mu[tp] \in G \rbrace
		\end{align*}
		\item If $P$ is $(P_1 \OpAND P_2)$, then $\evalOrigPG{P}{G} = \evalOrigPG{P_1}{G} \Join \evalOrigPG{P_2}{G}$.
		\item If $P$ is $(P_1 \OpUNION P_2)$, then $\evalOrigPG{P}{G} = \evalOrigPG{P_1}{G} \cup \evalOrigPG{P_2}{G}$.
		\item If $P$ is $(P_1 \OpOPT P_2)$, then $\evalOrigPG{P}{G} = \evalOrigPG{P_1}{G} \LJoin \evalOrigPG{P_2}{G}$.
		\item If $P$ is $(P' \OpFILTER R)$, then $\evalOrigPG{P}{G} = \sigma_R\bigl( \evalOrigPG{P'}{G} \bigr)$.
	\end{enumerate}
	Each valuation $\mu \in \evalOrigPG{P}{G}$ is called a \definedTerm{solution} for $P$ in $G$.
\end{definition}

\section{Proofs} \label{Appendix:Proofs}

\subsection{Additional References for the Proofs}
\vspace{1ex}
[Pap93] C.~H.~Papadimitriou. \textit{Computational Complexity}. Addison Wesley, 1993.

\subsection{Proof of Proposition~\ref{Proposition:FullWeb:SatAndMonoAsInSPARQL}, Case~\ref{Proposition:FullWeb:SatAndMonoAsInSPARQL:CaseSat}} \label{Proof:Proposition:FullWeb:SatAndMonoAsInSPARQL:CaseSat}
\noindent
For this proof we introduce a notion of lineage for valuations. Informally, the lineage of a valuation $\mu$ is the set of all {\triple}s that are required to construct $\mu$. Formally:
\begin{definition} \label{Definition:Lineage}
	Let $P$ be a SPARQL expression and $G$ be a (potentially infinite) set of {\triple}s such that $\evalOrigPG{P}{G} \neq \emptyset$. For each $\mu \in \evalOrigPG{P}{G}$ we define the \definedTerm{$(P,G)$-lineage of $\mu$}, denoted by $\mathrm{lin}^{P,G}(\mu)$, recursively as follows:
	\begin{enumerate}
		\item If $P$ is a triple pattern $tp$,
			then $\mathrm{lin}^{P,G}(\mu) = \big\lbrace \mu[tp] \big\rbrace$.
		\item If $P$ is $(P_1 \OpAND P_2)$,
			then
			\begin{equation*}
				\mathrm{lin}^{P,G}(\mu) = \mathrm{lin}^{P_1,G}(\mu_1) \cup \mathrm{lin}^{P_2,G}(\mu_2)
			\end{equation*}
			where $\mu_1 \in \evalOrigPG{P_1}{G}$ and $\mu_2 \in \evalOrigPG{P_2}{G}$ such that $\mu_1 \sim \mu_2$ and $\mu = \mu_1 \cup \mu_2$. Notice, $\mu_1$ and $\mu_2$ must exist because $\mu \in \evalOrigPG{P}{G}$.
		\item If $P$ is $(P_1 \OpUNION P_2)$,
			then
			\begin{equation*}
				\mathrm{lin}^{P,G}(\mu) = \begin{cases}
					\mathrm{lin}^{P_1,G}(\mu_1) & \text{ if } \exists \, \mu_1 \in \evalOrigPG{P_1}{G} : \mu_1 = \mu , \\
					\mathrm{lin}^{P_2,G}(\mu_2) & \text{ if } \exists \, \mu_2 \in \evalOrigPG{P_2}{G} : \mu_2 = \mu .
				\end{cases}
			\end{equation*}
			Notice, if $\mu_1$ does not exist then $\mu_2$ must exist because $\mu \in \evalOrigPG{P}{G}$.
		\item If $P$ is $(P_1 \OpOPT P_2)$,
			then
			\begin{equation*}
				\mathrm{lin}^{P,G}(\mu) = \begin{cases}
					\mathrm{lin}^{P_1,G}(\mu_1) \cup \mathrm{lin}^{P_2,G}(\mu_2) & \text{ if } \exists \, (\mu_1,\mu_2) \in \evalOrigPG{P_1}{G} \times \evalOrigPG{P_2}{G} : \bigl( \mu_1 \sim \mu_2 \land \mu = \mu_1 \cup \mu_2 \bigr), \\
					\mathrm{lin}^{P_1,G}(\mu') & \text{ if } \exists \, \mu' \in \evalOrigPG{P_1}{G} : \bigl( \mu' = \mu \, \land \, \forall \, \mu^* \in \evalOrigPG{P_2}{G} : \mu^* \not\sim \mu' \bigr) .
				\end{cases}
			\end{equation*}
			Notice, either $\mu_1$ and $\mu_2$ or $\mu'$ must exist because $\mu \in \evalOrigPG{P}{G}$.
		\item If $P$ is $(P' \OpFILTER R)$,
			then $\mathrm{lin}^{P,G}(\mu) = \mathrm{lin}^{P'\!,G}(\mu')$ where $\mu' \in \evalOrigPG{P'}{G}$ such that $\mu = \mu'$. Notice, $\mu'$ must exist because $\mu \in \evalOrigPG{P}{G}$.
	\end{enumerate}
\end{definition}

\noindent
For any SPARQL expression $P$, any (potentially infinite) set $G$ of {\triple}s, and any valuation $\mu \in \evalOrigPG{P}{G}$ it can be easily seen that i)~$G' = \mathrm{lin}^{P,G}(\mu)$ is finite and ii)~$\mu \in \evalOrigPG{P}{G'}$.
We now prove Proposition~\ref{Proposition:FullWeb:SatAndMonoAsInSPARQL}, case~\ref{Proposition:FullWeb:SatAndMonoAsInSPARQL:CaseSat}:

\vspace{1ex} \noindent
\textbf{If:} Let $P$ be a SPARQL expression that is satisfiable. Hence, there exists a set of {\triple}s $G$ such that $\evalOrigPG{P}{G} \neq \emptyset$. W.l.o.g., let 
$\mu$ be an arbitrary solution for $P$ in $G$, that is, $\mu \in \evalOrigPG{P}{G}$. Furthermore, let $G' = \mathrm{lin}^{P,G}(\mu)$ be the $(P,G)$-lineage of $\mu$. We use $G'$ to construct a Web of Linked Data $\WoD_\mu= ( D_\mu,\dataFct_\mu,\adocFct_\mu )$ which consists of a single {\LDdoc}. This document may be retrieved using any {\ID} and it contains the $(P,G)$-lineage of $\mu$ (recall that the lineage is guaranteed to be a finite). Formally:
\begin{align*}
	D_\mu = \lbrace d \rbrace &&
	\dataFct_\mu(d) = G' &&
	\forall \, \symURI \in \symAllURIs : \adocFct_\mu(\symURI) = d
\end{align*}
We now consider the result of {\LDSPARQLFull} query $\evalFullFctP{P}$ (which uses $P$) over $\WoD_\mu$. Obviously, $\fctAllDataName(\WoD_\mu) = G'$ and, thus, $\evalFullPW{P}{\WoD_\mu} = \evalOrigPG{P}{G'}$ (cf.~Definition~\ref{Definition:FullWebSemantics}). Since we know $\mu \in \evalOrigPG{P}{G'}$ it holds $\evalFullPW{P}{\WoD_\mu} \neq \emptyset$, which shows that $\evalFullFctP{P}$ is satisfiable.

\vspace{1ex} \noindent
\textbf{Only if:} Let $\evalFullFctP{P}$ be a satisfiable {\LDSPARQLFull} query that uses SPARQL expression $P$. Since $\evalFullFctP{P}$ is satisfiable, exists a Web of Linked Data $\WoD$ such that $\evalFullPW{P}{\WoD} \neq \emptyset$. Since $\evalFullPW{P}{\WoD} = \evalOrigPG{P}{\fctAllDataName(\WoD)}$ (cf.~Definition~\ref{Definition:FullWebSemantics}), we conclude that $P$ is satisfiable.

\subsection{Proof of Proposition~\ref{Proposition:FullWeb:SatAndMonoAsInSPARQL}, Case~\ref{Proposition:FullWeb:SatAndMonoAsInSPARQL:CaseNonTrivSat}} \label{Proof:Proposition:FullWeb:SatAndMonoAsInSPARQL:CaseNonTrivSat}
\noindent
We prove case~\ref{Proposition:FullWeb:SatAndMonoAsInSPARQL:CaseNonTrivSat} of Proposition~\ref{Proposition:FullWeb:SatAndMonoAsInSPARQL} using the same argumentation that we use in Section~\ref{Proof:Proposition:FullWeb:SatAndMonoAsInSPARQL:CaseSat} for case~\ref{Proposition:FullWeb:SatAndMonoAsInSPARQL:CaseSat}.

\vspace{1ex} \noindent
\textbf{If:} Let $P$ be a SPARQL expression that is nontrivially satisfiable. Hence, there exists a set of {\triple}s $G$ and a valuation $\mu$ such that i)~$\mu \in \evalOrigPG{P}{G}$ and ii)~$\fctDom{\mu} \neq \emptyset$. Let $G' = \mathrm{lin}^{P,G}(\mu)$ be the $(P,G)$-lineage of $\mu$. We use $G'$ to construct a Web of Linked Data $\WoD_\mu= ( D_\mu,\dataFct_\mu,\adocFct_\mu )$ which consists of a single {\LDdoc}. This document may be retrieved using any {\ID} and it contains the $(P,G)$-lineage of $\mu$ (recall that the lineage is guaranteed to be a finite). Formally:
\begin{align*}
	D_\mu = \lbrace d \rbrace &&
	\dataFct_\mu(d) = G' &&
	\forall \, \symURI \in \symAllURIs : \adocFct_\mu(\symURI) = d
\end{align*}
We now consider the result of {\LDSPARQLFull} query $\evalFullFctP{P}$ (which uses $P$) over $\WoD_\mu$. Obviously, $\fctAllDataName(\WoD_\mu) = G'$ and, thus, $\evalFullPW{P}{\WoD_\mu} = \evalOrigPG{P}{G'}$ (cf.~Definition~\ref{Definition:FullWebSemantics}). Since we know $\mu \in \evalOrigPG{P}{G'}$ and $\fctDom{\mu} \neq \emptyset$, we conclude that $\evalFullFctP{P}$ is nontrivially satisfiable.

\vspace{1ex} \noindent
\textbf{Only if:} Let $\evalFullFctP{P}$ be a nontrivially satisfiable {\LDSPARQLFull} query that uses SPARQL expression $P$. Since $\evalFullFctP{P}$ is nontrivially satisfiable, exists a Web of Linked Data $\WoD$ and a valuation $\mu$ such that i)~$\mu \in \evalFullPW{P}{\WoD}$ and ii)~$\fctDom{\mu} \neq \emptyset$. Since $\evalFullPW{P}{\WoD} = \evalOrigPG{P}{\fctAllDataName(\WoD)}$ (cf.~Definition~\ref{Definition:FullWebSemantics}), we conclude that $P$ is nontrivially satisfiable.

\subsection{Proof of Proposition~\ref{Proposition:FullWeb:SatAndMonoAsInSPARQL}, Case~\ref{Proposition:FullWeb:SatAndMonoAsInSPARQL:CaseMono}} \label{Proof:Proposition:FullWeb:SatAndMonoAsInSPARQL:CaseMono}
\noindent
\textbf{If:}
\begin{proofdeclitems}
	\item $P$
		be a SPARQL expression that is monotonic;
	\item $\evalFullFctP{P}$
		be the {\LDSPARQLFull} query that uses $P$; and
	\item $\WoD_1,\WoD_2$
		be an arbitrary pair of Webs of Linked Data such that $\WoD_1$ is an induced subweb of $\WoD_2$.
\end{proofdeclitems}
To prove that $\evalFullFctP{P}$ is monotonic it suffices to show $\evalFullPW{P}{\WoD_1} \subseteq \evalFullPW{P}{\WoD_2}$.
According to Definition~\ref{Definition:FullWebSemantics} we have $\evalFullPW{P}{\WoD_1} = \evalOrigPG{P}{\fctAllDataName(\WoD_1)}$ and $\evalFullPW{P}{\WoD_2} = \evalOrigPG{P}{\fctAllDataName(\WoD_2)}$. Since $\WoD_1$ is an induced subweb of $\WoD_2$ it holds $\fctAllDataName(\WoD_1) \subseteq \fctAllDataName(\WoD_2)$. We may now use the monotonicity of $P$ to show $\evalOrigPG{P}{\fctAllDataName(\WoD_1)} \subseteq \evalOrigPG{P}{\fctAllDataName(\WoD_2)}$. Hence, $\evalFullPW{P}{\WoD_1} \subseteq \evalFullPW{P}{\WoD_2}$.

\vspace{1ex} \noindent
\textbf{Only if:}
\begin{proofdeclitems}
	\item $\evalFullFctP{P}$
		be a monotonic {\LDSPARQLFull} query that uses SPARQL expression $P$; and
	\item $G_1,G_2$
		be an arbitrary pair of set of {\triple}s such that $G_1 \subseteq G_2$.
\end{proofdeclitems}
We distinguish two cases: either $P$ is satisfiable or $P$ is not satisfiable. In the latter case $P$ is trivially monotonic. 
Hence, we only have to discuss the first case. To prove that (the satisfiable) $P$ is monotonic it suffices to show $\evalOrigPG{P}{G_1} \subseteq \evalOrigPG{P}{G_2}$.

Similar to the proof for the other direction, we aim to use $G_1$ and $G_2$ for constructing two Webs of Linked Data $\WoD_1$ and $\WoD_2$ (where $\WoD_1$ is an induced subweb of $\WoD_2$) and then use the monotonicity of $\evalFullFctP{P}$ to show the monotonicity of $P$. However, since $G_1$ and $G_2$ may be (countably) infinite we cannot simply construct Webs of Linked Data that consist of single {\LDdoc}s which contain all {\triple}s of $G_1$ and $G_2$ (recall, the data in each {\LDdoc} of a Web of Linked Data must be finite). As an alternative strategy we construct Webs of Linked Data that consists of as many {\LDdoc}s as we have {\triple}s in $G_1$ and $G_2$ (which may be infinitely many). However, since the data of each {\LDdoc} in a Web of Linked Data must use a unique set of blank nodes, we may lose certain solutions $\mu \in \evalOrigPG{P}{G_1}$ by distributing the {\triple}s from $G_1$ over multiple {\LDdoc}s; similarly for $G_2$. To avoid this issue we assume i)~a set $U_B \subset \symAllURIs$ of new {\ID}s not mentioned in $G_2$ (i.e.~$U_B \cap \fctTerms{G_2} = \emptyset$) and ii)~a bijective mapping $\varrho : \fctTerms{G_2} \rightarrow \bigl( U_B \cup \fctTerms{G_2} \cap (\symAllURIs \cup \symAllLiterals) \bigr)$ that, for any $x \in \fctTerms{G_2}$, is defined as follows:
\begin{equation*}
	\varrho(x) = \begin{cases}
		\varrho_B(x) & \text{if $x \in ( \fctTerms{G_2} \cap \symAllBNodes )$,} \\
		x & \text{else.} \\
	\end{cases}
\end{equation*}
where $\varrho_B : ( \fctTerms{G_2} \cap \symAllBNodes ) \rightarrow U_B$ is an arbitrary bijection that maps each blank node in $G_2$ to a new, unique {\ID} $\symURI \in U_B$.

The application of $\varrho$ to an arbitrary valuation $\mu$, denoted by $\varrho[\mu]$, results in a valuation $\mu'$ such that $\fctDom{\mu'}=\fctDom{\mu}$ and $\mu'(?v) = \varrho( \mu(?v) )$ for all $?v \in \fctDom{\mu}$.
Furthermore, the application of $\varrho$ to an arbitrary {\triple} $t = (x_1,x_2,x_3)$, denoted by $\varrho[t]$, results in an {\triple} $t' = (x_1',x_2',x_3')$ such that $x_i'=\varrho(x_i)$ for all $i \in \lbrace 1,2,3 \rbrace$.
We now let $G_1'=\lbrace \varrho[t] \,|\, t \in G_1 \rbrace$ and $G_2'=\lbrace \varrho[t] \,|\, t \in G_2 \rbrace$. The following facts are verified easily:

\begin{fact} \label{Fact:Proof:Proposition:FullWeb:SatAndMonoAsInSPARQL:CaseMono:Facts1}
	It holds $G_1' \subseteq G_2'$, $\left| G_1 \right| = \left| G_1' \right|$, and $\left| G_2 \right| = \left| G_2' \right|$.
\end{fact}

\begin{fact} \label{Fact:Proof:Proposition:FullWeb:SatAndMonoAsInSPARQL:CaseMono:Fact2}
	For all $j \in \lbrace 1,2 \rbrace$ it holds: Let $\mu$ be an arbitrary valuation, then $\mu' = \varrho[\mu]$ is a solution for $P$ in $G_j'$ \IFF $\mu$ is a solution for $P$ in $G_j$. More precisely:
		\begin{align*}
			\forall \, \mu \in \evalOrigPG{P}{G_j} &: \varrho[\mu] \in \evalOrigPG{P}{G_j'} &
			&\text{and} &
			\forall \, \mu' \in \evalOrigPG{P}{G_j'} &: \varrho^{-1}[\mu'] \in \evalOrigPG{P}{G_j}
		\end{align*}
		where $\varrho^{-1}$ denotes the inverse of the bijective mapping $\varrho$.
\end{fact}

\noindent
We now use $G_2'$ to construct a Web of Linked Data $\WoD_2 = ( D_2,\dataFct_2,\adocFct_2 )$ as follows: $D_2$ consists of $\left| G_2' \right|$ {\LDdoc}s, each of which contains a particular {\triple} from $G_2'$. Furthermore, we assume a set $U_2$ of {\ID}s, each of which corresponds to a particular {\triple} from $G_2'$; hence, $U_2 \subset \symAllURIs$ and $\left| U_2 \right| = \left| G_2' \right|$. These {\ID}s may be used to retrieve the {\LDdoc} for the corresponding {\triple}. For a formal definition let $d_{t_i}$ denote the {\LDdoc} for {\triple} $t_i \in G_2'$ and let $\symURI_{t_i}$ denote the {\ID} that corresponds to $t_i \in G_2'$. Then, we let:
\begin{align*}
	D_2 = \bigcup_{t_i \in G_2'} d_{t_i} &&
	\dataFct_2(d_{t_i}) = \lbrace t_i \rbrace &&
	\forall \, \symURI_{t_i} \in U_2 : \adocFct_2(\symURI_{t_i}) = d_{t_i}
\end{align*}
In addition to $\WoD_2$, we introduce a Web of Linked Data $\WoD_1 = ( D_1,\dataFct_1,\adocFct_1 )$ that is an induced subweb of $\WoD_2$ and that is defined by $D_1 = \lbrace d_{t_i} \in D_2 \,|\, t_i \in G_1' \rbrace$. Recall, any induced subweb is unambiguously defined by specifying its set of {\LDdoc}s.
It can be easily seen that $\fctAllDataName( \WoD_1 ) = G_1'$ and $\fctAllDataName( \WoD_2 ) = G_2'$.

We now use $\WoD_1$ and $\WoD_2$ and the monotonicity of $\evalFullFctP{P}$ to show $\evalOrigPG{P}{G_1} \subseteq \evalOrigPG{P}{G_2}$ (which proves that $P$ is monotonic). W.l.o.g., let $\mu$ be an arbitrary solution for $P$ in $G_1$, that is, $\mu \in \evalOrigPG{P}{G_1}$. Notice, such a $\mu$ must exist because we assume that $P$ is satisfiable (see before). To prove $\evalOrigPG{P}{G_1} \subseteq \evalOrigPG{P}{G_2}$ it suffices to show $\mu \in \evalOrigPG{P}{G_2}$.

Due to Fact~\ref{Fact:Proof:Proposition:FullWeb:SatAndMonoAsInSPARQL:CaseMono:Fact2} it holds $\varrho[\mu] \in \evalOrigPG{P}{G_1'}$; and with $\fctAllDataName( \WoD_1 ) = G_1'$ and Definition~\ref{Definition:FullWebSemantics} we have $\evalOrigPG{P}{G_1'} = \evalOrigPG{P}{\fctAllDataName( \WoD_1 )} = \evalFullPW{P}{\WoD_1}$. Since $\WoD_1$ is an induced subweb of $\WoD_2$ and $\evalFullFctP{P}$ is monotonic, it holds $\evalFullPW{P}{\WoD_1} \subseteq \evalFullPW{P}{\WoD_2}$. We now use $\fctAllDataName( \WoD_2 ) = G_2'$ to show $\varrho[\mu] \in \evalOrigPG{P}{G_2'}$. Finally, we use Fact~\ref{Fact:Proof:Proposition:FullWeb:SatAndMonoAsInSPARQL:CaseMono:Fact2} again and find $\varrho^{-1}\bigl[ \varrho[\mu] \bigr] = \mu \in \evalOrigPG{P}{G_2}$.

\subsection{Proof of Theorem~\ref{Theorem:FullWeb:ComputabilityDependsOnMonotonicity}} \label{Proof:Theorem:FullWeb:ComputabilityDependsOnMonotonicity}
\noindent
To prove the theorem we first show that not any satisfiable {\LDSPARQLFull} query is finitely computable. Next, we study {\LDSPARQLFull} queries that are not monotonic and show that these queries are not eventually computable (and, thus, not computable at all). Finally, we prove that satisfiable, monotonic {\LDSPARQLFull} queries are eventually computable.

\vspace{1ex} \noindent
To show that not any satisfiable {\LDSPARQLFull} query is finitely computable we use a contradiction, that is, we assume a satisfiable {\LDSPARQLFull} query $\evalFullFctP{P}$ which is finitely computable and show that this assumption must be false. If $\evalFullFctP{P}$ were finitely computable, there would be an LD machine that, for any Web of Linked Data $\WoD$ encoded on the Web tape, halts after a finite number of computation steps and produces a possible encoding of $\evalFullPW{P}{\WoD}$ on its output tape (cf.~Definition~\ref{Definition:FinitelyComputable}). To obtain a contradiction we show that such a machine does not exist. However, for the proof we assume $M$ were such a machine.

To compute $\evalFullFctP{P}$  over an arbitrary Web of Linked Data $\WoD = (D,\dataFct,\adocFct)$ (which is encoded on the Web tape of $M$), machine $M$ requires access to the data of all {\LDdoc}s $d \in D$ (Recall that $\evalFullPW{P}{\WoD} = \evalOrigPG{P}{\fctAllDataName(\WoD)}$ where
	$\fctAllDataName(\WoD)= \big\lbrace \dataFct(d) \,|\, d \in D \big\rbrace$%
). However, $M$ may only access an {\LDdoc} $d \in D$
(and its data) after entering the expand state with a corresponding {\ID} $\symURI \in \symAllURIs$ on the link traversal tape (i.e.~for $\symURI$ it must hold $\adocFct(\symURI)=d$). Initially, the machine has no information about which {\ID}(s) to use for accessing any $d \in D$. Hence, to ensure that all $d \in D$ have been accessed, $M$ must expand all $\symURI \in \symAllURIs$. Notice, a real query system for the WWW would have to perform a similar procedure: To guarantee that such a system sees all documents, it must enumerate and lookup all URIs. However, since $\symAllURIs$ is (countably) infinite, this process does not terminate, which is a contradiction to our assumption that $M$ halts after a finite number of computation steps. Hence,
satisfiable {\LDSPARQLFull} queries cannot be finitely computable.

\vspace{1ex} \noindent
We now show by contradiction that non-monotonic {\LDSPARQLFull} queries are not eventually computable. To obtain a contradiction we assume a (satisfiable) {\LDSPARQLFull} query $\evalFullFctP{P}$ that is not monotonic and an LD machine $M$ whose computation of $\evalFullFctP{P}$ on any Web of Linked Data $\WoD$
has the two properties given in Definition~\ref{Definition:EventuallyComputable}. Let $\WoD = (D,\dataFct,\adocFct)$ be a Web of Linked Data such that $\evalFullPW{P}{\WoD} \neq \emptyset$; such a Web exists because $\evalFullFctP{P}$ is satisfiable. Let $\WoD$ be encoded on the Web tape of $M$ and let $\mu$ be an arbitrary solution for $\evalFullFctP{P}$ in $\WoD$; i.e.~$\mu \in \evalFullPW{P}{\WoD}$. Based on our assumption, machine $M$ must write $\fctEnc{\mu}$ to its output tape after a finite number of computation steps (cf.~property~\ref{DefinitionRequirement:EventuallyComputable:All} in Definition~\ref{Definition:EventuallyComputable}). We argue that this is impossible: Since $\evalFullFctP{P}$ is not monotonic, $M$ cannot add $\mu$ to the output before it is guaranteed that all $d \in D$ have been accessed. As discussed before, such a guarantee requires expanding all $\symURI \in \symAllURIs$ because $M$ has no a-priory information about $\WoD$. However, expanding all $\symURI \in \symAllURIs$ is a non-terminating process (due to the infiniteness of $\symAllURIs$) and, thus, $M$ does not write $\mu$ to its output after a finite number of steps. As a consequence, the computation of $\evalFullPW{P}{\WoD}$ by $M$ does not have the properties given in Definition~\ref{Definition:EventuallyComputable}, which contradicts our initial assumption. This contradiction shows that non-monotonic {\LDSPARQLFull} queries are not eventually computable.

\vspace{1ex} \noindent
In the remainder, we prove that satisfiable, monotonic {\LDSPARQLFull} queries are eventually computable. For this proof we introduce specific LD machines which we call $(P)$-machines. Such a $(P)$-machine implements a generic (i.e.~input independent) computation of {\LDSPARQLFull} query $\evalFullFctP{P}$. We shall see that if a {\LDSPARQLFull} query $\evalFullFctP{P}$ is monotonic, the corresponding $(P)$-machine (eventually) computes $\evalFullFctP{P}$ over any Web of Linked Data. Formally, we define $(P)$-machines as follows:
\begin{definition} \label{Definition:FullQueryMachine}
	Let $P$ be SPARQL expression.
	The \definedTerm{$(P)$-ma\-chine}
	is an LD machine that implements Algorithm~\ref{Algorithm:FullQueryMachine}. This algorithm
		makes use of
	a
		special
	subroutine
		called
	\code{lookup}%
		, which,
	when called with {\ID} $\symURI \in \symAllURIs$,
	i)~writes $\fctEnc{\symURI}$ to the right end of the word on the link traversal tape,
	ii)~enters the expand state, and
	iii)~performs the expand procedure as specified in Definition~\ref{Definition:LDMachine}.
\end{definition}
\begin{algorithm}[tb]
	\caption{\, The program of the $(P)$-machine.} \label{Algorithm:FullQueryMachine}
	\begin{algorithmic}[1]
		\STATE {$j := 1$}
		\FOR {$\symURI \in \symAllURIs$} \label{Line:FullQueryMachine:LoopBegin}
			\STATE {Call \code{lookup} for $\symURI$.} \label{Line:FullQueryMachine:Step1_Lookup}
			\STATE {Let $T_j$ denote the set of all {\triple}s currently encoded on the link traversal tape. Use the work tape to enumerate the set $\evalOrigPG{P}{T_j}$.} \label{Line:FullQueryMachine:Step2_EnumPartialResult}
			\STATE {For each $\mu \in \evalOrigPG{P}{T_j}$ check whether $\mu$ is already encoded on the output tape; if not, then add $\fctEnc{\mu}$ to the output.} \label{Line:FullQueryMachine:Step3_OutputNewSolutions}
			\STATE {$j := j + 1$} \label{Line:FullQueryMachine:IncrementCounter}
		\ENDFOR \label{Line:FullQueryMachine:LoopEnd}
	\end{algorithmic}
\end{algorithm}

\noindent
Before we complete the proof we discuss important properties of $(P)$-machines.
As can be seen in Algorithm~\ref{Algorithm:FullQueryMachine}, any computation performed by $(P)$-machines enters a loop that iterates over the set $\symAllURIs$ of all possible {\ID}s.
	As discussed before,
expanding all $\symURI \in \symAllURIs$ is necessary
to guarantee completeness of the computed query result. However, since $\symAllURIs$ is (countably) infinite the algorithm does not terminate (which is not a requirement for eventual computability).

During each iteration of
	its main processing
loop,
	a $(P)$-%
machine generates valuations using all data that is currently encoded on the link traversal tape. The following lemma shows that these valuations are part of the corresponding query result (find the proof for Lemma~\ref{Lemma:FullWeb:Computability:Soundness} below in Section~\ref{Proof:Lemma:FullWeb:Computability:Soundness}):
\begin{lemma} \label{Lemma:FullWeb:Computability:Soundness}
	Let $\evalFullFctP{P}$ be a satisfiable {\LDSPARQLFull} query
		that is monotonic;
	let $M^{P}$ denote the $(P)$-machine for SPARQL expression $P$ used by $\evalFullFctP{P}$\!;
	and let $\WoD$ be an arbitrary Web of Linked Data encoded on the Web tape of $M^{P}$\!.
		During the execution of Algorithm~\ref{Algorithm:FullQueryMachine} by $M^{P}$ it holds $\forall \, j \in \lbrace 1,2,...\rbrace : \evalOrigPG{P}{T_j} \subseteq \evalFullPW{P}{\WoD}$.
\end{lemma}

\noindent
We now use the notion of $(P)$-machines to prove that satisfiable, monotonic {\LDSPARQLFull} queries are eventually computable. Let $\evalFullFctP{P}$ be a satisfiable {\LDSPARQLFull} query that is monotonic and let $\WoD$ be an arbitrary Web of Linked Data encoded on the Web tape of the $(P)$-machine for $\evalFullFctP{P}$\!; to denote this machine we write $M^{P}$.
W.l.o.g.~it suffices to show that the computation of $M^{P}$ on (Web) input $\fctEnc{\WoD}$ has the two properties given in Definition~\ref{Definition:EventuallyComputable}.

During the computation $M^{P}$ only writes to its output tape when it adds (encoded) valuations $\mu \in \evalOrigPG{P}{T_j}$ (for $j=1,2,...$). Since all these valuations are solutions for $\evalFullFctP{P}$ in $\WoD$ (cf.~Lemma~\ref{Lemma:FullWeb:Computability:Soundness}) and line~\ref{Line:FullQueryMachine:Step3_OutputNewSolutions} in Algorithm~\ref{Algorithm:FullQueryMachine} ensures that the output is free of duplicates, we see that the word on the output tape is always a prefix of a possible encoding of $\evalFullPW{P}{\WoD}$. Hence, the computation of $M^{P}$ has the first property specified in Definition~\ref{Definition:EventuallyComputable}.

To verify that the computation also has the second property it is important to note that Algorithm~\ref{Algorithm:FullQueryMachine} looks up no more than one {\ID} per iteration (cf.~line~\ref{Line:FullQueryMachine:Step1_Lookup}). Hence, $(P)$-machines prioritize result construction over data retrieval. This feature allows us to show that for each solution in a query result exists an iteration during which that solution is computed (find the proof for Lemma~\ref{Lemma:FullWeb:Computability:Completeness} below in Section~\ref{Proof:Lemma:FullWeb:Computability:Completeness}):
\begin{lemma} \label{Lemma:FullWeb:Computability:Completeness}
	Let $\evalFullFctP{P}$ be a satisfiable {\LDSPARQLFull} query
		that is monotonic;
	let $M^{P}$ denote the $(P)$-machine for SPARQL expression $P$ used by $\evalFullFctP{P}$\!;
	and let $\WoD$ be an arbitrary Web of Linked Data encoded on the Web tape of $M^{P}$\!.
		For each $\mu \in \evalFullPW{P}{\WoD}$ exists a $j_\mu \in \lbrace 1,2,...\rbrace$ such that during the execution of Algorithm~\ref{Algorithm:FullQueryMachine} by $M^{P}$ it holds $\forall \, j \in \lbrace j_\mu,j_\mu \!+\! 1,...\rbrace : \mu \in \evalOrigPG{P}{T_j}$.
\end{lemma}

\noindent
It remains to show that the computation of $M^{P}$ definitely reaches each iteration of the processing loop after a finite number of computation steps. To prove this property we show that each iteration of the loop finishes after a finite number of computation steps:
\begin{itemize}
	\item The call of the subroutine \code{lookup} (cf.~Definition~\ref{Definition:FullQueryMachine}) in line~\ref{Line:FullQueryMachine:Step1_Lookup} of Algorithm~\ref{Algorithm:FullQueryMachine} terminates because the encoding of $\WoD = (D,\dataFct,\adocFct)$ is ordered following the order of the {\ID}s in $\fctDom{\adocFct}$.
	\item
		At any point in the computation the word on the link traversal tape is
	finite because $M^{P}$ only gradually appends (encoded) {\LDdoc}s to the link traversal tape and the encoding of each document is finite (recall that the set of {\triple}s $\dataFct(d)$ for each {\LDdoc} $d$ is finite). Due to the finiteness of
		the word on the link traversal tape,
	each $\evalOrigPG{P}{T_j}$ (for $j=1,2,...$) is
	finite, resulting in a finite number of computation steps for line~\ref{Line:FullQueryMachine:Step2_EnumPartialResult} during any iteration.
	\item Finally, line~\ref{Line:FullQueryMachine:Step3_OutputNewSolutions} requires only a finite number of computation steps because
		the word on the link traversal tape
	is finite at any point in the computation; so is the word on the output tape.
\end{itemize}

\subsection{Proof of Lemma~\ref{Lemma:FullWeb:Computability:Soundness}} \label{Proof:Lemma:FullWeb:Computability:Soundness}
\begin{proofdeclitems}
	\item $\evalFullFctP{P}$ be a {\LDSPARQLFull} query that is satisfiable and monotonic;
	\item $M^{P}$ denote the $(P)$-machine for SPARQL expression $P$ used by $\evalFullFctP{P}$\!;
	\item $\WoD$ be a Web of Linked Data which is encoded on the Web tape of $M^{P}$\!.
\end{proofdeclitems}
To prove Lemma~\ref{Lemma:FullWeb:Computability:Soundness} we use the following result.
\begin{lemma} \label{SubLemma:FullWeb:Computability:Soundness}
	During the execution of Algorithm~\ref{Algorithm:FullQueryMachine} by $M^{P}$ on (Web) input $\fctEnc{\WoD}$ it holds $\forall \, j \in \lbrace 1,2,...\rbrace : T_j \subseteq \fctAllDataName( \WoD )$.
\end{lemma}

\begin{myproof}{of Lemma~\ref{SubLemma:FullWeb:Computability:Soundness}}
The computation of $M^{P}$ starts with an empty link traversal tape (cf.~Definition~\ref{Definition:LDMachine}). Let $w_j$ be the word on the link traversal tape of $M^{P}$ before $M^{P}$ executes line~\ref{Line:FullQueryMachine:Step2_EnumPartialResult} during the $j$-th iteration of the main processing loop in Algorithm~\ref{Algorithm:FullQueryMachine}. It can be easily seen that for each $w_j$ (where $j \in \lbrace 1,2,...\rbrace$) exists a finite sequence $\symURI_1 ,...\,,\symURI_{j}$ of $j$ different {\ID}s
such that
i)~$w_j$ is\footnote{We assume $\fctEnc{\adocFct(\symURI_i)}$ is the empty word if $\symURI_i \notin \fctDom{\adocFct}$.}
\begin{equation*}
	\fctEncName(\symURI_1) \, \fctEncName( \adocFct(\symURI_1) ) \, \sharp \, ... \, \sharp \, \fctEncName(\symURI_{j}) \, \fctEncName( \adocFct(\symURI_{j}) ) \, \sharp
\end{equation*}
and
ii)~for each $i \in [1,j]$ either $\symURI_i \notin \fctDom{\adocFct}$ or $\adocFct(\symURI_i) \in D$.
If $U_j$ is the set that contains all {\ID}s in this sequence $\symURI_1 ,...\,,\symURI_{j}$, it holds $T_j = \big\lbrace \dataFct\left( \adocFct(\symURI_i) \right) \,\big|\, \symURI_i \in U_j \text{ and } \symURI_i \in \fctDom{\adocFct} \big\rbrace$. Clearly, $T_j \subseteq \fctAllDataName( \WoD )$.
\end{myproof}

\noindent
Due to the monotonicity of $\evalFullFctP{P}$ it is trivial to show Lemma~\ref{Lemma:FullWeb:Computability:Soundness} using Lemma~\ref{SubLemma:FullWeb:Computability:Soundness} (recall $\evalFullPW{P}{\WoD} = \evalOrigPG{P}{\fctAllDataName(\WoD)}$).

\subsection{Proof of Lemma~\ref{Lemma:FullWeb:Computability:Completeness}} \label{Proof:Lemma:FullWeb:Computability:Completeness}
\begin{proofdeclitems}
	\item $\evalFullFctP{P}$ be a {\LDSPARQLFull} query that is satisfiable and monotonic;
	\item $M^{P}$ denote the $(P)$-machine for SPARQL expression $P$ used by $\evalFullFctP{P}$\!;
	\item $\WoD$ be a Web of Linked Data which is encoded on the Web tape of $M^{P}$\!.
\end{proofdeclitems}
To prove Lemma~\ref{Lemma:FullWeb:Computability:Completeness} we use the following result.
\begin{lemma} \label{SubLemma:FullWeb:Computability:Completeness}
	For each {\triple} $t \in \fctAllDataName\bigl( \WoD \bigr)$ exists a $j_t \in \lbrace 1,2,...\rbrace$ such that during the execution of Algorithm~\ref{Algorithm:FullQueryMachine} by $M^{P}$ on (Web) input $\fctEncName(\WoD)$ it holds $\forall \, j \in \lbrace j_t,j_t\!+\!1,...\rbrace : t \in T_j$.
\end{lemma}

\begin{myproof}{of Lemma~\ref{SubLemma:FullWeb:Computability:Completeness}}
W.l.o.g., let $t'$ be an arbitrary {\triple}
$t' \in \fctAllDataName\bigl( \WoD \bigr)$; hence, there exists an {\LDdoc} $d \in D$ such that $t' \in \dataFct(d)$. Let $d'$ be such an {\LDdoc}.
Since mapping $\adocFct$ is surjective (cf.~Definition~\ref{Definition:WebOfData}), exists a {\ID} $\symURI \in \symAllURIs$ such that $\adocFct(\symURI) = d'$. Let $\symURI'$ be such a {\ID}. Since $\symURI' \in \symAllURIs$ exists a $j' \in \lbrace 1,2,...\rbrace$ such that $M^{P}$ selects $\symURI'$ for processing in the $j'$-th iteration of the main loop in Algorithm~\ref{Algorithm:FullQueryMachine}. After completing the lookup of $\symURI'$ during this iteration (cf.~line~\ref{Line:FullQueryMachine:Step1_Lookup} in Algorithm~\ref{Algorithm:FullQueryMachine}), the word on the link traversal tape contains sub-word $\fctEnc{d'}$ (cf.~Definitions~\ref{Definition:FullQueryMachine} and~\ref{Definition:LDMachine}). Since $t' \in \dataFct(d')$, this word $\fctEnc{d'}$ contains sub-word $\fctEnc{t'}$ (cf.~Appendix~\ref{Appendix:Encoding}). Hence, $t' \in T_{j'}$. Since $(P)$-machines only append to (the right end of) the word on the link traversal tape, $M^{P}$ will never remove $\fctEnc{t'}$ from that tape and, thus, it holds $\forall \, j \in \lbrace j',j'\!+\!1,...\rbrace : t' \in T_j$.
\end{myproof}

\noindent
We now prove Lemma~\ref{Lemma:FullWeb:Computability:Completeness} by induction over the structure of possible SPARQL expressions.

\vspace{1ex} \noindent
\textbf{Base case}: Assume that SPARQL expression $P$ is a triple pattern $tp$.
W.l.o.g., let $\mu \in \evalFullPW{P}{\WoD}$. It holds $\fctDom{\mu} = \fctVarsName(tp)$ and $t = \mu[tp] \in \fctAllDataName(\WoD)$ (cf.~Definitions~\ref{Definition:FullWebSemantics} and~\ref{Definition:EvaluationSPARQL}). According to Lemma~\ref{SubLemma:FullWeb:Computability:Completeness} exists a $j_\mu \in \lbrace 1,2,...\rbrace$ such that $\forall \, j \in \lbrace j_\mu,j_\mu\!+\!1,...\rbrace : t \in T_j$. Since $\evalFullFctP{P}$ is monotonic we conclude $\forall \, j \in \lbrace j_\mu,j_\mu \!+\! 1,...\rbrace : \mu \in \evalOrigPG{P}{T_j}$.

\vspace{1ex} \noindent
\textbf{Induction step}: Our inductive hypothesis is that for SPARQL expressions $P_1$ and $P_2$ it holds:
\begin{enumerate}
	\item
		For each $\mu \in \evalFullPW{P_1}{\WoD}$ exists a $j_\mu \in \lbrace 1,2,...\rbrace$ such that during the execution of Algorithm~\ref{Algorithm:FullQueryMachine} by $M^{P}$ it holds
		$\forall \, j \in \lbrace j_\mu,j_\mu \!+\! 1,...\rbrace : \mu \in \evalOrigPG{P_1}{T_j}$; and
	\item
		For each $\mu \in \evalFullPW{P_2}{\WoD}$ exists a $j_\mu \in \lbrace 1,2,...\rbrace$ such that during the execution of Algorithm~\ref{Algorithm:FullQueryMachine} by $M^{P}$ it holds
		$\forall \, j \in \lbrace j_\mu,j_\mu \!+\! 1,...\rbrace : \mu \in \evalOrigPG{P_2}{T_j}$.
\end{enumerate}
\noindent
Based on this hypothesis we show that for any SPARQL expression $P$ that can be constructed using $P_1$ and $P_2$ it holds: For each $\mu \in \evalFullPW{P}{\WoD}$ exists a $j_\mu \in \lbrace 1,2,...\rbrace$ such that during the execution of Algorithm~\ref{Algorithm:FullQueryMachine} by $M^{P}$ it holds $\forall \, j \in \lbrace j_\mu,j_\mu \!+\! 1,...\rbrace : \mu \in \evalOrigPG{P}{T_j}$. W.l.o.g., let $\mu' \in \evalFullPW{P}{\WoD}$. According to Definition~\ref{Definition:ExpressionSPARQL} we distinguish the following cases:
\begin{itemize}
	\item $P$ is $(P_1 \OpAND P_2)$.
		In this case exist $\mu_1 \in \evalFullPW{P_1}{\WoD}$ and $\mu_2 \in \evalFullPW{P_2}{\WoD}$ such that $\mu' = \mu_1 \cup \mu_2$ and $\mu_1 \sim \mu_2$. According to our inductive hypothesis exist $j_{\mu_1}, j_{\mu_2} \in \lbrace 1,2,...\rbrace$ such that i)~$\forall \, j \in \lbrace j_{\mu_1},j_{\mu_1} \!+\! 1,...\rbrace : \mu_1 \in \evalOrigPG{P_1}{T_j}$ and ii)~$\forall \, j \in \lbrace j_{\mu_2},j_{\mu_2} \!+\! 1,...\rbrace : \mu_2 \in \evalOrigPG{P_2}{T_j}$. Let $j_{\mu'} = \max\bigl( \lbrace j_{\mu_1},j_{\mu_2} \rbrace \bigr)$. Due to the monotonicity of $\evalFullFctP{P}$ it holds $\forall \, j \in \lbrace j_{\mu'},j_{\mu'} \!+\! 1,...\rbrace : \mu' \in \evalOrigPG{P}{T_j}$.
	\item $P$ is $(P_1 \OpFILTER R)$.
		In this case exist $\mu^* \in \evalFullPW{P_1}{\WoD}$ such that $\mu' = \mu^*$. According to our inductive hypothesis exist $j_{\mu^*} \in \lbrace 1,2,...\rbrace$ such that $\forall \, j \in \lbrace j_{\mu^*},j_{\mu^*} \!+\! 1,...\rbrace : \mu^* \in \evalOrigPG{P_1}{T_j}$. Due to the monotonicity of $\evalFullFctP{P}$ it holds $\forall \, j \in \lbrace j_{\mu^*},j_{\mu^*} \!+\! 1,...\rbrace : \mu' \in \evalOrigPG{P}{T_j}$.
	\item $P$ is $(P_1 \OpOPT P_2)$.
		We distinguish two cases:
		\begin{enumerate}
			\item There exist $\mu_1 \in \evalFullPW{P_1}{\WoD}$ and $\mu_2 \in \evalFullPW{P_2}{\WoD}$ such that $\mu' = \mu_1 \cup \mu_2$ and $\mu_1 \sim \mu_2$. This case corresponds to the case where $P$ is $(P_1 \OpAND P_2)$ (see above).
			\item There exist $\mu_1 \in \evalFullPW{P_1}{\WoD}$ such that $\mu' = \mu_1$ and $\forall \, \mu_2 \in \evalFullPW{P_2}{\WoD} : \mu_1 \not\sim \mu_2$. According to our inductive hypothesis exist $j_{\mu_1} \in \lbrace 1,2,...\rbrace$ such that $\forall \, j \in \lbrace j_{\mu_1},j_{\mu_1} \!+\! 1,...\rbrace : \mu_1 \in \evalOrigPG{P_1}{T_j}$. Due to the monotonicity of $\evalFullFctP{P}$ it holds $\forall \, j \in \lbrace j_{\mu_1},j_{\mu_1} \!+\! 1,...\rbrace : \mu' \in \evalOrigPG{P}{T_j}$.
		\end{enumerate}
	\item $P$ is $(P_1 \OpUNION P_2)$.
		We distinguish two cases:
		\begin{enumerate}
			\item
				There exists $\mu^* \in \evalFullPW{P_1}{\WoD}$ such that $\mu' = \mu^*$. According to our inductive hypothesis exist $j_{\mu^*} \in \lbrace 1,2,...\rbrace$ such that
					\par
				$\forall \, j \in \lbrace j_{\mu^*},j_{\mu^*} \!+\! 1,...\rbrace : \mu^* \in \evalOrigPG{P_1}{T_j}$.
			\item
				There exists $\mu^* \in \evalFullPW{P_2}{\WoD}$ such that $\mu' = \mu^*$. According to our inductive hypothesis exist $j_{\mu^*} \in \lbrace 1,2,...\rbrace$ such that
					\par
				$\forall \, j \in \lbrace j_{\mu^*},j_{\mu^*} \!+\! 1,...\rbrace : \mu^* \in \evalOrigPG{P_2}{T_j}$.
		\end{enumerate}
		Due to the monotonicity of $\evalFullFctP{P}$ it holds for both cases: $\forall \, j \in \lbrace j_{\mu^*},j_{\mu^*} \!+\! 1,...\rbrace : \mu' \in \evalOrigPG{P}{T_j}$.
\end{itemize}

%
%

\subsection{Proof of Lemma~\ref{Lemma:FullWeb:NoTerminationWithVars}}
\noindent
We prove the lemma by contradiction, that is, we assume a nontrivially satisfiable {\LDSPARQLFull} query $\evalFullFctP{P}$ for which exists an LD machine that, for some Web of Linked Data $\WoD$ encoded on the Web tape, halts after a finite number of computation steps and produces an encoding of $\evalFullPW{P}{\WoD}$ on its output tape. To obtain a contradiction we show that such an LD machine and such a Web of Linked Data does not exist. However, for the proof we assume $M'$ were such a machine and $\WoD' = (D',\dataFct',\adocFct')$ were such a Web of Linked Data.

Since $\evalFullFctP{P}$ is nontrivially satisfiable it is possible that $\WoD'$ is a Web of Linked Data for which exist solutions in $\evalFullPW{P}{\WoD'}$ such that each of these solutions provides a binding for at least one variable. Hence, for computing $\evalFullFctP{P}$ over $\WoD'$ completely, machine $M'$ requires access to the data of all {\LDdoc}s $d \in D'$ (recall that $\evalFullPW{P}{\WoD'} = \evalOrigPG{P}{\fctAllDataName(\WoD')}$ where $\fctAllDataName(\WoD') = \big\lbrace \dataFct'(d) \,|\, d \in D' \big\rbrace$). 
However, $M'$ may access an {\LDdoc} $d \in D'$ (and its data) only after performing the expand procedure with a corresponding {\ID} $\symURI \in \symAllURIs$ on the link traversal tape (i.e.~for $\symURI$ it must hold $\adocFct'(\symURI)=d$). Initially, the machine has no information about which {\ID}(s) to use for accessing any $d \in D'$. Hence, to ensure that all $d \in D'$ have been accessed, $M'$ must expand all $\symURI \in \symAllURIs$. Notice, a real query system for the WWW would have to perform a similar procedure: To guarantee that such a system sees all documents, it must enumerate and lookup all URIs. However, since $\symAllURIs$ is (countably) infinite, this process does not terminate, which is a contradiction to our assumption that $M'$ halts after a finite number of computation steps.

\subsection{Proof of Theorem~\ref{Theorem:FullWeb:Problem:Termination}} \label{Proof:Theorem:FullWeb:Problem:Termination}

\noindent
We formally define the termination problem for {\LDSPARQLFull} as follows:

\vspace{3mm}
\webproblem{Termination(\LDSPARQLFull)}
{a Web of Linked Data $\WoD$}
{a satisfiable {\LDSPARQLFull} query $\evalFullFctP{P}$}
{Does an LD machine exist that computes $\evalFullPW{P}{\WoD}$ and halts?}
\vspace{2mm}

\noindent
We show that \problemName{Termination(\LDSPARQLFull)} is not LD machine decidable by reducing the halting problem to \problemName{Termination(\LDSPARQLFull)}.

The halting problem asks whether a given Turing machine (TM) halts on a given input. 
For the reduction we assume an infinite Web of Linked Data $\WoD_\mathsf{TMs}$ which we define in the following. Informally, $\WoD_\mathsf{TMs}$ describes all possible computations of all TMs. For a formal definition of $\WoD_\mathsf{TMs}$ we adopt the usual approach to unambiguously describe TMs and their input by finite words over the (finite) alphabet of a universal TM
	(e.g.~[Pap93]).
Let $\mathcal{W}$ be the countably infinite set of all words that describe TMs. For each $w \in \mathcal{W}$ let $M(w)$ denote the machine described by $w$, let $c^{w,x}$ denote the computation of $M(w)$ on input $x$, and let $\symURI^{w,x}$ denote a {\ID} that identifies $c^{w,x}$. Furthermore, let $\symURI_i^{w,x}$ denote a {\ID} that identifies the $i$-th step in computation $c^{w,x}$. To denote the (infinite) set of all such {\ID}s we write $\symAllURIs_\mathsf{TMsteps}$. Using the {\ID}s $\symAllURIs_\mathsf{TMsteps}$ we may unambiguously identify each step in each possible computation of any TM on any given input.
However, if a {\ID} $\symURI \in \symAllURIs$ could potentially identify a computation step of a TM on some input (because $\symURI$ adheres to the pattern used for such {\ID}s) but the corresponding step may never exist, then $\symURI \notin \symAllURIs_\mathsf{TMsteps}$. For instance, if the computation of a particular TM $M(w_j)$ on a particular input $x_k$ halts 
with the $i'$-th step, then $\forall \, i \in \lbrace 1 , ... \, , i' \rbrace :\symURI_{i}^{w_j,x_k} \in \symAllURIs_\mathsf{TMsteps}$ and $\forall \, i \in \lbrace i'\!+\!1 , ... \rbrace :\symURI_{i}^{w_j,x_k} \not\in \symAllURIs_\mathsf{TMsteps}$.
Notice, while the set $\symAllURIs_\mathsf{TMsteps}$ is infinite, it is still countable because i)~$\mathcal{W}$ is countably infinite, ii)~the set of all possible input words for TMs is countably infinite, and iii)~$i$ is a natural number.

We now define $\WoD_\mathsf{TMs}$ as a Web of Linked Data $( D_\mathsf{TMs},\dataFct_\mathsf{TMs},$ $ \adocFct_\mathsf{TMs} )$ with the following elements:
$D_\mathsf{TMs}$ consists of $\left| \symAllURIs_\mathsf{TMsteps} \right|$ different {\LDdoc}s, each of which corresponds to one of the {\ID}s in $\symAllURIs_\mathsf{TMsteps}$ (and, thus, to a particular step in a particular computation of a particular TM).
Mapping $\adocFct_\mathsf{TMs}$ is bijective and maps each $\symURI_i^{w,x} \in \symAllURIs_\mathsf{TMsteps}$ to the corresponding $d_i^{w,x} \in  D_\mathsf{TMs}$ ($\fctDom{\adocFct_\mathsf{TMs}} = \symAllURIs_\mathsf{TMsteps}$). We emphasize that mapping $\adocFct_\mathsf{TMs}$ is (Turing) computable because a universal TM may determine by simulation whether the computation of a particular TM on a particular input halts before a particular number of steps (i.e.~whether the $i$-th step in computation $c^{w,x}$ for a given {\ID} $\symURI_i^{w,x}$ may actually exist).
Finally, mapping $\dataFct_\mathsf{TMs}$ is defined as follows: The set $\dataFct_\mathsf{TMs}\bigl( d_i^{w,x} \bigr)$ of {\triple}s for an {\LDdoc} $d_i^{w,x}$ is empty if computation $c^{w,x}$ does not halt with the $i$-th computation step. Otherwise, $\dataFct_\mathsf{TMs}\bigl( d_i^{w,x} \bigr)$ contains a single {\triple} $( \symURI^{w,x}, \mathsf{type} , \mathsf{TerminatingComputation} )$ where $\mathsf{type} \in \symAllURIs$ and $\mathsf{TerminatingComputation} \in \symAllURIs$. Formally:

\begin{equation*}
	\dataFct_\mathsf{TMs}\bigl( d_i^{w,x} \bigr) =
	\begin{cases}
		\big\lbrace ( \symURI^{w,x}, \mathsf{type} , \mathsf{TerminatingComputation} ) \big\rbrace & \text{if computation $c^{w,x}$} \\
		& \text{halts with the $i$-th} \\
		& \text{step,} \\
		\emptyset & \text{else.}
	\end{cases}
\end{equation*}

\noindent
Mapping $\dataFct_\mathsf{TMs}$ is computable because a universal TM may determine by simulation whether the computation of a particular TM on a particular input halts after a given number of steps.

We now reduce the halting problem to \problemName{Termination(\LDSPARQLFull)}. The input to the halting problem is a pair $(w,x)$ consisting of a TM description $w$ and a possible input word $x$. For the reduction we need a computable mapping $f$ that, given such a pair $(w,x)$, produces a tuple $(\WoD,\evalFullFctP{P})$ as input for \problemName{Termination(\LDSPARQLFull)}. We define $f$ as follows: Let
	$(w,x)$ be an input to the halting problem,
then $f( w,x ) = \bigl( \WoD_\mathsf{TMs}, \evalFullFctP{P_{w,x}} \bigr)$ with $P_{w,x} = ( \symURI^{w,x}, \mathsf{type} , \mathsf{TerminatingComputation} )$. Given that $\WoD_\mathsf{TMs}$ is independent of $(w,x)$, it is easy to see that $f$ is computable by TMs (including LD machines).

We emphasize that for any possible $\evalFullFctP{P_{w,x}}$ it holds:
\begin{equation*}
	\evalFullPW{P_{w,x}}{\WoD_\mathsf{TMs}} =
	\begin{cases}
		\lbrace \mu_\emptyset \rbrace & \text{if the computation of TM $M(w)$ on input $x$ halts,} \\
		& \text{step,} \\
		\emptyset & \text{else.}
	\end{cases}
\end{equation*}
where $\mu_\emptyset$ is the empty valuation with $\fctDom{\mu_\emptyset}=\emptyset$. Hence, any $\evalFullFctP{P_{w,x}}$ is satisfiable but not nontrivially satisfiable.

To show that \problemName{Termination(\LDSPARQLFull)} is not LD machine decidable, suppose it were LD machine decidable. In such a case an LD machine could answer the halting problem for any input $(w,x)$ \removable{as follows}: $M(w)$ halts on $x$ if and only if an LD machine exists that computes $\evalFullPW{P_{w,x}}{\WoD_\mathsf{TMs}} = \lbrace \mu_\emptyset \rbrace$ and halts. However, we know the halting problem is undecidable for TMs (which includes LD machines). Hence, we have a contradiction and\removable{, thus,} \problemName{Termination(\LDSPARQLFull)} cannot be LD machine decidable.


\subsection{Proof of Proposition~\ref{Proposition:FullWeb:FiniteWeb}}
\noindent
Let $\WoD = (D,\dataFct,\adocFct)$ be a finite Web of Linked Data.
For each {\LDSPARQLFull} query $\evalFullFctP{P}$ it holds $\evalFullFctP{P}(\WoD) = \evalOrigPG{P}{\fctAllDataName(\WoD)}$ (cf.~Definition~\ref{Definition:FullWebSemantics}). To prove the proposition it suffices to show $\evalOrigPG{P}{\fctAllDataName(\WoD)}$ is finite for any possible SPARQL expression $P$. We use induction over the structure of SPARQL expressions for this proof:

\vspace{1ex} \noindent
\textbf{Base case}: Assume that SPARQL expression $P$ is a triple pattern $tp$. In this case (cf.~Definition~\ref{Definition:EvaluationSPARQL})
\begin{align*}
	\evalOrigPG{P}{\fctAllDataName(\WoD)} =
	\big\lbrace \mu \,\big|\, \mu \text{ is a valuation with }
		& \fctDom{\mu} = \fctVars{tp} \text{ and } \\
		& \mu[tp] \in \fctAllDataName(\WoD)
	\big\rbrace
\end{align*}
Since $\WoD$ (and, thus, $D$) is finite and for all $d \in D$ the set $\dataFct(d)$ is finite, there exist only a finite number of {\triple}s in $\fctAllDataName( \WoD ) = \bigcup_{d \in D} \dataFct(d)$. Hence, there can be only a finite number of different valuations $\mu$ with $\mu[tp] \in \fctAllDataName(\WoD)$ and, thus, $\evalOrigPG{P}{\fctAllDataName(\WoD)}$ must be finite.

\vspace{1ex} \noindent
\textbf{Induction step}: Our inductive hypothesis is that for SPARQL expressions $P_1$ and $P_2$, $\evalOrigPG{P_1}{\fctAllDataName(\WoD)}$ and $\evalOrigPG{P_2}{\fctAllDataName(\WoD)}$ is finite, respectively. Based on this hypothesis we show that for any SPARQL expression $P$ that can be constructed using $P_1$ and $P_2$ (cf.~Definition~\ref{Definition:ExpressionSPARQL}), it holds $\evalOrigPG{P}{\fctAllDataName(\WoD)}$ is finite. According to Definition~\ref{Definition:ExpressionSPARQL} we distinguish the following cases:
\begin{itemize}
	\item $P$ is $(P_1 \OpAND P_2)$. In this case $\evalOrigPG{P}{\fctAllDataName(\WoD)} = \evalOrigPG{P_1}{\fctAllDataName(\WoD)} \Join \evalOrigPG{P_2}{\fctAllDataName(\WoD)}$. The result of the join may contain at most $\left|\evalOrigPG{P_1}{\fctAllDataName(\WoD)}\right| \cdot \left|\evalOrigPG{P_2}{\fctAllDataName(\WoD)}\right|$ elements, which is a finite number because $\evalOrigPG{P_1}{\fctAllDataName(\WoD)}$ and $\evalOrigPG{P_2}{\fctAllDataName(\WoD)}$ are finite.
	\item $P$ is $(P_1 \OpUNION P_2)$. In this case $\evalOrigPG{P}{\fctAllDataName(\WoD)} = \evalOrigPG{P_1}{\fctAllDataName(\WoD)} \cup \evalOrigPG{P_2}{\fctAllDataName(\WoD)}$. The result of the union may contain at most $\left|\evalOrigPG{P_1}{\fctAllDataName(\WoD)}\right| + \left|\evalOrigPG{P_2}{\fctAllDataName(\WoD)}\right|$ elements, which is a finite number because $\evalOrigPG{P_1}{\fctAllDataName(\WoD)}$ and $\evalOrigPG{P_2}{\fctAllDataName(\WoD)}$ are finite.
	\item $P$ is $(P_1 \OpOPT P_2)$. In this case $\evalOrigPG{P}{\fctAllDataName(\WoD)} = \evalOrigPG{P_1}{\fctAllDataName(\WoD)} \LJoin \evalOrigPG{P_2}{\fctAllDataName(\WoD)}$. The result of the left outer join contains at most $\left|\evalOrigPG{P_1}{\fctAllDataName(\WoD)}\right| \cdot \left|\evalOrigPG{P_2}{\fctAllDataName(\WoD)}\right|$ elements, which is a finite number because $\evalOrigPG{P_1}{\fctAllDataName(\WoD)}$ and $\evalOrigPG{P_2}{\fctAllDataName(\WoD)}$ are finite.
	\item $P$ is $(P_1 \OpFILTER R)$. In this case $\evalOrigPG{P}{\fctAllDataName(\WoD)} = \sigma_R\bigl( \evalOrigPG{P_1}{\fctAllDataName(\WoD)} \bigr)$. The result of the selection may contain at most $\left|\evalOrigPG{P_1}{\fctAllDataName(\WoD)}\right|$ elements, which is a finite number because $\evalOrigPG{P_1}{\fctAllDataName(\WoD)}$ is finite.
\end{itemize}

\subsection{Proof of Theorem~\ref{Theorem:FullWeb:Problem:Finiteness}} \label{Proof:Theorem:FullWeb:Problem:Finiteness}
\noindent
We formally define the finiteness problem for {\LDSPARQLFull} as follows:

\vspace{3mm}
\webproblem{Finiteness(\LDSPARQLFull)}
{a (potentially infinite) Web of Linked Data $\WoD$}
{a satisfiable SPARQL expression $P$}
{Is
	the result of (the satisfiable) {\LDSPARQLFull} query $\evalFullFctP{P}$ over $\WoD$
finite?}
\vspace{2mm}

\noindent
We show that \problemName{Finiteness(\LDSPARQLFull)} is not LD machine decidable by reducing the halting problem to \problemName{Finiteness(\LDSPARQLFull)}.
While this proof is similar to the proof of Theorem~\ref{Theorem:FullWeb:Problem:Termination} (cf.~Section~\ref{Proof:Theorem:FullWeb:Problem:Termination}), we now use a Web of Linked Data $\WoD_\mathsf{TMs2}$ which differs from $\WoD_\mathsf{TMs}$ in the way it describes all possible computations of all Turing machines (TM).

For the proof we use the same symbols as in Section~\ref{Proof:Theorem:FullWeb:Problem:Termination}. That is, $\mathcal{W}$ denotes the countably infinite set of all words that describe TMs. $M(w)$ denote the machine described by $w$ (for all $w \in \mathcal{W}$); $c^{w,x}$ denotes the computation of $M(w)$ on input $x$; $\symURI_i^{w,x} \in \symAllURIs$ identifies the $i$-th step in computation $c^{w,x}$. The set of all these identifiers is denoted by $\symAllURIs_\mathsf{TMsteps}$ (recall that, although $\symAllURIs_\mathsf{TMsteps}$ is infinite, it is countable).

We now define $\WoD_\mathsf{TMs2}$ as a Web of Linked Data $( D_\mathsf{TMs2},\dataFct_\mathsf{TMs2}, \adocFct_\mathsf{TMs2} )$ similar to the Web $\WoD_\mathsf{TMs}$ used in Section~\ref{Proof:Theorem:FullWeb:Problem:Termination}: $D_\mathsf{TMs2}$ and $\adocFct_\mathsf{TMs2}$ are the same is in $\WoD_\mathsf{TMs}$. That is, $D_\mathsf{TMs2}$ consists of $\left| \symAllURIs_\mathsf{TMsteps} \right|$ different {\LDdoc}s, each of which corresponds to one of the {\ID}s in $\symAllURIs_\mathsf{TMsteps}$. Mapping $\adocFct_\mathsf{TMs2}$ is bijective and maps each $\symURI_i^{w,x} \in \symAllURIs_\mathsf{TMsteps}$ to the corresponding $d_i^{w,x} \in D_\mathsf{TMs2}$.
Mapping $\dataFct_\mathsf{TMs2}$ for $\WoD_\mathsf{TMs2}$ is different from the corresponding mapping for $\WoD_\mathsf{TMs}$:  The set $\dataFct_\mathsf{TMs2}\bigl( d_i^{w,x} \bigr)$ of {\triple}s for \emph{each} {\LDdoc} $d_i^{w,x}$ contains a single {\triple} $( \symURI_i^{w,x} \!, \mathsf{first} , \symURI_{1}^{w,x} )$ which associates the corresponding computation step $\symURI_i^{w,x}$ with the first step of the corresponding computation $c^{w,x}$ ($\mathsf{first} \in \symAllURIs$ denotes a {\ID} for this relationship).

Before we come to the reduction we highlight a property of $\WoD_\mathsf{TMs2}$ that is important for our proof. Each {\triple}
$( \symURI_i^{w,x} \!, \mathsf{first} , \symURI_{1}^{w,x} )$ establishes a data link from $d_i^{w,x}$ to $d_{1}^{w,x}$. Hence, the link graph of $\WoD_\mathsf{TMs2}$ consists of an infinite number of separate subgraphs, each of which corresponds to a particular computation $c^{w,x}$, is weakly connected, and has a star-like form where the corresponding $d_{1}^{w,x}$ is in the center of the star. More precisely, for subgraph $(V^{w_j,x_k},E^{w_j,x_k})$ that corresponds to computation $c^{w_j,x_k}$ it holds
	\begin{align*}
		V^{w_j,x_k} &= \big\lbrace d_i^{w,x} \in D_\mathsf{TMs2} \,\big|\, w = w_j \text{ and } x = x_k \big\rbrace
		\intertext{and}
 		E^{w_j,x_k} &=  V^{w_j,x_k} \times \big\lbrace d_{1}^{w_j,x_k} \big\rbrace .
	\end{align*}
Each of these subgraphs is infinitely large (i.e.~has an infinite number of vertices) if and only if the corresponding computation halts.

For the reduction we use mapping $f$ which is defined as follows: Let
	$w$ be the description of a TM, let $x$ be a possible input word for $M(w)$,
and let $?v \in \symAllVariables$ be a query variable, then $f( w,x ) = \bigl( \WoD_\mathsf{TMs2}, P_{w,x} \bigr)$ with $P_{w,x} = ( ?v, \mathsf{first} ,\symURI_1^{w,x} )$. Given that $\WoD_\mathsf{TMs2}$ is independent of $(w,x)$, it is easy to see that $f$ is computable by TMs (including LD machines).

To show that \problemName{Finiteness(\LDSPARQLFull)} is not LD machine decidable, suppose it were LD machine decidable. In such a case an LD machine could answer the halting problem for any input $(w,x)$ \removable{as follows}: $M(w)$ halts on $x$ if and only if $\evalFullPW{P_{w,x}}{\WoD_\mathsf{TMs2}}$ is finite. However, we know the halting problem is undecidable for TMs (which includes LD machines). Hence, we have a contradiction and\removable{, thus,} \problemName{Finiteness(\LDSPARQLFull)} cannot be LD machine decidable.
\subsection{Proof of Proposition~\ref{Proposition:Reach:ComparisonToFullWeb}, Case~\ref{Proposition:Reach:ComparisonToFullWeb:Case1}}
\begin{proofdeclitems}
	\item $\evalFullFctP{P}$ be a {\LDSPARQLFull} query that is monotonic;
	\item $\evalReachFctPSc{P}{S}{c}$ be a {\LDSPARQLReach} query that uses the same SPARQL expression $P$ as $\evalFullFctP{P}$ (and an arbitrary reachability criterion $c$ and (finite) set $S \subset \symAllURIs$ of seed {\ID}s);
	\item $\WoD$ be a Web of Linked Data; and
	\item $\ReachPartScPW{S}{c}{P}{\WoD}$ denote the $(S,c,P)$-reach\-able part of $\WoD$.
\end{proofdeclitems}
W.l.o.g.~it suffices to show: $\evalReachPScW{P}{S}{c}{\WoD} \subseteq \evalFullPW{P}{\WoD}$.

\vspace{1ex} \noindent
Since $\ReachPartScPW{S}{c}{P}{\WoD}$ is an induced subweb of $\WoD$ (cf.~Definition~\ref{Definition:ReachablePart}) and $\evalFullFctP{P}$ is monotonic, it holds $\evalFullPW{P}{ \ReachPartScPW{S}{c}{P}{\WoD} } \subseteq \evalFullPW{P}{\WoD}$.
	Furthermore, we have $\evalReachPScW{P}{S}{c}{\WoD} = \evalFullPW{P}{ \ReachPartScPW{S}{c}{P}{\WoD} }$ (cf.~Proposition~\ref{Proposition:Reach:ComparisonToFullWeb}, case~\ref{Proposition:Reach:ComparisonToFullWeb:Case2}).
Hence, $\evalReachPScW{P}{S}{c}{\WoD} \subseteq \evalFullPW{P}{\WoD}$.

\subsection{Proof of Proposition~\ref{Proposition:Reach:ComparisonToFullWeb}, Case~\ref{Proposition:Reach:ComparisonToFullWeb:Case2}}
\begin{proofdeclitems}
	\item $\evalReachFctPSc{P}{S}{c}$ be a {\LDSPARQLReach} query (that uses SPARQL expression $P$, reachability criterion $c$, and (finite) set $S \subset \symAllURIs$ of seed {\ID}s);
	\item $\evalFullFctP{P}$ be a {\LDSPARQLFull} query that uses the same SPARQL expression $P$ as $\evalReachFctPSc{P}{S}{c}$;
	\item $\WoD$ be an arbitrary Web of Linked Data; and
	\item $\ReachPartScPW{S}{c}{P}{\WoD}$ denote the $(S,c,P)$-reach\-able part of $\WoD$.
\end{proofdeclitems}
It holds:
\begin{itemize}
	\item $\evalReachFctPSc{P}{S}{c}(\WoD) = \evalOrigPG{P}{\fctAllDataName( \ReachPartScPW{S}{c}{P}{\WoD} )}$ (cf.~Definition~\ref{Definition:ReachSemantics}) and
	\item $\evalFullFctP{P}( \ReachPartScPW{S}{c}{P}{\WoD} ) = \evalOrigPG{P}{\fctAllDataName( \ReachPartScPW{S}{c}{P}{\WoD} )}$ (cf.~Definition~\ref{Definition:FullWebSemantics}).
\end{itemize}
Hence, $\evalReachFctPSc{P}{S}{c}(\WoD) = \evalFullPW{P}{ \ReachPartScPW{S}{c}{P}{\WoD} }$.

\subsection{Proof of Proposition~\ref{Proposition:Reach:FiniteWeb}} \label{Proof:Proposition:Reach:FiniteWeb}
\begin{proofdeclitems}
	\item $\evalReachFctPSc{P}{S}{c}$ be a {\LDSPARQLReach} query (that uses SPARQL expression $P$, reachability criterion $c$, and (finite) set $S \subset \symAllURIs$ of seed {\ID}s);
	\item $\WoD = (D,\dataFct,\adocFct)$ be a \emph{finite} Web of Linked Data; and
	\item $\ReachPartScPW{S}{c}{P}{\WoD} = (\Reach{D},\Reach{\dataFct},\Reach{\adocFct})$ be the $(S,c,P)$-reach\-able part of $\WoD$.
\end{proofdeclitems}
W.l.o.g.~it suffices to show: $\evalReachPScW{P}{S}{c}{\WoD}$ is finite and $\ReachPartScPW{S}{c}{P}{\WoD}$ is finite.

\vspace{1ex} \noindent
$\WoD$ is finite, which means $D$ is finite. Therefore, any subset of $D$ must also be finite; this includes $\Reach{D} \subseteq D$ because $\ReachPartScPW{S}{c}{P}{\WoD}$ is an induced subweb of $\WoD$ (cf.~Definition~\ref{Definition:ReachablePart}). Hence, $\ReachPartScPW{S}{c}{P}{\WoD}$ is finite.

The finiteness of $\evalReachPScW{P}{S}{c}{\WoD}$ follows directly from the finiteness of $\ReachPartScPW{S}{c}{P}{\WoD}$ (and is independent of the finiteness of $\WoD$) as the following lemma shows.
\begin{lemma} \label{SubLemma:Reach:FiniteWeb}
	For any {\LDSPARQLReach} query $\evalReachFctPSc{P}{S}{c}$ and any (potentially infinite) Web of Linked Data $\WoD$ it holds:
	If $\ReachPartScPW{S}{c}{P}{\WoD}$ is finite, then $\evalReachPScW{P}{S}{c}{\WoD}$ is finite.
\end{lemma}

\begin{myproof}{of Lemma~\ref{SubLemma:Reach:FiniteWeb}}
The lemma follows directly from Propositions~\ref{Proposition:Reach:ComparisonToFullWeb} (case~\ref{Proposition:Reach:ComparisonToFullWeb:Case2}) and~\ref{Proposition:FullWeb:FiniteWeb}.
\end{myproof}

\subsection{Proof of Proposition~\ref{Proposition:Reach:InfiniteneWebFindings}}
\begin{proofdeclitems}
	\item $P$ be a SPARQL expression;
	\item $c$ and $c'$ be reachability criteria;
	\item $S \subset \symAllURIs$ be a finite but nonempty set of seed {\ID}s;
	\item $\WoD = (D,\dataFct,\adocFct)$ be an \emph{infinite} Web of Linked Data.
\end{proofdeclitems}

\noindent
\textbf{\ref{Proposition:Reach:InfiniteneWebFindings:Case1}. $\ReachPartScPW{S}{\cNone}{P}{\WoD}$ is always finite; so is $\evalReachPScW{P}{S}{\cNone}{\WoD}$.}
\\
Let $\Reach{D}$ denote the set of all {\LDdoc}s in $\ReachPartScPW{S}{\cNone}{P}{\WoD}$. Since $\cNone$ always returns $\false$ it is easily verified that there is no {\LDdoc} $d \in D$ that satisfies case~\ref{DefinitionCase:QualifiedReachability:IndStep} in Definition~\ref{Definition:QualifiedReachability}. Hence, it must hold $\Reach{D} = \lbrace d \in D \,|\, \symURI \in S \text{ and } \adocFct(\symURI) = d \rbrace$ (cf.~case~\ref{DefinitionCase:QualifiedReachability:IndBegin} in Definition~\ref{Definition:QualifiedReachability}). Since $S$ is finite we see that $\Reach{D}$ is guaranteed to be finite (and so is $\ReachPartScPW{S}{\cNone}{P}{\WoD}$).
The finiteness of $\evalReachPScW{P}{S}{\cNone}{\WoD}$ can then be shown using Lemma~\ref{SubLemma:Reach:FiniteWeb} (cf.~Section~\ref{Proof:Proposition:Reach:FiniteWeb}).


\vspace{1ex} \noindent
\textbf{\ref{Proposition:Reach:InfiniteneWebFindings:Case3}. If $\ReachPartScPW{S}{c}{P}{\WoD}$ is finite, then $\evalReachPScW{P}{S}{c}{\WoD}$ is finite.}
\\
See Lemma~\ref{SubLemma:Reach:FiniteWeb} in Section~\ref{Proof:Proposition:Reach:FiniteWeb}.

\vspace{1ex} \noindent
\textbf{\ref{Proposition:Reach:InfiniteneWebFindings:Case4}. If $\evalReachPScW{P}{S}{c}{\WoD}$ is infinite, then $\ReachPartScPW{S}{c}{P}{\WoD}$ is infinite.}
\\
Let $\evalReachPScW{P}{S}{c}{\WoD}$ be infinite. We prove by contradiction that $\ReachPartScPW{S}{c}{P}{\WoD}$ is infinite: Suppose $\ReachPartScPW{S}{c}{P}{\WoD}$ were finite. In this case $\evalReachPScW{P}{S}{c}{\WoD}$ would be finite (cf.~Lemma~\ref{SubLemma:Reach:FiniteWeb} in Section~\ref{Proof:Proposition:Reach:FiniteWeb}), which is a contradiction to our premise. Hence, $\ReachPartScPW{S}{c}{P}{\WoD}$ must be infinite.

\vspace{1ex} \noindent
\textbf{\ref{Proposition:Reach:InfiniteneWebFindings:Case5}. If $c$ is less restrictive than $c'$ and $\ReachPartScPW{S}{c}{P}{\WoD}$ is finite, then $\ReachPartScPW{S}{c'}{P}{\WoD}$ is finite.}
\\
If $\ReachPartScPW{S}{c}{P}{\WoD}$ is finite, then exist finitely many {\LDdoc}s $d\in D$ that are $(c,P)$-reach\-able from $S$ in $\WoD$. A subset of them is also $(c',P)$-reach\-able from $S$ in $\WoD$ because $c$ is less restrictive than $c'$. Hence, $\ReachPartScPW{S}{c'}{P}{\WoD}$ must also be finite.

\vspace{1ex} \noindent
\textbf{\ref{Proposition:Reach:InfiniteneWebFindings:Case6}. If $c'$ is less restrictive than $c$ and $\ReachPartScPW{S}{c}{P}{\WoD}$ is infinite, then $\ReachPartScPW{S}{c'}{P}{\WoD}$ is infinite.}
\\
If $\ReachPartScPW{S}{c}{P}{\WoD}$ is infinite, then exist infinitely many {\LDdoc}s $d\in D$ that are $(c,P)$-reach\-able from $S$ in $\WoD$. Each of them is also $(c',P)$-reach\-able from $S$ in $\WoD$ because $c'$ is less restrictive than $c$. Hence, $\ReachPartScPW{S}{c'}{P}{\WoD}$ must be infinite.

\subsection{Proof of Theorem~\ref{Theorem:Reach:Problem:Finiteness}}
\noindent
We formally define the decision problems \problemName{Fi\-nite\-nessReachablePart} and \problemName{Fi\-nite\-ness(\LDSPARQLReach)} as follows:

\vspace{3mm}
\webproblem{FinitenessReachablePart}
{a (potentially infinite) Web of Linked Data $\WoD$}
{a finite but nonempty set $S \subset \symAllURIs$ \removable{of seed {\ID}s} \par
a reachability criterion $c$ that is less restrictive than $\cNone$ \par
a SPARQL expression $P$}
{Is the $(S,c,P)$-reach\-able part of $\WoD$ finite?}

\vspace{2mm}

\webproblem{Finiteness(\LDSPARQLReach)}
{a (potentially infinite) Web of Linked Data $\WoD$}
{a finite but nonempty set $S \subset \symAllURIs$ \removable{of seed {\ID}s} \par
a reachability criterion $c$ that is less restrictive than $\cNone$ \par
a SPARQL expression $P$}
{Is the result of {\LDSPARQLReach} query $\evalReachFctPSc{P}{S}{c}$ over $\WoD$ finite?}
\vspace{2mm}

\noindent
We
	now prove Theorem~\ref{Theorem:Reach:Problem:Finiteness}
by reducing the halting problem to \problemName{FinitenessReachablePart} and \problemName{Finiteness(\LDSPARQLReach)}. While this proof is similar to the proofs of Theorem~\ref{Theorem:FullWeb:Problem:Termination} (cf.~Section~\ref{Proof:Theorem:FullWeb:Problem:Termination}) and Theorem~\ref{Theorem:FullWeb:Problem:Finiteness} (cf.~Section~\ref{Proof:Theorem:FullWeb:Problem:Finiteness}), we now use a Web of Linked Data $\WoD_\mathsf{TMs3}$ which (again) differs from $\WoD_\mathsf{TMs}$ and $\WoD_\mathsf{TMs2}$ in the way it describes all possible computations of all Turing machines (TM).

For the proof we use the same symbols as in Section~\ref{Proof:Theorem:FullWeb:Problem:Termination}. That is, $\mathcal{W}$ denotes the countably infinite set of all words that describe TMs. $M(w)$ denote the machine described by $w$ (for all $w \in \mathcal{W}$); $c^{w,x}$ denotes the computation of $M(w)$ on input $x$; $\symURI_i^{w,x} \in \symAllURIs$ identifies the $i$-th step in computation $c^{w,x}$. The set of all these identifiers is denoted by $\symAllURIs_\mathsf{TMsteps}$ (recall that, although $\symAllURIs_\mathsf{TMsteps}$ is infinite, it is countable).

We now define $\WoD_\mathsf{TMs3}$ as a Web of Linked Data $( D_\mathsf{TMs3},\dataFct_\mathsf{TMs3}, \adocFct_\mathsf{TMs3} )$ similar to the Web $\WoD_\mathsf{TMs}$ used in Section~\ref{Proof:Theorem:FullWeb:Problem:Termination}: $D_\mathsf{TMs3}$ and $\adocFct_\mathsf{TMs3}$ are the same is in $\WoD_\mathsf{TMs}$. That is, $D_\mathsf{TMs3}$ consists of $\left| \symAllURIs_\mathsf{TMsteps} \right|$ different {\LDdoc}s, each of which corresponds to one of the {\ID}s in $\symAllURIs_\mathsf{TMsteps}$. Mapping $\adocFct_\mathsf{TMs3}$ is bijective and maps each $\symURI_i^{w,x} \in \symAllURIs_\mathsf{TMsteps}$ to the corresponding $d_i^{w,x} \in D_\mathsf{TMs3}$.
Mapping $\dataFct_\mathsf{TMs3}$ for $\WoD_\mathsf{TMs3}$ is different from the corresponding mapping for $\WoD_\mathsf{TMs}$: The set $\dataFct_\mathsf{TMs3}\bigl( d_i^{w,x} \bigr)$ of {\triple}s for an {\LDdoc} $d_i^{w,x}$ is empty if computation $c^{w,x}$ halts with the $i$-th computation step. Otherwise, $\dataFct_\mathsf{TMs3}\bigl( d_i^{w,x} \bigr)$ contains a single {\triple} $( \symURI_i^{w,x} \!, \mathsf{next} , \symURI_{i+1}^{w,x} )$ which associates the computation step $\symURI_i^{w,x}$ with the next step in $c^{w,x}$ ($\mathsf{next} \in \symAllURIs$ denotes a {\ID} for this relationship).
	Formally:
	\begin{equation*}
		\dataFct_\mathsf{TMs3}\bigl( d_i^{w,x} \bigr) \!=\!
		\begin{cases}
			\emptyset & \text{if computation $c^{w,x}$ halts} \\
			& \text{with the $i$-th step,} \\
			\lbrace ( \symURI_i^{w,x} \!, \mathsf{next} , \symURI_{i+1}^{w,x} ) \rbrace & \text{else.}
		\end{cases}
	\end{equation*}
Mapping $\dataFct_\mathsf{TMs3}$ is (Turing) computable because a universal TM may determine by simulation whether the computation of a particular TM on a particular input halts after a given number of steps.

Before we come to the reduction we highlight a property of $\WoD_\mathsf{TMs3}$ that is important for our proof. Each {\triple}
$( \symURI_i^{w,x} \!, \mathsf{next} , \symURI_{i+1}^{w,x} )$ establishes a data link from $d_i^{w,x}$ to $d_{i+1}^{w,x}$. Due to such links we recursively may reach all {\LDdoc}s about all steps in a particular computation of any TM. Hence, for each possible computation $c^{w,x}$ of any TM $M(w)$ we have a (potentially infinite) simple path $\left( d_1^{w,x} \!, ... \, , d_i^{w,x} \!, ... \right)$ in the link graph of $\WoD_\mathsf{TMs3}$. Each of these paths is finite if and only if the corresponding computation halts. Finally, we note that each of these paths forms a separate subgraph of the link graph of $\WoD_\mathsf{TMs3}$ because
	we use a separate set of step {\ID}s for each computation and the {\triple}s in the corresponding {\LDdoc}s only mention steps from the same computation.

For the reduction we use mapping $f$ which is defined as follows: Let
	$w$ be the description of a TM, let $x$ be a possible input word for $M(w)$,
and let $?a,?b \in \symAllVariables$ be two distinct query variables,
then $f( w,x ) = \bigl( \WoD_\mathsf{TMs3}, S_{w,x}, \cMatch, P_{w,x} \bigr)$ with $S_{w,x} = \big\lbrace \symURI_1^{w,x} \big\rbrace$ and $P_{w,x} = (?a, \mathsf{next} , ?b)$. Given that $\cMatch$ and $\WoD_\mathsf{TMs3}$ are independent of $(w,x)$, it can be easily seen that $f$ is computable by TMs (including LD machines).

To show that \problemName{FinitenessReachablePart} is not LD machine decidable, suppose it were LD machine decidable. In such a case an LD machine could answer the halting problem for any input $(w,x)$ \removable{as follows}: $M(w)$ halts on $x$ if and only if the $(S_{w,x},\cMatch,P_{w,x})$-reach\-able part of $\WoD_\mathsf{TMs3}$ is finite. However, we know the halting problem is undecidable for TMs (which includes LD machines). Hence, we have a contradiction and\removable{, thus,} \problemName{FinitenessReachablePart} cannot be LD machine decidable.

The proof that \problemName{Finiteness(\LDSPARQLReach)} is undecidable is similar to that for \problemName{FinitenessReachablePart}. Hence, we only outline the idea: Instead of reducing the halting problem to \problemName{FinitenessReachablePart} based on mapping $f$ we now reduce the halting problem to \problemName{FinitenessQueryResult} using the same mapping. If \problemName{Finiteness(\LDSPARQLReach)} were decidable then we could answer the halting problem for any $(w,x)$: $M(w)$ halts on $x$ if and only if $\evalReachPScW{P_{w,x}}{S_{w,x}}{\cMatch}{\WoD_\mathsf{TMs3}}$ is finite.

\subsection{Proof of Proposition~\ref{Proposition:Reach:SatAndMonoAsInSPARQL}, Case~\ref{Proposition:Reach:SatAndMonoAsInSPARQL:CaseSat}} \label{Proof:Proposition:Reach:SatAndMonoAsInSPARQL:CaseSat}
\noindent
This proof is similar to the proof of Proposition~\ref{Proposition:FullWeb:SatAndMonoAsInSPARQL}, case~\ref{Proposition:FullWeb:SatAndMonoAsInSPARQL:CaseSat} (cf.~Section~\ref{Proof:Proposition:FullWeb:SatAndMonoAsInSPARQL:CaseSat}).

\vspace{1ex} \noindent
\textbf{If:}
Let $P$ be a SPARQL expression that is satisfiable and let $\evalReachFctPSc{P}{S}{c}$ be an arbitrary {\LDSPARQLReach} query that uses $P$, a nonempty \removable{set} $S \subset \symAllURIs$ \removable{of seed {\ID}s}, and an arbitrary reachability criterion $c$. W.l.o.g.~it suffices to show that $\evalReachFctPSc{P}{S}{c}$ is satisfiable.

For this proof we use the notion of $(P,G)$-lineage of valuations that we introduced before (cf.~Definition~\ref{Definition:Lineage} in Section~\ref{Proof:Proposition:FullWeb:SatAndMonoAsInSPARQL:CaseSat}). Recall that for any SPARQL expression $P$, any (potentially infinite) set $G$ of {\triple}s, and any valuation $\mu \in \evalOrigPG{P}{G}$ it holds: i)~$G' = \mathrm{lin}^{P,G}(\mu)$ is finite and ii)~$\mu \in \evalOrigPG{P}{G'}$.

Due to the satisfiability of $P$ exists a set of {\triple}s $G$ such that $\evalOrigPG{P}{G} \neq \emptyset$. W.l.o.g., let $\mu$ be an arbitrary solution for $P$ in $G$, that is, $\mu \in \evalOrigPG{P}{G}$. Furthermore, let $G' = \mathrm{lin}^{P,G}(\mu)$ be the $(P,G)$-lineage of $\mu$. We use $G'$ to construct a Web of Linked Data $\WoD_\mu= ( D_\mu,\dataFct_\mu,\adocFct_\mu )$ which consists of a single {\LDdoc}. This document may be retrieved using any {\ID} from the (nonempty) set $S$ and it contains the $(P,G)$-lineage of $\mu$ (recall that the lineage is guaranteed to be a finite). Formally:
\begin{align*}
	D_\mu = \lbrace d \rbrace &&
	\dataFct_\mu(d) = G' &&
	\forall \, \symURI \in S : \adocFct_\mu(\symURI) = d
\end{align*}
Due to our construction it holds $\fctAllDataName(\WoD_\mu) = \fctAllDataName( \WoD_\mathfrak{R} ) = G'$ where $\WoD_\mathfrak{R}$ denotes the $(S,c,P)$-reach\-able part of $\WoD_\mu$. Thus, we have $\evalReachPScW{P}{S}{c}{\WoD_\mu} = \evalOrigPG{P}{G'}$ (cf.~Definition~\ref{Definition:ReachSemantics}). Since we know $\mu \in \evalOrigPG{P}{G'}$ it holds $\evalReachPScW{P}{S}{c}{\WoD_\mu} \neq \emptyset$, which shows that $\evalReachFctPSc{P}{S}{c}$ is satisfiable.

\vspace{1ex} \noindent
\textbf{Only if:} 
Let $\evalReachFctPSc{P}{S}{c}$ be a satisfiable {\LDSPARQLReach} query that uses SPARQL expression $P$, a nonempty \removable{set} $S \subset \symAllURIs$ \removable{of seed {\ID}s}, and an arbitrary reachability criterion $c$. Since $\evalReachFctPSc{P}{S}{c}$ is satisfiable, exists a Web of Linked Data $\WoD$ such that $\evalReachPScW{P}{S}{c}{\WoD} \neq \emptyset$. According to Definition~\ref{Definition:ReachSemantics} we also have $\evalReachPScW{P}{S}{c}{\WoD} = \evalOrigPG{P}{\fctAllDataName(\WoD)}$. Thus, we conclude that $P$ is satisfiable.

\subsection{Proof of Proposition~\ref{Proposition:Reach:SatAndMonoAsInSPARQL}, Case~\ref{Proposition:Reach:SatAndMonoAsInSPARQL:CaseNonTrivSat}} \label{Proof:Proposition:Reach:SatAndMonoAsInSPARQL:CaseNonTrivSat}
\noindent
This proof is similar to the proof of Proposition~\ref{Proposition:FullWeb:SatAndMonoAsInSPARQL}, case~\ref{Proposition:FullWeb:SatAndMonoAsInSPARQL:CaseNonTrivSat} (cf.~Section~\ref{Proof:Proposition:FullWeb:SatAndMonoAsInSPARQL:CaseNonTrivSat}).

\vspace{1ex} \noindent
\textbf{If:}
Let $P$ be a SPARQL expression that is nontrivially satisfiable and let $\evalReachFctPSc{P}{S}{c}$ be an arbitrary {\LDSPARQLReach} query that uses $P$, a nonempty \removable{set} $S \subset \symAllURIs$ \removable{of seed {\ID}s}, and an arbitrary reachability criterion $c$. W.l.o.g.~it suffices to show that $\evalReachFctPSc{P}{S}{c}$ is nontrivially satisfiable.

Due to the nontrivial satisfiability of $P$ exists a set of {\triple}s $G$ and a valuation $\mu$ such that i)~$\mu \in \evalOrigPG{P}{G}$ and ii)~$\fctDom{\mu} \neq \emptyset$. Let $G' = \mathrm{lin}^{P,G}(\mu)$ be the $(P,G)$-lineage of $\mu$. We use $G'$ to construct a Web of Linked Data $\WoD_\mu= ( D_\mu,\dataFct_\mu,\adocFct_\mu )$ which consists of a single {\LDdoc}. This document may be retrieved using any {\ID} from the (nonempty) set $S$ and it contains the $(P,G)$-lineage of $\mu$ (recall that the lineage is guaranteed to be a finite). Formally:
\begin{align*}
	D_\mu = \lbrace d \rbrace &&
	\dataFct_\mu(d) = G' &&
	\forall \, \symURI \in S : \adocFct_\mu(\symURI) = d
\end{align*}
Due to our construction it holds $\fctAllDataName(\WoD_\mu) = \fctAllDataName( \WoD_\mathfrak{R} ) = G'$ where $\WoD_\mathfrak{R}$ denotes the $(S,c,P)$-reach\-able part of $\WoD_\mu$. Thus, we have $\evalReachPScW{P}{S}{c}{\WoD_\mu} = \evalOrigPG{P}{G'}$ (cf.~Definition~\ref{Definition:ReachSemantics}). Since we know $\mu \in \evalOrigPG{P}{G'}$ and $\fctDom{\mu} \neq \emptyset$, we conclude that $\evalReachFctPSc{P}{S}{c}$ is nontrivially satisfiable.

\vspace{1ex} \noindent
\textbf{Only if:} 
Let $\evalReachFctPSc{P}{S}{c}$ be a nontrivially satisfiable {\LDSPARQLReach} query that uses SPARQL expression $P$, a nonempty \removable{set}
	$S \subset \symAllURIs$,
and an arbitrary reachability criterion $c$. Since $\evalReachFctPSc{P}{S}{c}$ is nontrivially satisfiable, exists a Web of Linked Data $\WoD$ and a valuation $\mu$ such that i)~$\mu \in \evalReachPScW{P}{S}{c}{\WoD}$ and ii)~$\fctDom{\mu} \neq \emptyset$. According to Definition~\ref{Definition:ReachSemantics} we also have $\evalReachPScW{P}{S}{c}{\WoD} = \evalOrigPG{P}{\fctAllDataName(\WoD)}$. Thus, we conclude that $P$ is nontrivially satisfiable.

\subsection{Proof of Proposition~\ref{Proposition:Reach:SatAndMonoAsInSPARQL}, Case~\ref{Proposition:Reach:SatAndMonoAsInSPARQL:CaseMono}} \label{Proof:Proposition:Reach:SatAndMonoAsInSPARQL:CaseMono}
\noindent
\textbf{If:}
\begin{proofdeclitems}
	\item $P$
		be a SPARQL expression that is monotonic;
	\item $\evalReachFctPSc{P}{S}{c}$
		be an arbitrary {\LDSPARQLReach} query that uses $P$, a nonempty \removable{set} $S \subset \symAllURIs$ \removable{of seed {\ID}s}, and an arbitrary reachability criterion $c$; and
	\item $\WoD_1,\WoD_2$
		be an arbitrary pair of Webs of Linked Data such that $\WoD_1$ is an induced subweb of $\WoD_2$; and
	\item $\WoD_{\mathfrak{R}1} = (D_{\mathfrak{R}1},\dataFct_{\mathfrak{R}1},\adocFct_{\mathfrak{R}1})$ and $\WoD_{\mathfrak{R}2} = (D_{\mathfrak{R}2},\dataFct_{\mathfrak{R}2},\adocFct_{\mathfrak{R}2})$
		denote the $(S,c,P)$-reach\-able part of $\WoD_1$ and of $\WoD_2$, respectively.
\end{proofdeclitems}
To prove that $\evalReachFctPSc{P}{S}{c}$ is monotonic it suffices to show $\evalReachPScW{P}{S}{c}{\WoD_1} \subseteq \evalReachPScW{P}{S}{c}{\WoD_2}$.

Any {\LDdoc} that is $(c,P)$-reach\-able from $S$ in $\WoD_1$ is also $(c,P)$-reach\-able from $S$ in $\WoD_2$ because $\WoD_1$ is an induced subweb of $\WoD_2$. Hence, $D_{\mathfrak{R}1} \subseteq D_{\mathfrak{R}2}$ and, thus, $\fctAllDataName(\WoD_{\mathfrak{R}1}) \subseteq \fctAllDataName(\WoD_{\mathfrak{R}2})$.
Furthermore,
$\evalReachPScW{P}{S}{c}{\WoD_1} = \evalOrigPG{P}{\fctAllDataName(\WoD_{\mathfrak{R}1})}$ and $\evalReachPScW{P}{S}{c}{\WoD_2} = \evalOrigPG{P}{\fctAllDataName(\WoD_{\mathfrak{R}2})}$ (cf.~Definition~\ref{Definition:ReachSemantics}).
Due to the monotonicity of $P$ it also holds $\evalOrigPG{P}{\fctAllDataName(\WoD_{\mathfrak{R}1})} \subseteq \evalOrigPG{P}{\fctAllDataName(\WoD_{\mathfrak{R}2})}$. Hence, $\evalReachPScW{P}{S}{c}{\WoD_1} \subseteq \evalReachPScW{P}{S}{c}{\WoD_2}$.

%
%

\subsection{Proof of Proposition~\ref{Proposition:Reach:MonotonicityCNone}}
\begin{proofdeclitems}
	\item $\evalReachFctPSc{P}{S}{\cNone}$ be a {\LDSPARQLReach} query (under $\cNone$-semantics) such that $\left|S\right|=1$;
	\item $\WoD_1 = ( D_1,\dataFct_1, \adocFct_1 )$ and $\WoD_2 = ( D_2,\dataFct_2, \adocFct_2 )$ be two Webs of Linked Data such that $\WoD_1$ is an induced subweb of $\WoD_2$; and
	\item $\WoD^\mathcal{R}_1$ and $\WoD^\mathcal{R}_2$ denote the $(S,\cNone,P)$-reach\-able part of $\WoD_1$ and of $\WoD_2$, respectively.
\end{proofdeclitems}
W.l.o.g.~it suffices to show $\evalReachPScW{P}{S}{\cNone}{\WoD_1} \subseteq \evalReachPScW{P}{S}{\cNone}{\WoD_2}$.
We distinguish the following three cases for $\symURI \in S = \lbrace \symURI \rbrace$:
\begin{enumerate}
	\item $\symURI \notin \fctDom{\adocFct_1}$ and $\symURI \notin \fctDom{\adocFct_2}$. \par
		In this case $\WoD^\mathcal{R}_1$ and $\WoD^\mathcal{R}_2$ are equal to the empty Web (which contains no {\LDdoc}s). Hence, $\evalReachPScW{P}{S}{\cNone}{\WoD_1} = \evalReachPScW{P}{S}{\cNone}{\WoD_2} = \emptyset$.
	\item $\symURI \notin \fctDom{\adocFct_1}$ and $\adocFct_2(\symURI) = d$ where $d \in D_2$. \par
		In this case $\WoD^\mathcal{R}_1$ is equal to the empty Web, whereas $\WoD^\mathcal{R}_2$ contains a single {\LDdoc}, namely $d$. Hence, $\evalReachPScW{P}{S}{\cNone}{\WoD_1} = \emptyset$ and $\evalReachPScW{P}{S}{\cNone}{\WoD_2} = \evalOrigPG{P}{\fctAllDataName(\WoD^\mathcal{R}_2)} = \evalOrigPG{P}{\dataFct_2(d)}$ and, thus, $\evalReachPScW{P}{S}{\cNone}{\WoD_1} \subseteq \evalReachPScW{P}{S}{\cNone}{\WoD_2}$.
	\item $\adocFct_1(\symURI) = d$ and $\adocFct_2(\symURI) = d$ where $d \in D_1 \subseteq D_2$. \par
		In this case both Webs, $\WoD^\mathcal{R}_1$ and $\WoD^\mathcal{R}_2$, contain a single {\LDdoc}, namely $d$. Hence, $\evalReachPScW{P}{S}{\cNone}{\WoD_1} = \evalReachPScW{P}{S}{\cNone}{\WoD_2} = \evalOrigPG{P}{\dataFct_2(d)}$ (recall, in this case holds $\dataFct_1(d) = \dataFct_2(d)$).
	\item $\symURI \in \fctDom{\adocFct_1}$ and $\symURI \notin \fctDom{\adocFct_2}$. \par
		This case is impossible because $\WoD_1$ is an induced subweb of $\WoD_2$.
\end{enumerate}

\subsection{Proof of Lemma~\ref{Lemma:Reach:TerminationCriterion}} \label{Proof:Lemma:Reach:TerminationCriterion}
\noindent
For proving Lemma~\ref{Lemma:Reach:TerminationCriterion} we introduce specific LD machines for {\LDSPARQLReach} queries. We call these machines $(P,S,c)'$-machines. The $(P,S,c)'$-machine for {\LDSPARQLReach} query $\evalReachFctPSc{P}{S}{c}$ implements a generic (i.e.~input independent) computation of $\evalReachFctPSc{P}{S}{c}$. 
For any nontrivially satisfiable {\LDSPARQLReach} query $\evalReachFctPSc{P}{S}{c}$ we shall see that if and only if the $(S,c,P)$-reach\-able part of an arbitrary Web of Linked Data $\WoD$ is finite, the corresponding $(P,S,c)'$-machine computes $\evalReachPScW{P}{S}{c}{\WoD}$ and halts. Formally, we define $(P,S,c)'$-machines as follows:

\begin{definition} \label{Definition:ReachFiniteQueryMachine}
	Let $S \subset \symAllURIs$ be a finite set of \removable{seed} {\ID}s;
	let $c$ be a reachability criterion; and
	let $P$ be a SPARQL expression.
	The \definedTerm{$(P,S,c)'$-machine} is an LD machine that implements Algorithm~\ref{Algorithm:ReachFiniteQueryMachine}. This algorithm
		makes use of
	a
	subroutine
		called
	\code{lookup}. This subroutine, when called with a {\ID} $\symURI \in \symAllURIs$,
	i)~writes $\fctEnc{\symURI}$ to the right end of the word on the link traversal tape,
	ii)~enters the expand state, and
	iii)~performs the expand procedure as specified in Definition~\ref{Definition:LDMachine}.
\end{definition}

\begin{algorithm}[tb]
	\caption{\, The program of the $(P,S,c)'$-machine.} \label{Algorithm:ReachFiniteQueryMachine}
	\begin{algorithmic}[1]
			\STATE {Call \code{lookup} for each $\symURI \in S$.}
		\label{Line:ReachFiniteQueryMachine:Init}

			\medskip
		\STATE {$expansionCompleted := \false$}
		\WHILE { $expansionCompleted = \false$ } \label{Line:ReachFiniteQueryMachine:LoopBegin}
			\STATE {Scan the link traversal tape for an {\triple} $t$ and a {\ID} $\symURI \in \fctIDs{t}$ such that i)~$c(t,\symURI,P)=\true$ and ii)~the word on the link traversal tape neither contains \,$\fctEnc{\symURI} \, \fctEnc{ \adocFct(\symURI) } \, \sharp$\, nor \,$\fctEnc{\symURI} \, \sharp$. If such $t$ and $\symURI$ exist, call \code{lookup} for $\symURI$; otherwise $expansionCompleted := \true$.} \label{Line:ReachFiniteQueryMachine:LoopStep}
		\ENDWHILE \label{Line:ReachFiniteQueryMachine:LoopEnd}

			\medskip
		\STATE {Let $G$ denote the set of all {\triple}s currently encoded on the link traversal tape. For each $\mu \in \evalOrigPG{P}{G}$ add $\fctEnc{\mu}$ to the output.} \label{Line:ReachFiniteQueryMachine:OutputResult}
	\end{algorithmic}
\end{algorithm}

\noindent
Before we complete the proof of Lemma~\ref{Lemma:Reach:TerminationCriterion} we discuss properties of any $(P,S,c)'$-machine as they are relevant for the proof.
The computation of each $(P,S,c)'$-machine
	(with a Web of Linked Data $\WoD$ encoded on its input tape)
starts with an initialization (cf.~line~\ref{Line:ReachFiniteQueryMachine:Init} in Algorithm~\ref{Algorithm:ReachFiniteQueryMachine}). After the initialization, the machine enters a (potentially non-terminating) loop that recursively discovers (i.e.~expands) all {\LDdoc}s of the corresponding reach\-able part of $\WoD$. The following lemma shows that for each such document exists an iteration of the loop during which the machine copies that document to its link traversal tape.
\begin{lemma} \label{SubLemma1:Lemma:Reach:TerminationCriterion}
	\begin{proofdeclitems}
		\vspace{-1mm}
		\item $M^{(P,S,c)'}$ be the $(P,S,c)'$-machine for a SPARQL expression $P$, a finite set $S \subset \symAllURIs$, and a reachability criterion $c$;
		\item $\WoD = (D,\dataFct,\adocFct)$ be a (potentially infinite) Web of Linked Data encoded on the Web tape of $M^{(P,S,c)'}$\!; and
		\item $d \in D$ be an {\LDdoc} that is $(c,P)$-reach\-able from $S$ in $\WoD$.
	\end{proofdeclitems}
	\vspace{-1mm}
	During the execution of Algorithm~\ref{Algorithm:ReachFiniteQueryMachine} by $M^{(P,S,c)'}$
	exists an iteration of the loop (lines~\ref{Line:ReachFiniteQueryMachine:LoopBegin} to~\ref{Line:ReachFiniteQueryMachine:LoopEnd}) after which the word on the link traversal tape of $M^{(P,S,c)'}$ (permanently) contains $\fctEnc{d}$.
\end{lemma}
\begin{myproof}{of Lemma~\ref{SubLemma1:Lemma:Reach:TerminationCriterion}}
	To prove the lemma we first emphasize that $M^{(P,S,c)'}$ only appends to the word on its link traversal tape. Hence, $M^{(P,S,c)'}$ never removes $\fctEnc{d}$ from that word once it has been added. The same holds for the encoding of any other {\LDdoc} $d' \in D$.

	Since $d$ is $(c,P)$-reach\-able from $S$ in $\WoD$, the link graph for $\WoD$ contains at least one finite path $(d_0, ... \, , d_n)$ of {\LDdoc}s $d_i$ where
	i)~$n \in \lbrace 0,1,... \rbrace$,
	ii)~$d_n = d$,
	iii)~$\exists \, \symURI \in S : \adocFct(\symURI) = d_0$,
	and iv)~for each $i \in \lbrace 1,... \,,n \rbrace$ it holds:

	\begin{equation} \label{Equation:Proof:SubLemma1:Lemma:Reach:TerminationCriterion}
		\exists \, t \in \dataFct(d_{i-1}) : \Bigl( \exists \, \symURI \in \fctIDsName(t) : \bigl( \adocFct(\symURI)=d_{i} \text{ and } c(t,\symURI,P) = \true \bigr) \Bigr)
	\end{equation}

	\noindent
	Let $(d_0^*, ... \,, d_n^*)$ be such a path. We use this path to prove the lemma. More precisely, we show by induction over $i \in \lbrace 0,...\,,n \rbrace$ that there exists an iteration after which the word on the link traversal tape of $M^{(P,S,c)'}$ contains \,$\fctEnc{d_n^*}$\, (which is the same as \,$\fctEnc{d}$\, because $d_n^* = d$).

	\vspace{1ex} \noindent
	\textit{Base case} ($i=0$): Since $\exists \, \symURI \in S : \adocFct(\symURI) = d_0^*$ it is easy to verify that after the $0$-th iteration (i.e.~before the first iteration) the word on the link traversal tape of $M^{(P,S,c)'}$ contains \,$\fctEnc{d_0^*}$\, (cf.~line~\ref{Line:ReachFiniteQueryMachine:Init} in Algorithm~\ref{Algorithm:ReachFiniteQueryMachine}).

	\vspace{1ex} \noindent
	\textit{Induction step} ($i>0$): Our inductive hypothesis is: There exists an iteration after which the word on the link traversal tape of $M^{(P,S,c)'}$ contains \,$\fctEnc{d_{i-1}^*}$. Let this be the $j$-th iteration. Based on the hypothesis we show that there exists an iteration after which the word on the link traversal tape of $M^{(P,S,c)'}$ contains \,$\fctEnc{d_{i}^*}$. We distinguish two cases: after the $j$-th iteration the word on the link traversal tape either already contains \,$\fctEnc{d_{i}^*}$\, or it does not contain \,$\fctEnc{d_{i}^*}$. We have to discuss the latter case only.
	Due to (\ref{Equation:Proof:SubLemma1:Lemma:Reach:TerminationCriterion}) exist $t^* \in \dataFct(d_{i-1}^*)$ and $\symURI^* \in \fctIDsName(t^*)$ such that $\adocFct(\symURI^*)=d_{i}^*$ and $c(t^*,\symURI^*,P) = \true$. Hence, there exists a $\delta \in \mathbb{N}^+$ such that $M^{(P,S,c)'}$ finds $t^*$ and $\symURI^*$ in the ($j$+$\delta$)-th iteration. Since $M^{(P,S,c)'}$ calls \code{lookup} for $\symURI^*$ in that iteration (cf.~line~\ref{Line:ReachFiniteQueryMachine:LoopStep} in Algorithm~\ref{Algorithm:ReachFiniteQueryMachine}), the link traversal tape contains \,$\fctEnc{d_{i}^*}$\, after that iteration.
\end{myproof}

\noindent
While Lemma~\ref{SubLemma1:Lemma:Reach:TerminationCriterion} shows that Algorithm~\ref{Algorithm:ReachFiniteQueryMachine} discovers all reach\-able {\LDdoc}s, the following lemma verifies that the algorithm does not copy data from unreach\-able documents to the link traversal tape.

\begin{lemma} \label{SubLemma2:Lemma:Reach:TerminationCriterion}
	\begin{proofdeclitems}
		\vspace{-1mm}
		\item $M^{(P,S,c)'}$ be the $(P,S,c)'$-machine for a SPARQL expression $P$, a finite set $S \subset \symAllURIs$, and a reachability criterion $c$;
		\item $\WoD = (D,\dataFct,\adocFct)$ be a (potentially infinite) Web of Linked Data encoded on the Web tape of $M^{(P,S,c)'}$\!; and
		\item $\ReachPartScPW{S}{c}{P}{\WoD}$ denotes the $(S,c,P)$-reach\-able part of $\WoD$.
	\end{proofdeclitems}
	\vspace{-1mm}
	For any {\triple} $t$ encoded on the link traversal tape of $M^{(P,S,c)'}$ it holds (at any point of the computation): $t \in \fctAllDataName( \ReachPartScPW{S}{c}{P}{\WoD} )$.
\end{lemma}
\begin{myproof}{of Lemma~\ref{SubLemma2:Lemma:Reach:TerminationCriterion}}
	Let $w_j$ denote the word on the link traversal tape of $M^{(P,S,c)'}$ when $M^{(P,S,c)'}$ finishes the $j$-th iteration of the loop in Algorithm~\ref{Algorithm:ReachFiniteQueryMachine}; $w_0$ denotes the corresponding word before the first iteration.
	To prove the lemma it is sufficient to show for each $w_j$ (where $j \in \lbrace 0,1,...\rbrace$) exists a finite sequence $\symURI_1 ,...\,,\symURI_{n_j}$ of $n_j$ different {\ID}s $\symURI_i \in \symAllURIs$ (for all $i \in [1,n_j]$) such that i)~$w_j$ is\footnote{We assume $\fctEncName(\adocFct(\symURI_i))$ is the empty word if $\adocFct(\symURI_i)$ is undefined (i.e.~$\symURI_i \notin \fctDom{\adocFct}$).}
	\begin{equation*}
		\fctEncName(\symURI_1) \, \fctEncName( \adocFct(\symURI_1) ) \, \sharp \, ... \, \sharp \, \fctEncName(\symURI_{n_j}) \, \fctEncName( \adocFct(\symURI_{n_j}) ) \, \sharp
	\end{equation*}
	and
	ii)~for each $i \in [1,n_j]$ either $\symURI_i \notin \fctDom{\adocFct}$ (and, thus, $\adocFct(\symURI_i)$ is undefined) or $\adocFct(\symURI_i)$ is an {\LDdoc} which is $(c,P)$-reach\-able from $S$ in $\WoD$.
	We use an induction over $j$ for this proof.

	\vspace{1ex} \noindent
	\textit{Base case} ($j=0$): The computation of $M^{(P,S,c)'}$ starts with an empty link traversal tape (cf.~Definition~\ref{Definition:LDMachine}). Due to the initialization, $w_0$ is a concatenation of sub-words \,$\fctEncName(\symURI) \, \fctEncName( \adocFct(\symURI) ) \, \sharp$\, for all $\symURI \in S$ (cf.~line~\ref{Line:ReachFiniteQueryMachine:Init} in Algorithm~\ref{Algorithm:ReachFiniteQueryMachine}). Hence, we have a corresponding sequence $\symURI_1 ,...\,,\symURI_{n_0}$ where $n_0=\left| S \right|$ and $\forall \,i \in [1,n_0]: \symURI_i \in S$.
	The order of the {\ID}s in that sequence depends on the order in which they have been looked up and is irrelevant for our proof.
	For all $\symURI \in S$ it holds either $\symURI_i \notin \fctDom{\adocFct}$ or $\adocFct(\symURI)$ is $(c,P)$-reach\-able from $S$ in $\WoD$ (cf.~case~\ref{DefinitionCase:QualifiedReachability:IndBegin} in Definition~\ref{Definition:QualifiedReachability}).

	\vspace{1ex} \noindent
	\textit{Induction step} ($j>0$): Our inductive hypothesis is that there exists a finite sequence $\symURI_1 ,...\,,\symURI_{n_{j-1}}$ of $n_{j-1}$ different {\ID}s ($\forall \,i \in [1,n_{j-1}]: \symURI_i \in \symAllURIs$) such that i)~$w_{j-1}$ is
	\begin{equation*}
		\fctEncName(\symURI_1) \, \fctEncName( \adocFct(\symURI_1) ) \, \sharp \, ... \, \sharp \, \fctEncName(\symURI_{n_{j-1}}) \, \fctEncName( \adocFct(\symURI_{n_{j-1}}) ) \, \sharp
	\end{equation*}
	and ii)~for each $i \in [1,n_{j-1}]$ either $\symURI_i \notin \fctDom{\adocFct}$ or $\adocFct(\symURI_i)$ is $(c,P)$-reach\-able from $S$ in $\WoD$.
	In the $j$-th iteration $M^{(P,S,c)'}$ finds an {\triple} $t$ encoded as part of $w_{j-1}$ such that $\exists \, \symURI \in \fctIDsName(t) : c(t,\symURI,P) = \true$ and \code{lookup} has not been called for $\symURI$. The machine calls \code{lookup} for $\symURI$, which changes the word on the link traversal tape to $w_j$. Hence, $w_j$ is equal to \,$w_{j-1} \, \fctEncName(\symURI) \, \fctEncName( \adocFct(\symURI) ) \, \sharp\,$ and, thus, our sequence of {\ID}s for $w_j$ is $\symURI_1 ,...\,,\symURI_{n_{j-1}}, \symURI$. It remains to show that if $\symURI \in \fctDom{\adocFct}$ then $\adocFct(\symURI)$ is $(c,P)$-reach\-able from $S$ in $\WoD$.

	Assume $\symURI \in \fctDom{\adocFct}$. Since {\triple} $t$ is encoded as part of $w_{j-1}$ we know, from our inductive hypothesis, that $t$ must be contained in the data of an {\LDdoc} $d^*$ that is $(c,P)$-reach\-able from $S$ in $\WoD$ (and for which exists $i \in [1,n_{j-1}]$ such that $\adocFct(\symURI_i) = d^*$). Therefore, $t$ and $\symURI$ satisfy the requirements as given in case~\ref{DefinitionCase:QualifiedReachability:IndStep} of Definition~\ref{Definition:QualifiedReachability} and, thus, $\adocFct(\symURI)$ is $(c,P)$-reach\-able from $S$ in $\WoD$.
\end{myproof}

\noindent
After verifying that Algorithm~\ref{Algorithm:ReachFiniteQueryMachine} is sound (cf.~Lemma~\ref{SubLemma2:Lemma:Reach:TerminationCriterion}) and complete (cf.~Lemma~\ref{SubLemma1:Lemma:Reach:TerminationCriterion}) w.r.t.~discovering reach\-able {\LDdoc}s, we now show that an execution of the algorithm terminates if the corresponding reach\-able part of the input Web is finite.
\begin{lemma} \label{SubLemma3:Lemma:Reach:TerminationCriterion}
	\begin{proofdeclitems}
		\vspace{-1mm}
		\item $M^{(P,S,c)'}$ be the $(P,S,c)'$-machine for a SPARQL expression $P$, a finite set $S \subset \symAllURIs$, and a reachability criterion $c$;
		\item $\WoD = (D,\dataFct,\adocFct)$ be a (potentially infinite) Web of Linked Data encoded on the Web tape of $M^{(P,S,c)'}$\!; and
		\item $\ReachPartScPW{S}{c}{P}{\WoD}$ denotes the $(S,c,P)$-reach\-able part of $\WoD$.
	\end{proofdeclitems}
	\vspace{-1mm}
	The computation of $M^{(P,S,c)'}$
	halts after a finite number of steps if $\ReachPartScPW{S}{c}{P}{\WoD}$ is finite.
\end{lemma}
\begin{myproof}{of Lemma~\ref{SubLemma3:Lemma:Reach:TerminationCriterion}}
	Let $\ReachPartScPW{S}{c}{P}{\WoD}$ be finite.
	To show that the computation of $M^{(P,S,c)'}$ on (Web) input $\fctEncName(\WoD)$ halts after a finite number of steps we emphasize the following facts:
	\begin{enumerate}
		\item Each call of subroutine \code{lookup} by $M^{(P,S,c)'}$ terminates because the encoding of $\WoD$ is ordered following the order of the {\ID}s in $\fctDom{\adocFct}$.
		\item $M^{(P,S,c)'}$ completes the initialization in line~\ref{Line:ReachFiniteQueryMachine:Init} of Algorithm~\ref{Algorithm:ReachFiniteQueryMachine} after a finite number of steps because $S$ is finite.
		\item 
			At any point in the computation the word on the link traversal tape of $M^{(P,S,c)'}$ is
		finite because $M^{(P,S,c)'}$ only gradually appends (encoded) {\LDdoc}s to that tape (one document per iteration) and the encoding of each document is finite (recall that the set of {\triple}s $\dataFct(d)$ for each {\LDdoc} $d \in D$ is finite).
		\item During each iteration of the loop in Algorithm~\ref{Algorithm:ReachFiniteQueryMachine}, $M^{(P,S,c)'}$ completes the scan of its link traversal tape (cf.~line~\ref{Line:ReachFiniteQueryMachine:LoopStep}) after a finite number of computation steps because
			the word on that tape
		is always finite. Thus, $M^{(P,S,c)'}$ finishes each iteration of the loop after a finite number of steps.
		\item $M^{(P,S,c)'}$ considers only those {\ID}s for a call of subroutine \code{lookup} that i)~have not been considered before and that ii)~are mentioned in ({\triple}s of) {\LDdoc}s from $\ReachPartScPW{S}{c}{P}{\WoD}$ (cf.~line~\ref{Line:ReachFiniteQueryMachine:LoopStep}). Since $\ReachPartScPW{S}{c}{P}{\WoD}$ is finite there is only a finite number of such {\ID}s and, thus, the loop in Algorithm~\ref{Algorithm:ReachFiniteQueryMachine} as performed by $M^{(P,S,c)'}$ has a finite number of iterations.
		\item Due to the finiteness of
			the word on the link traversal tape
		the set $G$ used in line~\ref{Line:ReachFiniteQueryMachine:OutputResult} of Algorithm~\ref{Algorithm:ReachFiniteQueryMachine} is finite and, thus, $\evalOrigPG{P}{G}$ is finite. As a consequence $M^{(P,S,c)'}$ requires only a finite number of computation steps for executing line~\ref{Line:ReachFiniteQueryMachine:OutputResult}.
	\end{enumerate}
	\noindent
	Altogether, these facts prove Lemma~\ref{SubLemma3:Lemma:Reach:TerminationCriterion}.
\end{myproof}

\noindent
We now prove Lemma~\ref{Lemma:Reach:TerminationCriterion}. \begin{proofdeclitems}
		\vspace{-1mm}
		\item $\evalReachFctPSc{P}{S}{c}$ be a {\LDSPARQLReach} query that is nontrivially satisfiable;
		\item $\WoD$ be a (potentially infinite) Web of Linked Data; and
		\item $\ReachPartScPW{S}{c}{P}{\WoD} = (\Reach{D},\Reach{\dataFct},\Reach{\adocFct} )$ denote the $(S,c,P)$-reach\-able part of $\WoD$.
	\end{proofdeclitems}

\noindent
\textbf{If:} Let $\ReachPartScPW{S}{c}{P}{\WoD}$ be finite.
We have to show that there exists an LD machine that computes $\evalReachPScW{P}{S}{c}{\WoD}$ and halts after a finite number of computation steps. Based on Lemmas~\ref{SubLemma1:Lemma:Reach:TerminationCriterion} to~\ref{SubLemma3:Lemma:Reach:TerminationCriterion} it is easy to verify that the $(P,S,c)'$-machine (for $P$, $S$ and $c$ as used by $\evalReachFctPSc{P}{S}{c}$) is such a machine.

\vspace{1ex} \noindent
\textbf{Only if:} Let $M$ be an LD machine (not necessarily a $(P,S,c)'$-machine) that computes $\evalReachPScW{P}{S}{c}{\WoD}$ and halts after a finite number of computation steps. We have to show that $\ReachPartScPW{S}{c}{P}{\WoD}$ is finite. We show this by contradiction, that is, we assume $\ReachPartScPW{S}{c}{P}{\WoD}$ is infinite. In this case $\Reach{D}$ is infinite.
Since $\evalReachFctPSc{P}{S}{c}$ is nontrivially satisfiable it is possible that $\WoD$ is a Web of Linked Data for which exist solutions in $\ReachPartScPW{S}{c}{P}{\WoD}$ such that each of these solutions provides a binding for at least one variable. Hence, for computing $\evalReachFctPSc{P}{S}{c}$ over $\WoD$ completely, machine $M$ must (recursively) expand the word on its link traversal tape until it contains the encodings of (at least) each {\LDdoc} in $\Reach{D}$. Such an expansion is necessary to ensure that the computed query result is complete. Since $\Reach{D}$ is infinite the expansion requires infinitely many computing steps. However, we know that $M$ halts after a finite number of computation steps. Hence, we have a contradiction and, thus, $\ReachPartScPW{S}{c}{P}{\WoD}$ must be finite.

\subsection{Proof of Proposition~\ref{Proposition:Reach:ComputabilityFiniteness}} \label{Proof:Proposition:Reach:ComputabilityFiniteness}
\noindent
Let $c_{e\!f}$ be a reachability criterion that ensures finiteness.
To prove that all {\LDSPARQLReach} queries under $c_{e\!f}$-semantics are finitely computable we have to show for each such query exists an LD machine that computes the query over any Web of Linked Data and halts after a finite number of computation steps (with an encoding of the complete query result on its output tape). W.l.o.g., let $\evalReachFctPSc{P}{S}{c_{e\!f}}$ be such a {\LDSPARQLReach} query (under $c_{e\!f}$-semantics). Based on Lemmas~\ref{SubLemma1:Lemma:Reach:TerminationCriterion} to~\ref{SubLemma3:Lemma:Reach:TerminationCriterion} (cf.~Section~\ref{Proof:Lemma:Reach:TerminationCriterion}) it is easy to verify that the $(P,S,c_{e\!f})'$-machine (for $P$, $S$ and $c_{e\!f}$ as used by $\evalReachFctPSc{P}{S}{c_{e\!f}}$) is such a machine (notice, Lemmas~\ref{SubLemma1:Lemma:Reach:TerminationCriterion} to~\ref{SubLemma3:Lemma:Reach:TerminationCriterion} are not restricted to {\LDSPARQLReach} queries that are nontrivially satisfiable).

\subsection{Proof of Theorem~\ref{Theorem:Reach:ComputabilityDependsOnMonotonicity}}

\todo{Notice, the (corrected) theorem is still not entirely correct -- see the two TODOs within the following proof.}

\noindent
Let $c_{n\!f}$ be a reachability criterion that does not ensure finiteness.
To prove Theorem~\ref{Theorem:Reach:ComputabilityDependsOnMonotonicity} we distinguish three cases for a satisfiable {\LDSPARQLReach} query $\evalReachFctPSc{P}{S}{c_{n\!f}}$ under $c_{n\!f}$-semantics:
\begin{enumerate}
	\item The $(S,c_{n\!f},P)$-reach\-able part of \emph{any} Web of Linked Data is finite (which is possible even if $c_{n\!f}$ does not ensure finiteness for all {\LDSPARQLReach} queries under $c_{n\!f}$-semantics).
	\item The $(S,c_{n\!f},P)$-reach\-able part of some Web of Linked Data is infinite and $\evalReachFctPSc{P}{S}{c_{n\!f}}$ is monotonic.
	\item The $(S,c_{n\!f},P)$-reach\-able part of some Web of Linked Data is infinite and $\evalReachFctPSc{P}{S}{c_{n\!f}}$ is not monotonic.
\end{enumerate}
\noindent
In the following we discuss each of these cases.

\vspace{1ex} \noindent
\textbf{Case (1):} Let $\evalReachFctPSc{P'}{S'}{c_{n\!f}}$ be a satisfiable {\LDSPARQLReach} query (under $c_{n\!f}$-semantics) such that the $(S',c_{n\!f},P')$-reach\-able part of \emph{any} Web of Linked Data is finite. We claim that in this case $\evalReachFctPSc{P'}{S'}{c_{n\!f}}$ is finitely computable (independent of its monotonicity). To prove this claim we use the same argument that we use for proving Proposition~\ref{Proposition:Reach:ComputabilityFiniteness} in Section~\ref{Proof:Proposition:Reach:ComputabilityFiniteness}:
Based on Lemmas~\ref{SubLemma1:Lemma:Reach:TerminationCriterion} to~\ref{SubLemma3:Lemma:Reach:TerminationCriterion} (cf.~Section~\ref{Proof:Lemma:Reach:TerminationCriterion}) and the fact that
the $(S',c_{n\!f},P')$-reach\-able part of \emph{any} Web of Linked Data is finite, it is easy to verify that the $(P',S',c_{n\!f})'$-machine (for $P'$, $S'$ and $c_{n\!f}$ as used by $\evalReachFctPSc{P'}{S'}{c_{n\!f}}$) is an LD machine that computes $\evalReachFctPSc{P'}{S'}{c_{n\!f}}$ over any Web of Linked Data $\WoD$ and halts after a finite number of computation steps (with an encoding of $\evalReachPScW{P'}{S'}{c_{n\!f}}{\WoD}$ on its output tape). Hence, the $(P',S',c_{n\!f})'$-machine satisfies the requirements in Definition~\ref{Definition:FinitelyComputable} and, thus, $\evalReachFctPSc{P'}{S'}{c_{n\!f}}$ is finitely computable.

\vspace{1ex} \noindent
\textbf{Case (2):} Let $\evalReachFctPSc{P'}{S'}{c_{n\!f}}$ be a satisfiable, monotonic {\LDSPARQLReach} query (under $c_{n\!f}$-se\-man\-tics) for which exists a Web of Linked Data $\WoD$ such that the $(S',c_{n\!f},P')$-reach\-able part of $\WoD$ is infinite.
%
%
%
%
To show that $\evalReachFctPSc{P'}{S'}{c_{n\!f}}$ is
	(at least)
eventually computable we introduce specific LD machines for {\LDSPARQLReach} queries. We call these machines $(P,S,c)$-machines. The $(P,S,c)$-machine for a {\LDSPARQLReach} query $\evalReachFctPSc{P}{S}{c}$ implements a generic (i.e.~input independent) computation of $\evalReachFctPSc{P}{S}{c}$. We shall see that if a {\LDSPARQLReach} query $\evalReachFctPSc{P}{S}{c}$ is monotonic, the corresponding $(P,S,c)$-machine (eventually) computes $\evalReachFctPSc{P}{S}{c}$ over any Web of Linked Data. We emphasize that $(P,S,c)$-machines differ from the $(P,S,c)'$-machines that we use for proving Lemma~\ref{Lemma:Reach:TerminationCriterion} (cf.~Section~\ref{Proof:Lemma:Reach:TerminationCriterion}).
Formally, we define $(P,S,c)$-machines as follows:

\begin{definition} \label{Definition:ReachQueryMachine}
	Let $S \subset \symAllURIs$ be a finite set of \removable{seed} {\ID}s;
	let $c$ be a reachability criterion; and
	let $P$ be a SPARQL expression.
	The \definedTerm{$(P,S,c)$-machine} is an LD machine that implements Algorithm~\ref{Algorithm:ReachQueryMachine}. This algorithm
		makes use of
	a
	subroutine
		called
	\code{lookup}. This subroutine, when called with a {\ID} $\symURI \in \symAllURIs$,
	i)~writes $\fctEnc{\symURI}$ to the right end of the word on the link traversal tape,
	ii)~enters the expand state, and
	iii)~performs the expand procedure as specified in Definition~\ref{Definition:LDMachine}.
\end{definition}

\begin{algorithm}[bt]
	\caption{\, The program of the $(P,S,c)$-machine.} \label{Algorithm:ReachQueryMachine}
	\begin{algorithmic}[1]
			\STATE {Call \code{lookup} for each $\symURI \in S$.}
		\label{Line:ReachQueryMachine:Init}

			\medskip
		\FOR { $j = 1, 2, ...$} \label{Line:ReachQueryMachine:LoopBegin}
			\STATE {Let $T_j$ denote the set of all {\triple}s currently encoded on the link traversal tape. Use the work tape to enumerate the set $\evalOrigPG{P}{T_j}$.} \label{Line:ReachQueryMachine:Step1_EnumPartialResult}
			\STATE {For each $\mu \in \evalOrigPG{P}{T_j}$ check whether $\mu$ is already encoded on the output tape; if \removable{this is} not\removable{~the case}, then add $\fctEnc{\mu}$ to the output.} \label{Line:ReachQueryMachine:Step2_OutputNewSolutions}
			\STATE {Scan the link traversal tape for an {\triple} $t$ that contains a {\ID} $\symURI \in \fctIDs{t}$ such that i)~$c(t,\symURI,P)=\true$ and ii)~the word on the link traversal tape neither contains \,$\fctEncName(\symURI) \, \fctEncName( \adocFct(\symURI) ) \, \sharp$\, nor \,$\fctEncName(\symURI) \, \sharp$. If such $t$ and $\symURI$ exist, call \code{lookup} for $\symURI$; otherwise halt the computation.} \label{Line:ReachQueryMachine:Step3_LookupOrHalt}
		\ENDFOR \label{Line:ReachQueryMachine:LoopEnd}
	\end{algorithmic}
\end{algorithm}

\noindent
As can be seen in Algorithm~\ref{Algorithm:ReachQueryMachine}, the computation of each $(P,S,c)$-machine
	(with a Web of Linked Data $\WoD$ encoded on its input tape)
starts with an initialization (cf.~line~\ref{Line:ReachQueryMachine:Init}). After the initialization, the machine enters a (potentially non-terminating) loop. During each iteration of this loop, the machine generates valuations using all data that is currently encoded on the link traversal tape. The following proposition shows that these valuations are part of the corresponding query result (find the proof for Proposition~\ref{SubProposition:Reach:Computability:Soundness} below in Section~\ref{Proof:SubProposition:Reach:Computability:Soundness}):
\begin{proposition} \label{SubProposition:Reach:Computability:Soundness}
	\begin{proofdeclitems}
		\vspace{-1mm}
		\item $\evalReachFctPSc{P}{S}{c}$ be a {\LDSPARQLReach} query that is monotonic;
		\item $M^{(P,S,c)}$ denote the $(P,S,c)$-machine for $P$, $S$, and $c$ as used by $\evalReachFctPSc{P}{S}{c}$\!; and
		\item $\WoD$ be an arbitrary Web of Linked Data encoded on the Web tape of $M^{(P,S,c)}$\!.
	\end{proofdeclitems}
	\vspace{-1mm}
	During the execution of Algorithm~\ref{Algorithm:ReachQueryMachine} by $M^{(P,S,c)}$
	it holds:
	\begin{equation*}
		\forall \, j \in \lbrace 1,2,...\rbrace : \evalOrigPG{P}{T_j} \subseteq \evalReachPScW{P}{S}{c}{\WoD}
	\end{equation*}
\end{proposition}

\noindent
Proposition~\ref{SubProposition:Reach:Computability:Soundness} presents the basis to prove the soundness of (monotonic) query results computed by Algorithm~\ref{Algorithm:ReachQueryMachine}.
To verify the completeness of these results it is important to note that $(P,S,c)$-machines look up no more than one {\ID} per iteration (cf.~line~\ref{Line:ReachQueryMachine:Step3_LookupOrHalt} in Algorithm~\ref{Algorithm:ReachQueryMachine}). Hence, $(P,S,c)$-machines prioritize result construction over data retrieval. Due to this feature we show that for each solution in a query result exists an iteration during which that solution is computed (find the proof for Proposition~\ref{SubProposition:Reach:Computability:Completeness} below in Section~\ref{Proof:SubProposition:Reach:Computability:Completeness}):
\begin{proposition} \label{SubProposition:Reach:Computability:Completeness}
	\begin{proofdeclitems}
		\vspace{-1mm}
		\item $\evalReachFctPSc{P}{S}{c}$ be a {\LDSPARQLReach} query that is monotonic;
		\item $M^{(P,S,c)}$ denote the $(P,S,c)$-machine for $P$, $S$, and $c$ as used by $\evalReachFctPSc{P}{S}{c}$\!; and
		\item $\WoD$ be an arbitrary Web of Linked Data encoded on the Web tape of $M^{(P,S,c)}$\!.
	\end{proofdeclitems}
	\vspace{-1mm}
	For each
	$\mu \in \evalReachPScW{P}{S}{c}{\WoD}$ exists a $j_\mu \in \lbrace 1,2,...\rbrace$ such that during the execution of Algorithm~\ref{Algorithm:ReachQueryMachine} by $M^{(P,S,c)}$
	it holds:
	\begin{equation*}
		\forall \, j \in \lbrace j_\mu,j_\mu \!+\!1,...\rbrace : \mu \in \evalOrigPG{P}{T_j}
	\end{equation*}
\end{proposition}

\noindent
So far our results verify that i)~the set of query solutions computed after any iteration is sound and ii)~that this set is complete after a particular (potentially infinite) number of iterations. We now show that each iteration definitely finishes after a finite number of computation steps (find the proof for Proposition~\ref{SubProposition:Reach:Computability:Finiteness} below in Section~\ref{Proof:SubProposition:Reach:Computability:Finiteness}):
\begin{proposition} \label{SubProposition:Reach:Computability:Finiteness}
	\begin{proofdeclitems}
		\vspace{-1mm}
		\item $M^{(P,S,c)}$ be the $(P,S,c)$-machine for a SPARQL expression $P$, a finite set $S \subset \symAllURIs$, and a reachability criterion $c$; and
		\item $\WoD$ be a (potentially infinite) Web of Linked Data encoded on the Web tape of $M^{(P,S,c)}$\!.
	\end{proofdeclitems}
	\vspace{-1mm}
	During the execution of Algorithm~\ref{Algorithm:ReachQueryMachine}, $M^{(P,S,c)}$ finishes each iteration of the loop in that algorithm after a finite number of computation steps.
\end{proposition}

\noindent
Altogether, Propositions~\ref{SubProposition:Reach:Computability:Soundness} to~\ref{SubProposition:Reach:Computability:Finiteness} conclude the discussion of case (2), that is, based on these propositions it is easy to verify that the $(P',S',c_{n\!f})$-machine for our query $\evalReachFctPSc{P'}{S'}{c_{n\!f}}$ satisfies the requirements in Definition~\ref{Definition:EventuallyComputable} and, thus, $\evalReachFctPSc{P'}{S'}{c_{n\!f}}$ is eventually computable.

\vspace{1ex} \noindent
\textbf{Case (3):} Let $\evalReachFctPSc{P'}{S'}{c_{n\!f}}$ be a satisfiable, non-monotonic {\LDSPARQLReach} query (under $c_{n\!f}$-se\-man\-tics) for which exists a Web of Linked Data $\WoD$ such that the $(S',c_{n\!f},P')$-reach\-able part of $\WoD$ is infinite.
To show that $\evalReachFctPSc{P'}{S'}{c_{n\!f}}$ may not even be eventually computable, we assume $\evalReachPScW{P'}{S'}{c_{n\!f}}{\WoD} \neq \emptyset$.

	For the prove
we use the same argument that we use in the corresponding discussion for non-monotonic {\LDSPARQLFull} queries (see the proof of Theorem~\ref{Theorem:FullWeb:ComputabilityDependsOnMonotonicity} in Section~\ref{Proof:Theorem:FullWeb:ComputabilityDependsOnMonotonicity}).
Hence, we show a contradiction by assuming $\evalReachFctPSc{P'}{S'}{c_{n\!f}}$ were (at least) eventually computable, that is, we assume an LD machine $M$ (which is not necessarily a $(P,S,c)$-machine) whose computation of $\evalReachFctPSc{P'}{S'}{c_{n\!f}}$ on any Web of Linked Data
has the two properties given in Definition~\ref{Definition:EventuallyComputable}. To obtain a contradiction we show that such a machine does not exist.

Let $\WoD$ be a Web of Linked Data such that the $(S',c_{n\!f},P')$-reach\-able part of $\WoD$ is infinite; such a Web exists for case~(3). In the remainder of this proof we write $\WoD_\mathfrak{R}$ to denote the $(S',c_{n\!f},P')$-reach\-able part of $\WoD$.

Let $\WoD$ be encoded on the Web tape of $M$ and let $\mu$ be an arbitrary solution for $\evalReachFctPSc{P'}{S'}{c_{n\!f}}$ in $\WoD$; i.e.~$\mu \in \evalReachPScW{P'}{S'}{c_{n\!f}}{\WoD}$. Based on our assumption, machine $M$ must write $\fctEnc{\mu}$ to its output tape after a finite number of computation steps (cf.~property~\ref{DefinitionRequirement:EventuallyComputable:All} in Definition~\ref{Definition:EventuallyComputable}). We argue that this is impossible: Since $\evalReachFctPSc{P'}{S'}{c_{n\!f}}$ is not monotonic, $M$ cannot add $\mu$ to the output before $M$ has accessed all {\LDdoc}s in $\WoD_\mathfrak{R}$ (i.e.~all {\LDdoc}s that are $(c_{n\!f},P')$-reach\-able from $S'$ in $\WoD$). However, due to the infiniteness of $\WoD_\mathfrak{R}$, there is an infinite number of such documents. Therefore, accessing all these documents is a non-terminating process and, thus, $M$ cannot write $\mu$ to its output after a finite number of computation steps. As a consequence, the computation of $\evalReachFctPSc{P'}{S'}{c_{n\!f}}$ (over $\WoD$) by $M$ does not have the properties given in Definition~\ref{Definition:EventuallyComputable}, which contradicts our initial assumption. Due to this contradiction we may conclude that $\evalReachFctPSc{P'}{S'}{c_{n\!f}}$ is not eventually computable.

\subsection{Proof of Proposition~\ref{SubProposition:Reach:Computability:Soundness}} \label{Proof:SubProposition:Reach:Computability:Soundness}
\begin{proofdeclitems}
	\item $\evalReachFctPSc{P}{S}{c}$ be a {\LDSPARQLReach} query that is monotonic;
	\item $M^{(P,S,c)}$ denote the $(P,S,c)$-machine for $P$, $S$, and $c$ as used by $\evalReachFctPSc{P}{S}{c}$\!; and
	\item $\WoD = (D,\dataFct,\adocFct)$ be an arbitrary Web of Linked Data encoded on the Web tape of $M^{(P,S,c)}$\!.
\end{proofdeclitems}
To prove Proposition~\ref{SubProposition:Reach:Computability:Soundness} we use the following lemma.
\begin{lemma} \label{SubLemma:Reach:Computability:Soundness}
	During the execution of Algorithm~\ref{Algorithm:ReachQueryMachine} by $M^{(P,S,c)}$ on (Web) input $\fctEncName(\WoD)$ it holds $\forall \, j \in \lbrace 1,2,...\rbrace : T_j \subseteq \fctAllDataName\bigl( \ReachPartScPW{S}{c}{P}{\WoD} \bigr)$.
\end{lemma}
\begin{myproof}{of Lemma~\ref{SubLemma:Reach:Computability:Soundness}}
	This proof resembles the proof of the corresponding lemma for $M^{(P,S,c)'}$ machines (cf.~Lemma~\ref{SubLemma2:Lemma:Reach:TerminationCriterion} in Section~\ref{Proof:Lemma:Reach:TerminationCriterion}).
	Let $w_j$ be the word on the link traversal tape of $M^{(P,S,c)}$ when $M^{(P,S,c)}$ starts the $j$-th iteration of the main processing loop in Algorithm~\ref{Algorithm:ReachQueryMachine} (i.e.~before line~\ref{Line:ReachQueryMachine:Step1_EnumPartialResult}).

	To prove $\forall \, j \in \lbrace 1,2,...\rbrace : T_j \subseteq \fctAllDataName\bigl( \ReachPartScPW{S}{c}{P}{\WoD} \bigr)$ it is sufficient to show for each $w_j$ (where $j \in \lbrace 1,2,...\rbrace$) exists a finite sequence $\symURI_1 ,...\,,\symURI_{n_j}$ of $n_j$ different {\ID}s $\symURI_i \in \symAllURIs$ (where $i \in [1,n_j]$) such that
	i)~$w_j$ is\footnote{We, again, assume \,$\fctEncName(\adocFct(\symURI_i))$\, is the empty word if $\symURI_i \notin \fctDom{\adocFct}$.}
	\begin{equation*}
		\fctEncName(\symURI_1) \, \fctEncName( \adocFct(\symURI_1) ) \, \sharp \, ... \, \sharp \, \fctEncName(\symURI_{n_j}) \, \fctEncName( \adocFct(\symURI_{n_j}) ) \, \sharp
	\end{equation*}
	and ii)~for each $i \in [1,n_j]$ either $\symURI_i \notin \fctDom{\adocFct}$ (and, thus, $\adocFct(\symURI_i)$ is undefined) or $\adocFct(\symURI_i)$ is an {\LDdoc} which is $(c,P)$-reach\-able from $S$ in $\WoD$.
	We use an induction over $j$ for this proof.

	\vspace{1ex} \noindent
	\textit{Base case} ($j=1$): The computation of $M^{(P,S,c)}$ starts with an empty link traversal tape (cf.~Definition~\ref{Definition:LDMachine}). Due to the initialization, $w_1$ is a concatenation of sub-words \,$\fctEncName(\symURI) \, \fctEncName( \adocFct(\symURI) ) \, \sharp$\, for all $\symURI \in S$ (cf.~line~\ref{Line:ReachQueryMachine:Init} in Algorithm~\ref{Algorithm:ReachQueryMachine}). Hence, we have a corresponding sequence $\symURI_1 ,...\,,\symURI_{n_1}$ where $n_1=\left| S \right|$ and $\forall \,i \in [1,n_1]: \symURI_i \in S$.
	The order of the {\ID}s in that sequence depends on the order in which they have been looked up and is irrelevant for our proof.
	For all $\symURI \in S$ it holds either $\symURI \notin \fctDom{\adocFct}$ or $\adocFct(\symURI)$ is $(c,P)$-reach\-able from $S$ in $\WoD$ (cf.~case~\ref{DefinitionCase:QualifiedReachability:IndBegin} in Definition~\ref{Definition:QualifiedReachability}).

	\vspace{1ex} \noindent
	\textit{Induction step} ($j>1$): Our inductive hypothesis is that there exists a finite sequence $\symURI_1 ,...\,,\symURI_{n_{j-1}}$ of $n_{j-1}$ different {\ID}s ($\forall \,i \in [1,n_{j-1}]: \symURI_i \in \symAllURIs$) such that i)~$w_{j-1}$ is
	\begin{equation*}
		\fctEncName(\symURI_1) \, \fctEncName( \adocFct(\symURI_1) ) \, \sharp \, ... \, \sharp \, \fctEncName(\symURI_{n_{j-1}}) \, \fctEncName( \adocFct(\symURI_{n_{j-1}}) ) \, \sharp
	\end{equation*}
	and ii)~for each $i \in [1,n_{j-1}]$ either $\symURI_i \notin \fctDom{\adocFct}$ or $\adocFct(\symURI_i)$ is $(c,P)$-reach\-able from $S$ in $\WoD$.
	In the ($j$-1)-th iteration $M^{(P,S,c)}$ finds an {\triple} $t$ encoded as part of $w_{j-1}$ such that $\exists \, \symURI \in \fctIDsName(t) : c(t,\symURI,P) = \true$ and \code{lookup} has not been called for $\symURI$. The machine calls \code{lookup} for $\symURI$, which changes the word on the link traversal tape to $w_j$. Hence, $w_j$ is equal to \,$w_{j-1} \, \fctEncName(\symURI) \, \fctEncName( \adocFct(\symURI) ) \, \sharp\,$ and, thus, our sequence of {\ID}s for $w_j$ is $\symURI_1 ,...\,,\symURI_{n_{j-1}}, \symURI$. It remains to show that if $\symURI \in \fctDom{\adocFct}$ then $\adocFct(\symURI)$ is $(c,P)$-reach\-able from $S$ in $\WoD$.

	Assume $\symURI \in \fctDom{\adocFct}$. Since {\triple} $t$ is encoded as part of $w_{j-1}$ we know, from our inductive hypothesis, that $t$ must be contained in the data of an {\LDdoc} $d^*$ that is $(c,P)$-reach\-able from $S$ in $\WoD$ (and for which exists $i \in [1,n_{j-1}]$ such that $\adocFct(\symURI_i) = d^*$). Therefore, $t$ and $\symURI$ satisfy the requirements as given in case~\ref{DefinitionCase:QualifiedReachability:IndStep} of Definition~\ref{Definition:QualifiedReachability} and, thus, $\adocFct(\symURI)$ is $(c,P)$-reach\-able from $S$ in $\WoD$.
\end{myproof}

\noindent
Due to the monotonicity of $\evalReachFctPSc{P}{S}{c}$ it is trivial to show Proposition~\ref{SubProposition:Reach:Computability:Soundness} using Lemma~\ref{SubLemma:Reach:Computability:Soundness} (recall, $\evalReachPScW{P}{S}{c}{\WoD} = \evalOrigPG{P}{\fctAllDataName\bigl(\ReachPartScPW{S}{c}{P}{\WoD}\bigr)}$).

\subsection{Proof of Proposition~\ref{SubProposition:Reach:Computability:Completeness}} \label{Proof:SubProposition:Reach:Computability:Completeness}
\begin{proofdeclitems}
	\item $\evalReachFctPSc{P}{S}{c}$ be a {\LDSPARQLReach} query that is monotonic;
	\item $M^{(P,S,c)}$ denote the $(P,S,c)$-machine for $P$, $S$, and $c$ as used by $\evalReachFctPSc{P}{S}{c}$\!; and
	\item $\WoD = (D,\dataFct,\adocFct)$ be an arbitrary Web of Linked Data encoded on the Web tape of $M^{(P,S,c)}$\!.
\end{proofdeclitems}
To prove Proposition~\ref{SubProposition:Reach:Computability:Completeness} we use the following lemma.
\begin{lemma} \label{SubLemma:Reach:Computability:Completeness}
	For each {\triple} $t \in \fctAllDataName\bigl( \ReachPartScPW{S}{c}{P}{\WoD} \bigr)$ exists a $j_t \in \lbrace 1,2,...\rbrace$ such that during the execution of Algorithm~\ref{Algorithm:ReachQueryMachine} by $M^{(P,S,c)}$ on (Web) input $\fctEncName(\WoD)$ it holds $\forall \, j \in \lbrace j_t,j_t\!+\!1,...\rbrace : t \in T_j$.
\end{lemma}
\begin{myproof}{of Lemma~\ref{SubLemma:Reach:Computability:Completeness}}
	Let $w_j$ be the word on the link traversal tape of $M^{(P,S,c)}$ when $M^{(P,S,c)}$ starts the $j$-th iteration of the main processing loop in Algorithm~\ref{Algorithm:ReachQueryMachine} (i.e.~before line~\ref{Line:ReachQueryMachine:Step1_EnumPartialResult}).

	W.l.o.g., let $t'$ be an arbitrary {\triple} $t' \in \fctAllDataName\bigl( \ReachPartScPW{S}{c}{P}{\WoD} \bigr)$. There must exist an {\LDdoc} $d \in D$ such that i)~$t' \in \dataFct(d)$ and ii)~$d$ is $(c,P)$-reach\-able from $S$ in $\WoD$. Let $d'$ be such a document.
	Since $M^{(P,S,c)}$ only appends to its link traversal tape we prove that there exists a $j_{t'} \in \lbrace 1,2,...\rbrace$ with $\forall \, j \in \lbrace j_{t'},j_{t'}\!+\!1,...\rbrace : t' \in T_j$ by showing that there exists $j_{t'} \in \lbrace 1,2,...\rbrace$ such that $w_{j_{t'}}$ contains the sub-word \,$\fctEnc{d'}$.
	This proof resembles the proof of the corresponding lemma for $M^{(P,S,c)'}$ machines (cf.~Lemma~\ref{SubLemma1:Lemma:Reach:TerminationCriterion} in Section~\ref{Proof:Lemma:Reach:TerminationCriterion}).

	Since $d'$ is $(c,P)$-reach\-able from $S$ in $\WoD$, the link graph for $\WoD$ contains at least one finite path $(d_0, ... \, , d_n)$ of {\LDdoc}s $d_i$ where
	i)~$n \in \lbrace 0,1,... \rbrace$,
	ii)~$d_n = d'$,
	iii)~$\exists \, \symURI \in S : \adocFct(\symURI) = d_0$,
	and iv)~for each $i \in \lbrace 1,... \,,n \rbrace$ it holds:
	\begin{equation} \label{Equation:Proof:SubLemma:Reach:Computability:Completeness}
		\exists \, t \in \dataFct(d_{i-1}) : \Bigl( \exists \, \symURI \in \fctIDsName(t) : \bigl( \adocFct(\symURI)=d_{i} \text{ and } c(t,\symURI,P) = \true \bigr) \Bigr)
	\end{equation}
	Let $(d_0^*, ... \,, d_n^*)$ be such a path.
		\removable{We use this path for our proof. More precisely,}
	we show by induction over $i \in \lbrace 0,...,\,n \rbrace$ that there exists $j_t \in \lbrace 1,2,...\rbrace$ such that $w_{j_t}$ contains the sub-word \,$\fctEncName( d_n^* )$\, (which is the same as \,$\fctEncName( d' )$ because $d_n^* = d'$).

	\vspace{1ex} \noindent
	\textit{Base case} ($i=0$): Since $\exists \, \symURI \in S : \adocFct(\symURI) = d_0^*$ it is easy to verify that $w_1$ contains the sub-word \,$\fctEncName( d_0^* )$\, (cf.~line~\ref{Line:ReachQueryMachine:Init} in Algorithm~\ref{Algorithm:ReachQueryMachine}).

	\vspace{1ex} \noindent
	\textit{Induction step} ($i>0$): Our inductive hypothesis is: There exists $j \in \lbrace 1,2,...\rbrace$ such that $w_{j}$ contains sub-word \,$\fctEncName( d_{i-1}^* )$. Based on the hypothesis we show that there exists a $j' \in \lbrace j,j\!+\!1,...\rbrace$ such that $w_{j'}$ contains the sub-word \,$\fctEncName( d_{i}^* )$. We distinguish two cases: either \,$\fctEncName( d_{i}^* )$\, is already contained in $w_{j}$ or it is not contained in $w_{j}$. In the first case we have $j'=j$; in the latter case we have $j'>j$. We have to discuss the latter case only.
	Due to (\ref{Equation:Proof:SubLemma:Reach:Computability:Completeness}) exist $t^* \in \dataFct(d_{i-1}^*)$ and $\symURI^* \in \fctIDsName(t^*)$ such that $\adocFct(\symURI^*)=d_{i}^*$ and $c(t^*,\symURI^*,P) = \true$. Hence, there exists a $\delta \in \mathbb{N}^0$ such that $M^{(P,S,c)}$ finds $t^*$ and $\symURI^*$ in the ($j$+$\delta$)-th iteration. Since $M^{(P,S,c)}$ calls \code{lookup} for $\symURI^*$ in that iteration (cf.~line~\ref{Line:ReachQueryMachine:Step3_LookupOrHalt} in Algorithm~\ref{Algorithm:ReachQueryMachine}), it holds that $w_{j+\delta+1}$ contains \,$\fctEncName( d_{i}^* )$\, and, thus, $j'=j+\delta+1$.
\end{myproof}

\noindent
We now prove Proposition~\ref{SubProposition:Reach:Computability:Completeness} by induction over the structure of possible SPARQL expressions. This proof resembles the proof of Lemma~\ref{Lemma:FullWeb:Computability:Completeness} (cf.~Section~\ref{Proof:Lemma:FullWeb:Computability:Completeness}).

\vspace{1ex} \noindent
\textbf{Base case}: Assume that SPARQL expression $P$ is a triple pattern $tp$.
W.l.o.g., let $\mu \in \evalReachPScW{P}{S}{c}{\WoD}$. It holds $\fctDom{\mu} = \fctVarsName(tp)$ and $t = \mu[tp] \in \fctAllDataName\bigl( \ReachPartScPW{S}{c}{P}{\WoD} \bigr)$ (cf.~Definitions~\ref{Definition:ReachSemantics} and~\ref{Definition:EvaluationSPARQL}). According to Lemma~\ref{SubLemma:Reach:Computability:Completeness} exists a $j_\mu \in \lbrace 1,2,...\rbrace$ such that $\forall \, j \in \lbrace j_\mu,j_\mu\!+\!1,...\rbrace : t \in T_j$. Since $\evalReachFctPSc{P}{S}{c}$ is monotonic we conclude $\forall \, j \in \lbrace j_\mu,j_\mu \!+\! 1,...\rbrace : \mu \in \evalOrigPG{P}{T_j}$.

\vspace{1ex} \noindent
\textbf{Induction step}: Our inductive hypothesis is that for SPARQL expressions $P_1$ and $P_2$ it holds:
\begin{enumerate}
	\item
		For each $\mu \in \evalReachPScW{P_1}{S}{c}{\WoD}$ exists a $j_\mu \in \lbrace 1,2,...\rbrace$ such that during the execution of Algorithm~\ref{Algorithm:ReachQueryMachine} by $M^{(P,S,c)}$ it holds
		$\forall \, j \in \lbrace j_\mu,j_\mu \!+\! 1,...\rbrace : \mu \in \evalOrigPG{P_1}{T_j}$; and
	\item
		For each $\mu \in \evalReachPScW{P_2}{S}{c}{\WoD}$ exists a $j_\mu \in \lbrace 1,2,...\rbrace$ such that during the execution of Algorithm~\ref{Algorithm:ReachQueryMachine} by $M^{(P,S,c)}$ it holds
		$\forall \, j \in \lbrace j_\mu,j_\mu \!+\! 1,...\rbrace : \mu \in \evalOrigPG{P_2}{T_j}$.
\end{enumerate}
\noindent
Based on this hypothesis we show that for any SPARQL expression $P$ that can be constructed using $P_1$ and $P_2$ it holds: For each $\mu \in \evalReachPScW{P}{S}{c}{\WoD}$ exists a $j_\mu \in \lbrace 1,2,...\rbrace$ such that during the execution of Algorithm~\ref{Algorithm:ReachQueryMachine} by $M^{(P,S,c)}$ it holds $\forall \, j \in \lbrace j_\mu,j_\mu \!+\! 1,...\rbrace : \mu \in \evalOrigPG{P}{T_j}$. W.l.o.g., let $\mu' \in \evalReachPScW{P}{S}{c}{\WoD}$. According to Definition~\ref{Definition:ExpressionSPARQL} we distinguish the following cases:
\begin{itemize}
	\item $P$ is $(P_1 \OpAND P_2)$.
		In this case exist $\mu_1 \in \evalReachPScW{P_1}{S}{c}{\WoD}$ and $\mu_2 \in \evalReachPScW{P_2}{S}{c}{\WoD}$ such that $\mu' = \mu_1 \cup \mu_2$ and $\mu_1 \sim \mu_2$. According to our inductive hypothesis exist $j_{\mu_1}, j_{\mu_2} \in \lbrace 1,2,...\rbrace$ such that i)~$\forall \, j \in \lbrace j_{\mu_1},j_{\mu_1} \!+\! 1,...\rbrace : \mu_1 \in \evalOrigPG{P_1}{T_j}$ and ii)~$\forall \, j \in \lbrace j_{\mu_2},j_{\mu_2} \!+\! 1,...\rbrace : \mu_2 \in \evalOrigPG{P_2}{T_j}$. Let $j_{\mu'} = \max\bigl( \lbrace j_{\mu_1},j_{\mu_2} \rbrace \bigr)$. Due to the monotonicity of $\evalReachFctPSc{P}{S}{c}$ it holds $\forall \, j \in \lbrace j_{\mu'},j_{\mu'} \!+\! 1,...\rbrace : \mu' \in \evalOrigPG{P}{T_j}$.
	\item $P$ is $(P_1 \OpFILTER R)$.
		In this case exist $\mu^* \in \evalReachPScW{P_1}{S}{c}{\WoD}$ such that $\mu' = \mu^*$. According to our inductive hypothesis exist $j_{\mu^*} \in \lbrace 1,2,...\rbrace$ such that $\forall \, j \in \lbrace j_{\mu^*},j_{\mu^*} \!+\! 1,...\rbrace : \mu^* \in \evalOrigPG{P_1}{T_j}$. Due to the monotonicity of $\evalReachFctPSc{P}{S}{c}$ it holds $\forall \, j \in \lbrace j_{\mu^*},j_{\mu^*} \!+\! 1,...\rbrace : \mu' \in \evalOrigPG{P}{T_j}$.
	\item $P$ is $(P_1 \OpOPT P_2)$.
		We distinguish two cases:
		\begin{enumerate}
			\item There exist $\mu_1 \in \evalReachPScW{P_1}{S}{c}{\WoD}$ and $\mu_2 \in \evalReachPScW{P_2}{S}{c}{\WoD}$ such that $\mu' = \mu_1 \cup \mu_2$ and $\mu_1 \sim \mu_2$. This case corresponds to the case where $P$ is $(P_1 \OpAND P_2)$ (see above).
			\item There exist $\mu_1 \in \evalReachPScW{P_1}{S}{c}{\WoD}$ such that $\mu' = \mu_1$ and $\forall \, \mu_2 \in \evalReachPScW{P_2}{S}{c}{\WoD} : \mu_1 \not\sim \mu_2$. According to our inductive hypothesis exist $j_{\mu_1} \in \lbrace 1,2,...\rbrace$ such that $\forall \, j \in \lbrace j_{\mu_1},j_{\mu_1} \!+\! 1,...\rbrace : \mu_1 \in \evalOrigPG{P_1}{T_j}$. Due to the monotonicity of $\evalReachFctPSc{P}{S}{c}$ it holds $\forall \, j \in \lbrace j_{\mu_1},j_{\mu_1} \!+\! 1,...\rbrace : \mu' \in \evalOrigPG{P}{T_j}$.
		\end{enumerate}
	\item $P$ is $(P_1 \OpUNION P_2)$.
		We distinguish two cases:
		\begin{enumerate}
			\item
				There exists $\mu^* \in \evalReachPScW{P_1}{S}{c}{\WoD}$ such that $\mu' = \mu^*$. According to our inductive hypothesis exist $j_{\mu^*} \in \lbrace 1,2,...\rbrace$ such that
				$\forall \, j \in \lbrace j_{\mu^*},j_{\mu^*} \!+\! 1,...\rbrace : \mu^* \in \evalOrigPG{P_1}{T_j}$.
			\item
				There exists $\mu^* \in \evalReachPScW{P_2}{S}{c}{\WoD}$ such that $\mu' = \mu^*$. According to our inductive hypothesis exist $j_{\mu^*} \in \lbrace 1,2,...\rbrace$ such that
				$\forall \, j \in \lbrace j_{\mu^*},j_{\mu^*} \!+\! 1,...\rbrace : \mu^* \in \evalOrigPG{P_2}{T_j}$.
		\end{enumerate}
		Due to the monotonicity of $\evalReachFctPSc{P}{S}{c}$ it holds for both cases: $\forall \, j \in \lbrace j_{\mu^*},j_{\mu^*} \!+\! 1,...\rbrace : \mu' \in \evalOrigPG{P}{T_j}$.
\end{itemize}

\subsection{Proof of Proposition~\ref{SubProposition:Reach:Computability:Finiteness}} \label{Proof:SubProposition:Reach:Computability:Finiteness}
\begin{proofdeclitems}
	\item $M^{(P,S,c)}$ be the $(P,S,c)$-machine for a SPARQL expression $P$, a finite set $S \subset \symAllURIs$, and a reachability criterion $c$; and
	\item $\WoD$ be a (potentially infinite) Web of Linked Data encoded on the Web tape of $M^{(P,S,c)}$\!.
\end{proofdeclitems}
To prove that $M^{(P,S,c)}$ finishes each iteration of the loop in Algorithm~\ref{Algorithm:ReachQueryMachine} after a finite number of computation steps, we first emphasize the following facts:
\begin{enumerate}
	\item Each call of subroutine \code{lookup} by $M^{(P,S,c)}$ terminates because the encoding of $\WoD$ is ordered following the order of the {\ID}s in $\fctDom{\adocFct}$.
	\item $M^{(P,S,c)}$ completes the initialization in line~\ref{Line:ReachQueryMachine:Init} of Algorithm~\ref{Algorithm:ReachQueryMachine} after a finite number of steps because $S$ is finite.
	\item At any point in the computation the word on the link traversal tape of $M^{(P,S,c)}$ is finite because $M^{(P,S,c)}$ only gradually appends (encoded) {\LDdoc}s to that tape (one document per iteration) and the encoding of each document is finite (recall that the set of {\triple}s $\dataFct(d)$ for each {\LDdoc} $d$ is finite). 
\end{enumerate}
\noindent
It remains to show that each iteration of the loop also only requires a finite number of computation steps:
Due to the finiteness of the word on the link traversal tape, each $\evalOrigPG{P}{T_j}$ (for $j=1,2,...$) is
finite, resulting in a finite number of computation steps for lines~\ref{Line:ReachQueryMachine:Step1_EnumPartialResult} and~\ref{Line:ReachQueryMachine:Step2_OutputNewSolutions} during any iteration. The scan in line~\ref{Line:ReachQueryMachine:Step3_LookupOrHalt} also finishes after a finite number of computation steps because of the finiteness of the word on the link traversal tape.

\subsection{Proof of Theorem~\ref{Theorem:Reach:Problem:Termination}}
\noindent
We formally define the termination problem for {\LDSPARQLReach} as follows:

\vspace{3mm}
\webproblem{Termination(\LDSPARQLReach)}
{a (potentially infinite) Web of Linked Data $\WoD$}
{a finite but nonempty set $S \subset \symAllURIs$ \removable{of seed {\ID}s} \par
a reachability criterion $c_{n\!f}$ that does not ensure finiteness \par
a SPARQL expression $P$}
{Does an LD machine exist that computes $\evalReachPScW{P}{S}{c_{n\!f}}{\WoD}$ and halts?}
\vspace{2mm}

\noindent
To prove that \problemName{Termination(\LDSPARQLReach)} is not LD machine decidable we reduce the halting problem to \problemName{Termination(\LDSPARQLReach)}. For this reduction we use the same argumentation, including the same Web of Linked Data, that we use for proving Theorem~\ref{Theorem:FullWeb:Problem:Termination} (cf.~Section~\ref{Proof:Theorem:FullWeb:Problem:Termination}).

We define the mapping from input for the halting problem to input for \problemName{Termina\-tion(\LDSPARQLReach)} as follows: Let $(w,x)$ be an input to the halting problem, that is, $w$ is the description of a Turing machine $M(w)$ and $x$ is a possible input word for $M(w)$); then $f( w,x ) = \bigl( \WoD_\mathsf{TMs}, S_{w,x}, \cAll, P_{w,x} \bigr)$ where:
\begin{itemize}
	\item $\WoD_\mathsf{TMs}$ is the Web of Linked Data defined in Section~\ref{Proof:Theorem:FullWeb:Problem:Termination},
	\item $S_{w,x} = \big\lbrace \symURI_1^{w,x} \big\rbrace$ (recall, $\symURI_1^{w,x}$ denotes a {\ID} that identifies the first step in the computation of $M(w)$ on input $x$), and
	\item $P_{w,x} = ( \symURI^{w,x}, \mathsf{type} , \mathsf{TerminatingComputation} )$.
\end{itemize}
As before, $f$ is computable by Turing machines (including LD machines).

To show that \problemName{Termination(\LDSPARQLReach)} is not LD machine decidable, suppose it were LD machine decidable. In such a case an LD machine could answer the halting problem for any input $(w,x)$ \removable{as follows}: $M(w)$ halts on $x$ if and only if an LD machine exists that computes $\evalReachPScW{P}{S}{c_{n\!f}}{\WoD_\mathsf{TMs}}$ and halts. However, we know the halting problem is undecidable for TMs (which includes LD machines). Hence, we have a contradiction and\removable{, thus,} \problemName{Termination(\LDSPARQLReach)} cannot be LD machine decidable.

\subsection{Proof of Proposition~\ref{Proposition:Reach:Mono2AsInSPARQL}}
\begin{proofdeclitems}
	\item $\evalReachFctPSc{P}{S}{c_{n\!f}}$
		be a {\LDSPARQLReach} query that uses a finite, nonempty
			\removable{set} $S \subset \symAllURIs$ \removable{of seed {\ID}s}
		and a reachability criterion $c_{n\!f}$ which does not ensure finiteness; and
	\item $G_1,G_2$
		be an arbitrary pair of set of {\triple}s such that $G_1 \subseteq G_2$.
\end{proofdeclitems}

\noindent
Assume $\evalReachFctPSc{P}{S}{c_{n\!f}}$ is monotonic. We have to show that the SPARQL expression $P$ (used by $\evalReachFctPSc{P}{S}{c_{n\!f}}$ ) is monotonic as well. We distinguish two cases: either $P$ is satisfiable or $P$ is not satisfiable. In the latter case $P$ is trivially monotonic. 
Hence, we only have to discuss the first case. To prove that (the satisfiable) $P$ is monotonic it suffices to show $\evalOrigPG{P}{G_1} \subseteq \evalOrigPG{P}{G_2}$.
For this proof we construct two Webs of Linked Data $\WoD_1$ and $\WoD_2$ such that i)~$\WoD_1$ is an induced subweb of $\WoD_2$ and ii)~the data of $G_1$ and $G_2$ is distributed over $\WoD_1$ and $\WoD_2$, respectively. Using $\WoD_1$ and $\WoD_2$ we show the monotonicity of $P$ based on the monotonicity of $\evalReachFctPSc{P}{S}{c_{n\!f}}$.

To construct $\WoD_1$ and $\WoD_2$ we have to address two problems: First, we cannot simply construct $\WoD_1$ and $\WoD_2$ as Webs of Linked Data that consist of single {\LDdoc}s which contain all {\triple}s of $G_1$ and $G_2$ because $G_1$ and $G_2$ may be (countably) infinite, whereas the data in each {\LDdoc} of a Web of Linked Data must be finite. Recall the corresponding proof for {\LDSPARQLFull} where we have the same problem (cf.~Section~\ref{Proof:Proposition:FullWeb:SatAndMonoAsInSPARQL:CaseMono}); we shall use the same strategy for solving that problem in this proof. The second problem, however, is specific to the case of reachability-based semantics: The construction of $\WoD_1$ and $\WoD_2$ for {\LDSPARQLReach} queries
has to ensure that all {\LDdoc}s which contain {\triple}s of $G_1$ and $G_2$ are reach\-able. Due to this issue the construction is more complex than the corresponding construction for the full-Web semantics case.

To solve the first problem we construct $\WoD_1$ and $\WoD_2$ as Webs that contain an {\LDdoc} for each {\triple}s in $G_1$ and $G_2$, respectively. However, 
	by distributing the {\triple}s from (the potentially infinite) $G_1$ over multiple {\LDdoc}s in a constructed Web, we may lose certain solutions $\mu \in \evalOrigPG{P}{G_1}$ because the data of each {\LDdoc} in a Web of Linked Data must use a unique set of blank nodes. The same holds for $G_2$.
To avoid this issue we assume a mapping $\varrho$ that maps each blank node in $G_2$ to a new, unique {\ID}. To define $\varrho$ formally, we let $B$ denote the set of blank nodes in $G_2$, that is, $B = \fctTerms{G_2} \cap \symAllBNodes$. Furthermore, we assume a set $U_B \subset \symAllURIs$ such that $\left|U_B\right| = \left|B\right|$ and $U_B \cap \fctTerms{G_2} = \emptyset$. Now, $\varrho$ is a total, bijective mapping $\varrho : \bigl( (\symAllURIs \cup \symAllBNodes \cup \symAllLiterals) \setminus U_B \bigr) \rightarrow \bigl( (\symAllURIs \cup \symAllBNodes \cup \symAllLiterals) \setminus B \bigr)$ that, for any $x \in \bigl( (\symAllURIs \cup \symAllBNodes \cup \symAllLiterals) \setminus U_B \bigr)$, is defined as follows:
\begin{equation*}
	\varrho(x) = \begin{cases}
		\varrho_B(x) & \text{if $x \in B$,} \\
		x & \text{else.} \\
	\end{cases}
\end{equation*}
where $\varrho_B$ is an arbitrary bijection $\varrho_B : B \rightarrow U_B$.

The application of $\varrho$ to an arbitrary {\triple} $t = (x_1,x_2,x_3)$, denoted by $\varrho[t]$, results in an {\triple} $t' = (x_1',x_2',x_3')$ such that $x_i'=\varrho(x_i)$ for all $i \in \lbrace 1,2,3 \rbrace$.
Furthermore, the application of $\varrho$ to
	a
valuation $\mu$, denoted by $\varrho[\mu]$, results in a valuation $\mu'$ such that $\fctDom{\mu'}=\fctDom{\mu}$ and $\mu'(?v) = \varrho( \mu(?v) )$ for all $?v \in \fctDom{\mu}$.

We now let
	\begin{align*}
		G_1' &= \big\lbrace \varrho[t] \,\big|\, t \in G_1 \big\rbrace &&\text{and} & G_2' &= \big\lbrace \varrho[t] \,\big|\, t \in G_2 \big\rbrace
	\end{align*}
The following facts are verified easily:
\begin{fact} \label{Fact:Proof:Proposition:Reach:Mono2AsInSPARQL:Facts1}
	It holds: $G_1' \subseteq G_2'$, $\left| G_1 \right| = \left| G_1' \right|$, and $\left| G_2 \right| = \left| G_2' \right|$.
\end{fact}
\begin{fact} \label{Fact:Proof:Proposition:Reach:Mono2AsInSPARQL:Fact2}
	For all $j \in \lbrace 1,2 \rbrace$ it holds: Let $\mu$ be an arbitrary valuation, then $\varrho[\mu]$ is a solution for $P$ in $G_j'$ \IFF $\mu$ is a solution for $P$ in $G_j$. More precisely:
		\begin{align*}
			\forall \, \mu \in \evalOrigPG{P}{G_j} &: \varrho[\mu] \in \evalOrigPG{P}{G_j'} &
			&\text{and} &
			\forall \, \mu' \in \evalOrigPG{P}{G_j'} &: \varrho^{-1}[\mu'] \in \evalOrigPG{P}{G_j}
		\end{align*}
		where $\varrho^{-1}$ denotes the inverse of the bijective mapping $\varrho$.
\end{fact}

\noindent
We now address the second problem, that is, we construct $\WoD_1$ and $\WoD_2$ (using $G_1'$ and $G_2'$) in a way that all {\LDdoc}s which contain {\triple}s from $G_1'$ and $G_2'$ are reach\-able. To achieve this goal we use a reach\-able part of another Web of Linked Data for the construction. We emphasize that this reach\-able part must be infinite because $G_1$ and $G_2$ may be (countably) infinite. To find a Web of Linked Data with such a reach\-able part we make use of $c_{n\!f}$: Since $c_{n\!f}$ does not ensure finiteness, we know there exists a Web of Linked Data $\WoD^* = (D^*,\dataFct^*,\adocFct^*)$, a (finite, nonempty) set $S^* \subset \symAllURIs$ of seed {\ID}s, and a SPARQL expression $P^*$ such that the $(S^*,c_{n\!f},P^*)$-reach\-able part of $\WoD^*$ is infinite. Notice, $S^*$ and $P^*$ are not necessarily the same as $S$ and $P$.

While the $(S^*,c_{n\!f},P^*)$-reach\-able part of $\WoD^*$ presents the basis for our proof, we cannot use it directly because the data in that part may cause undesired side-effects for the evaluation of $P$. To avoid this issue we define an isomorphism $\sigma$ for $\WoD^*$, $S^*$, and $P^*$ such that the images of $\WoD^*$, $S^*$, and $P^*$ under $\sigma$ do not use any RDF term or query variable from $G_2'$ and $P$.

For the definition of $\sigma$ we write $U$, $L$, and $V$ to denote the sets of all {\ID}s, literals, and variables in $G_2'$ and $P$ (recall, neither $G_2'$ nor $P$ contain blank nodes). That is:
\begin{align*}
	U &= \bigl( \fctTerms{G_2'} \cup \fctTerms{P} \bigr) \cap \symAllURIs , \\
	L &= \bigl( \fctTerms{G_2'} \cup \fctTerms{P} \bigr) \cap \symAllLiterals , \text{ and} \\
	V &= \fctVars{P} \cup \fctVarsName_\mathsf{F}(P)
\end{align*}
where $\fctVarsName_\mathsf{F}(P)$ denotes the set of all variables in all filter conditions of $P$ (if any).
Similarly to $U$, $L$, and $V$, we write $U^*$, $L^*$, and $V^*$ to denote the sets of all {\ID}s, literals, and variables in $\WoD^*$, $S^*$, and $P^*$:
\begin{align*}
	U^* &= S^* \cup \fctTermsName\bigl( \fctAllDataName(\WoD^*) \bigr) \cap \symAllURIs , \\
	L^* &= \fctTermsName\bigl( \fctAllDataName(\WoD^*) \bigr) \cap \symAllLiterals , \hspace{5mm} \text{and} \\
	V^* &= \fctVars{P^*} \cup \fctVarsName_\mathsf{F}(P^*) .
\end{align*}
Moreover, we assume three new sets of {\ID}s, literals, and variables, denoted by $U_\mathsf{new}$, $L_\mathsf{new}$, and $V_\mathsf{new}$, respectively. For these sets it must hold:
\begin{align*}
	U_\mathsf{new} \subset \symAllURIs & \text{ such that } \left|U_\mathsf{new}\right|=\left|U\right| \text{ and } U_\mathsf{new} \cap ( U \cup U^* ) = \emptyset; \\
	L_\mathsf{new} \subset \symAllLiterals & \text{ such that } \left|L_\mathsf{new}\right|=\left|L\right| \text{ and } L_\mathsf{new} \cap ( L \cup L^* ) = \emptyset; \text{ and} \\
	V_\mathsf{new} \subset \symAllVariables & \text{ such that } \left|V_\mathsf{new}\right|=\left|V\right| \text{ and } V_\mathsf{new} \cap ( V \cup V^* ) = \emptyset.
\end{align*}
Furthermore, we assume three
	total, bijective mappings:
\begin{align*}
	\sigma_U &: U \rightarrow U_\mathsf{new} &
	\sigma_L &: L \rightarrow L_\mathsf{new} &
	\sigma_V &: V \rightarrow V_\mathsf{new}
\end{align*}
Now we define $\sigma$ as a total, bijective mapping
\begin{equation*}
	\sigma :
	\Bigl( \bigl(\symAllURIs \cup \symAllBNodes \cup \symAllLiterals \cup \symAllVariables \bigr) \setminus \bigl(U_\mathsf{new} \cup L_\mathsf{new} \cup V_\mathsf{new} \bigr) \Bigr)
	\rightarrow
	\Bigl( \bigl(\symAllURIs \cup \symAllBNodes \cup \symAllLiterals \cup \symAllVariables \bigr) \setminus \bigl(U \cup L \cup V \bigr) \Bigr)
\end{equation*}
such that for each $x \in \fctDom{\sigma}$ it holds:
\begin{equation*}
	\sigma(x) = \begin{cases}
		\sigma_U(x) & \text{if $x \in U$,} \\
		\sigma_L(x) & \text{if $x \in L$,} \\
		\sigma_V(x) & \text{if $x \in V$,} \\
		x & \text{else.} \\
	\end{cases}
\end{equation*}

\noindent
The application of $\sigma$ to an arbitrary valuation $\mu$ and to an arbitrary {\triple}
is defined in a way that corresponds to the application of $\varrho$ to $\mu$ and $t$, respectively.
An application of $\sigma$ to further, relevant structures is defined as follows:
\begin{itemize}
	\item The application of $\sigma$ to the aforementioned Web
	$\WoD^* = (D^*,\dataFct^*,\adocFct^*)$, denoted by $\sigma[\WoD^*]$, results in a Web of Linked Data $\WoD^*{}' = (D^*{}',\dataFct^*{}',\adocFct^*{}')$ such that $D^*{}'=D^*$ and mappings $\dataFct^*{}'$ and $\adocFct^*{}'$ are defined as follows:
	\begin{align*}
		\forall \, d \in D^*{}' &: \dataFct^*{}'(d) = \big\lbrace \sigma[t] \,\big|\, t \in \dataFct^*(d) \big\rbrace
		\\ 
		\forall \, \symURI \in \fctDom{\adocFct^*{}'} &: \adocFct^*{}'(\symURI) = \adocFct^* \bigl( \sigma^{-1}(\symURI) \bigr)
	\end{align*}
	where $\fctDom{\adocFct^*{}'} = \big\lbrace \sigma(\symURI) \,\big|\, \symURI \in \fctDom{\adocFct^*} \big\rbrace$ and $\sigma^{-1}$ is the inverse of
	$\sigma$.
	\item The application of $\sigma$ to an arbitrary (SPARQL) filter condition $R$, denoted by $\sigma[R]$, results in a filter condition that is defined as follows:
	\begin{inparaenum} [i)]
		\item If $R$ is $?x=c$, $?x = ?y$, or $\fctBound{?x}$, then $\sigma[R]$ is $?x'=c'$, $?x' = ?y'$, and $\fctBound{?x'}$, respectively, where $?x'=\sigma(?x)$, $?y'=\sigma(?y)$, and $c'=\sigma(c)$; and
		\item If $R$ is $(\neg R_1)$, $(R_1 \land R_2)$, or, $(R_1 \lor R_2)$, then $\sigma[R]$ is $(\neg R_1')$, $(R_1' \land R_2')$, or, $(R_1' \lor R_2')$, respectively, where $R_1'=\sigma[R_1]$ and $R_2'=\sigma[R_2]$.
	\end{inparaenum}
	\item The application of $\sigma$ to an arbitrary SPARQL expression $P'$, denoted by $\sigma[P']$, results in a SPARQL expression that is defined as follows:
	\begin{inparaenum} [i)]
		\item If $P'$ is a triple pattern $\bigl( x_1',x_2',x_3' \bigr)$, then $\sigma[P']$ is $(x_1'',x_2'',x_3'')$ such that $x_i''=\sigma(x_i')$ for all $i \in \lbrace 1,2,3 \rbrace$; and
		\item If $P'$ is $(P_1' \OpAND P_2')$, $(P_1' \OpUNION P_2')$, $(P_1' \OpOPT P_2')$, or $(P_1' \OpFILTER R')$, then $\sigma[P']$ is $(P_1'' \OpAND P_2'')$, $(P_1'' \OpUNION P_2'')$, or $(P_1'' \OpOPT P_2'')$, and $(P_1'' \OpFILTER R'')$, respectively, where $P_1''=\sigma[P_1']$, $P_2''=\sigma[P_2']$, and $R''=\sigma[R']$.
	\end{inparaenum}
\end{itemize}

\noindent
We now introduce $\WoD^*{}'$\!, $S^*{}'$\!, and $P^{*}{}'$ as image of $\WoD^*$\!, $S^*$\!, and $P^*$ under $\sigma$, respectively:
\begin{align*}
	\WoD^*{}' &= \sigma[\WoD^*] &
	S^*{}' &= \big\lbrace \sigma(\symURI) \,\big|\, \symURI \in S^* \big\rbrace &
	P^*{}' = \sigma[P^*]
\end{align*}
$\WoD^*{}'$ is structurally identical to $\WoD^*$\!. Furthermore, the $(S^*{}',c_{n\!f},P^*{}')$-reach\-able part of $\WoD^*{}'$ is infinite because the $(S^*,c_{n\!f},P^*)$-reach\-able part of $\WoD^*$ is infinite. Hereafter, we write $\WoD_\mathfrak{R} = (\Reach{D},\Reach{\dataFct},\Reach{\adocFct})$ to denote the $(S^*{}',c_{n\!f},P^*{}')$-reach\-able part of $\WoD^*{}'$.

We now use $\WoD_\mathfrak{R}$ to construct Webs of Linked Data that contain all {\triple}s from $G_1'$ and $G_2'$, respectively. Since $\WoD_\mathfrak{R}$ is infinite, there exists at least one infinite path in the link graph of $\WoD_\mathfrak{R}$. Let $p=d_1,d_2,...$ be such a path. Hence, for all $i \in \lbrace 1,2,... \rbrace$ holds:
\begin{align*}
	d_i &\in \Reach{D} &
	& \text{and} &
	\exists \, t \in \Reach{\dataFct}(d_i) &: \Bigl( \exists \, \symURI \in \fctIDs{t} : \Reach{\adocFct}(\symURI)=d_{i+1} \Bigr)
\end{align*}
We may use this path for constructing Webs of Linked Data $\WoD_1$ and $\WoD_2$ from $\WoD_\mathfrak{R}$ such that $\WoD_1$ and $\WoD_2$ contain the data from $G_1'$ and $G_2'$, respectively. However, to allow us to use the monotonicity of {\LDSPARQLReach} queries in our proof, it is necessary to construct $\WoD_1$ and $\WoD_2$ such that $\WoD_1$ is an induced subweb of $\WoD_2$. To achieve this goal we assume a strict total order on $G_2'$ such that each $t \in G_1' \subseteq G_2'$ comes before any $t' \in G_2' \setminus G_1'$ in that order. Formally, we denote this order by infix $<$ and, thus, require $\forall \, (t,t') \in G_1' \times (G_2' \setminus G_1') : t<t'$. Furthermore, we assume a total, injective function $pdoc : G_2' \rightarrow \big\lbrace d \in \Reach{D} \,\big|\, d \text{ is on path } p \big\rbrace$ which is order-preserving, that is, for each pair $(t,t')\in G_2' \times G_2'$ holds: If $t<t'$ then {\LDdoc} $pdoc(t)$ comes before {\LDdoc} $pdoc(t')$ on path $p$.

We now use $pdoc$, $G_2'$, and $\WoD_\mathfrak{R} = (\Reach{D},\Reach{\dataFct},\Reach{\adocFct})$ to construct a Web of Linked Data $\WoD_2 = (D_2,\dataFct_2,\adocFct_2)$ as follows:
\begin{align*}
	D_2 &= \Reach{D} \\
	\forall \, d \in D_2 : \dataFct_2(d) &= \begin{cases}
		\Reach{\dataFct}(d) \cup \lbrace t \rbrace & \text{if } \exists \, t \in G_2' : pdoc(t) = d, \\
		\Reach{\dataFct}(d) & \text{else.}
	\end{cases} \\
	\forall \, \symURI \in \fctDom{\Reach{\adocFct}} : \adocFct_2(\symURI) &= \Reach{\adocFct}(\symURI)
\end{align*}
In addition to $\WoD_2$, we introduce a Web of Linked Data $\WoD_1 = ( D_1,\dataFct_1,\adocFct_1 )$ that is an induced subweb of $\WoD_2$ and that is defined by\footnote{Recall, any induced subweb is unambiguously defined by specifying its set of {\LDdoc}s.}:
\begin{equation*}
	D_1 = \big\lbrace d \in D_2 \,\big|\, \text{either } d \text{ is not on path } p \text{ or } \exists \, t \in G_1' : d = pdoc(t) \big\rbrace
\end{equation*}

\noindent
The following facts are verified easily:
\begin{fact} \label{Fact:Proof:Proposition:Reach:Mono2AsInSPARQL:Fact3}
For all $j \in \lbrace 1,2 \rbrace$ it holds: $G_j' \subset \fctAllDataName\bigl( \WoD_j \bigr) = G_j' \cup \fctAllDataName\bigl( \WoD_\mathfrak{R} \bigr)$.
\end{fact}
\begin{fact} \label{Fact:Proof:Proposition:Reach:Mono2AsInSPARQL:Fact4}
For all $j \in \lbrace 1,2 \rbrace$ it holds: The $(S^*{}',c_{n\!f},P^*{}')$-reach\-able part of $\WoD_j$ is $\WoD_j$ itself.
\end{fact}
\begin{fact} \label{Fact:Proof:Proposition:Reach:Mono2AsInSPARQL:Fact5}
For all $j \in \lbrace 1,2 \rbrace$ it holds: $\evalOrigPG{P}{G_j'} = \evalOrigPG{P}{\fctAllDataName(\WoD_j)}$.
\end{fact}

\noindent
We now consider a SPARQL expression $(P \OpUNION P^*{}')$. In the following we write $\tilde{P}$ to denote this expression.
Since $\fctTermsName\bigl( G_2' \bigr) \cap \fctTermsName\bigl( \fctAllDataName( \WoD_\mathfrak{R} ) \bigr) = \emptyset$ we conclude the following facts:
\begin{fact} \label{Fact:Proof:Proposition:Reach:Mono2AsInSPARQL:Facts7}
For all $j \in \lbrace 1,2 \rbrace$ it holds:
\begin{enumerate}
	\item The $(S^*{}',c_{n\!f},\tilde{P})$-reach\-able part of $\WoD_j$ is $\WoD_j$ itself.
	\item $\evalOrigPG{P}{\fctAllDataName(\WoD_j)} \cup \evalOrigPG{P^*{}'}{\fctAllDataName(\WoD_j)} = \evalOrigPG{\tilde{P}}{\fctAllDataName(\WoD_j)}$
	\item $\evalOrigPG{P}{\fctAllDataName(\WoD_j)} \cap \evalOrigPG{P^*{}'}{\fctAllDataName(\WoD_j)} = \emptyset$
\end{enumerate}
\end{fact}

\noindent
Since $\WoD_1$ is an induced subweb of $\WoD_2$, $\evalReachFctPSc{P}{S}{c_{n\!f}}$ is monotonic, and $\tilde{P}$ is $(P \OpUNION P^*{}')$, we conclude the following inclusion from Fact~\ref{Fact:Proof:Proposition:Reach:Mono2AsInSPARQL:Facts7} and Definition~\ref{Definition:ReachSemantics}:
\begin{equation} \label{Equation:Proof:Proposition:Reach:Mono2AsInSPARQL:Fact8}
	\bigl( \evalReachPScW{\tilde{P}}{S^*{}'}{c_{n\!f}}{\WoD_1} \setminus \evalOrigPG{P^*{}'}{\fctAllDataName(\WoD_1)} \bigr) \subseteq \bigl( \evalReachPScW{\tilde{P}}{S^*{}'}{c_{n\!f}}{\WoD_2} \setminus \evalOrigPG{P^*{}'}{\fctAllDataName(\WoD_2)} \bigr)
\end{equation}

\noindent
We now use $\WoD_1$ and $\WoD_2$ and the monotonicity of $\evalReachFctPSc{P}{S}{c_{n\!f}}$ to show $\evalOrigPG{P}{G_1} \subseteq \evalOrigPG{P}{G_2}$ (which proves that $P$ is monotonic). W.l.o.g., let $\mu$ be an arbitrary solution for $P$ in $G_1$, that is, $\mu \in \evalOrigPG{P}{G_1}$. Notice, such a $\mu$ must exist because we assume $P$ is satisfiable (see before). To prove $\evalOrigPG{P}{G_1} \subseteq \evalOrigPG{P}{G_2}$ it suffices to show $\mu \in \evalOrigPG{P}{G_2}$.

Due to Fact~\ref{Fact:Proof:Proposition:Reach:Mono2AsInSPARQL:Fact2} it holds
\begin{align*}
	\varrho[\mu] &\in \evalOrigPG{P}{G_1'}
\intertext{and with Facts~\ref{Fact:Proof:Proposition:Reach:Mono2AsInSPARQL:Fact5} and~\ref{Fact:Proof:Proposition:Reach:Mono2AsInSPARQL:Facts7} and Definition~\ref{Definition:ReachSemantics} we have}
	\varrho[\mu] &\in \bigl( \evalReachPScW{\tilde{P}}{S^*{}'}{c_{n\!f}}{\WoD_1} \setminus \evalOrigPG{P^*{}'}{\fctAllDataName(\WoD_1)} \bigr)
.
\intertext{According to (\ref{Equation:Proof:Proposition:Reach:Mono2AsInSPARQL:Fact8}) we also have}
	\varrho[\mu] &\in \bigl( \evalReachPScW{\tilde{P}}{S^*{}'}{c_{n\!f}}{\WoD_2} \setminus \evalOrigPG{P^*{}'}{\fctAllDataName(\WoD_2)} \bigr)
.
\intertext{We now use Definition~\ref{Definition:ReachSemantics} and Facts~\ref{Fact:Proof:Proposition:Reach:Mono2AsInSPARQL:Facts7} and~\ref{Fact:Proof:Proposition:Reach:Mono2AsInSPARQL:Fact5} again, to show}
	\varrho[\mu] &\in \evalOrigPG{P}{G_2'}
.
\intertext{Finally, we use Fact~\ref{Fact:Proof:Proposition:Reach:Mono2AsInSPARQL:Fact2} again and find}
	\varrho^{-1}\bigl[ \varrho[\mu] \bigr] &\in \evalOrigPG{P}{G_2}
.
\end{align*}
Since $\varrho^{-1}$ is the inverse of bijective mapping $\varrho$, it holds $\varrho^{-1}\bigl[ \varrho[\mu] \bigr] = \mu$ and, thus, we conclude $\mu \in \evalOrigPG{P}{G_2}$.
\section{Constant Reachability Criteria} \label{Appendix:ConstReachCriteria}
\noindent
This section discusses a particular class of reachability criteria which we call \emph{constant reachability criteria}. These criteria always only accept a given, constant set of data links. As a consequence, each of these criteria ensures finiteness. In the following we formally introduce constant reachability criteria and prove that they ensure finiteness.

The (fixed) set of data links that a constant reachability criterion accepts may be specified differently.
	Accordingly,
we distinguish four different types of constant reachability criteria. Formally, we define them as follows:

\begin{definition} \label{Definition:ConstReachCriteria}
	Let $U \subset \symAllURIs$ be a finite set {\ID}s and let $T \subset \mathcal{T}$ be a finite set of {\triple}s.
	The \definedTerm{$U$-con\-stant reachability criterion} $c^U$ is a reachability criterion that for each tuple $(t,\symURI,P) \in \mathcal{T} \times \symAllURIs \times \mathcal{P}$ is defined as follows:
	\begin{equation*}
		c^U \Bigl( t,\symURI, P \Bigr) = \begin{cases}
			\true & \text{if $\symURI \in U$}, \\
			\false & \text{else}.
		\end{cases}
	\end{equation*}
	The \definedTerm{$T$-con\-stant reachability criterion} $c^T$ is a reachability criterion that for each tuple $(t,\symURI,P) \in \mathcal{T} \times \symAllURIs \times \mathcal{P}$ is defined as follows:
	\begin{equation*}
		c^T \Bigl( t,\symURI, P \Bigr) = \begin{cases}
			\true & \text{if $t \in T$}, \\
			\false & \text{else}.
		\end{cases}
	\end{equation*}
	The \definedTerm{$(U \!\land\! T)$-con\-stant reachability criterion} $c^{U \land T}$ is a reachability criterion that for each tuple $(t,\symURI,P) \in \mathcal{T} \times \symAllURIs \times \mathcal{P}$ is defined as follows:
	\begin{equation*}
		c^{U \land T} \Bigl( t,\symURI, P \Bigr) = \begin{cases}
			\true & \text{if $\symURI \in U$ and $t \in T$}, \\
			\false & \text{else}.
		\end{cases}
	\end{equation*}
	The \definedTerm{$(U \!\lor\! T)$-con\-stant reachability criterion} $c^{U \lor T}$ is a reachability criterion that for each tuple $(t,\symURI,P) \in \mathcal{T} \times \symAllURIs \times \mathcal{P}$ is defined as follows:
	\begin{equation*}
		c^{U \land T} \Bigl( t,\symURI, P \Bigr) = \begin{cases}
			\true & \text{if $\symURI \in U$ or $t \in T$}, \\
			\false & \text{else}.
		\end{cases}
	\end{equation*}
\end{definition}

\noindent
As can be seen from the definition, a $U$-con\-stant reachability criterion uses a
	(finite)
set $U$ of
{\ID}s to specify the data links it accepts. Similarly, a $T$-con\-stant reachability criterion uses a
	(finite)
set $T$ of
{\triple}s. $(U \!\land\! T)$-con\-stant reachability criteria and $(U \!\lor\! T)$-con\-stant reachability criteria combine $U$-con\-stant reachability criteria and $T$-con\-stant reachability criteria in a conjunctive and disjunctive manner, respectively.
The reachability criterion $\cNone$ may be understood as a special case of $U$-con\-stant reachability criteria; it uses a $U$ which is empty. Similarly, $\cNone$ may be understood as the $T$-con\-stant reachability criterion for which $T$ is empty.

The following facts are trivial to verify:
\begin{fact} \label{Fact:ConstReachCriteria}
	Let $U \subset \symAllURIs$ and $U' \subset \symAllURIs$ be finite sets of
	{\ID}s such that $U' \subset U$. Similarly, let $T \subset \mathcal{T}$ and $T' \subset \mathcal{T}$ be finite sets of
	{\triple}s such that $T' \subset T$.
	Furthermore, let $c^U$\!, $c^{U'}$\!, $c^T$\!, and $c^{T'}$ denote the $U$-con\-stant reachability criterion, the $U'$-con\-stant reachability criterion, the $T$-con\-stant reachability criterion, and the $T'$-con\-stant reachability criterion, respectively. Moreover, $c^{U \land T}$\!, $c^{U \lor T}$\!, $c^{U' \land T'}$\!, and $c^{U' \lor T'}$ denote the $(U \!\land\! T)$-con\-stant reachability criterion, the $(U \!\lor\! T)$-con\-stant reachability criterion, the $(U' \!\land\! T')$-con\-stant reachability criterion, and the $(U' \!\lor\! T')$-con\-stant reachability criterion, respectively. It holds:
	\begin{enumerate}
		\item $c^{U \lor T}$ is less restrictive than $c^U$ and less restrictive than $c^T$. \label{Fact:ConstReachCriteria:Case1}
		\item $c^U$ and $c^T$ are less restrictive than $c^{U \land T}$, respectively. \label{Fact:ConstReachCriteria:Case2}
		\item $c^U$ is less restrictive than $c^{U'}$.
		\item $c^T$ is less restrictive than $c^{T'}$.
		\item $c^{U \land T}$ is less restrictive than $c^{U' \land T'}$.
		\item $c^{U \lor T}$ is less restrictive than $c^{U' \lor T'}$.
	\end{enumerate}
\end{fact}

\noindent
We now show that all constant reachability criteria ensure finiteness:

\begin{proposition} \label{Proposition:ConstReachCriteria}
	All $U$-con\-stant, $T$-con\-stant, $(U \!\land\! T)$-con\-stant, and $(U \!\lor\! T)$-con\-stant reachability criteria ensure finiteness.
\end{proposition}
\begin{myproof}{of Proposition~\ref{Proposition:ConstReachCriteria}}
	To prove that a reachability criterion $c$ ensures finiteness we have to show that for any Web of Linked Data $\WoD$, any (finite) set $S \subset \symAllURIs$ of seed {\ID}s, and any SPARQL expression $P$, the $(S,c,P)$-reachable part of $\WoD$ is finite.
	W.l.o.g., let $S' \subset \symAllURIs$ be an arbitrary (but finite) set of seed {\ID}s, let $P'$ be an arbitrary SPARQL expression. According to Definition~\ref{Definition:ReachablePart} we know that the $(S',c,P')$-reachable part of any Web of Linked Data $\WoD$ is finite if the number of {\LDdoc}s that are $(c,P')$-reachable from $S'$ in $\WoD$ is finite. Due to the finiteness of $S'$, it suffices to show that the set
	\begin{equation*}
		X(c,P')=
		\big\lbrace (t,\symURI,P) \in \mathcal{T} \times \symAllURIs \times \mathcal{P} \,\big|\, \symURI \in \fctIDs{t} \text{ and } P=P'\text{ and } c(t,\symURI,P)=\true \big\rbrace
	\end{equation*}
	is finite for any Web of Linked Data (cf.~Definition~\ref{Definition:QualifiedReachability}). Notice, the given set presents an upper bound for all tuples $(t,\symURI,P) \in \mathcal{T} \times \symAllURIs \times \mathcal{P}$ based on which {\LDdoc}s may be reached by applying Definition~\ref{Definition:QualifiedReachability} recursively. Hence, it is not necessarily the case that all these tuples are discovered (and used) during such a recursive application in a particular Web of Linked Data.

	We now focus on $U$-con\-stant, $T$-con\-stant, $(U \!\land\! T)$-con\-stant, and $(U \!\lor\! T)$-con\-stant reachability criteria. W.l.o.g., let $U' \subset \symAllURIs$ be an arbitrary, finite set of
	{\ID}s and let $T' \subset \mathcal{T}$ be an arbitrary, finite set of
	{\triple}s. Furthermore, let $c^{U'}$\!, $c^{T'}$\!, $c^{U' \land T'}$\!, and $c^{U' \lor T'}$ denote the $U'$-con\-stant reachability criterion, the $T'$-con\-stant reachability criterion, the $(U' \!\land\! T')$-con\-stant reachability criterion, and the $(U' \!\lor\! T')$-con\-stant reachability criterion, respectively.

	For $c^{U' \lor T'}$ it holds
	\begin{equation*}
		\bigl| X(c^{U' \lor T'},P') \bigr| \leq \left| U'\right| + \left| T'\right|
	\end{equation*}
	and, thus,
		\removable{the set}
	$X(c^{U' \lor T'},P')$ is finite (recall, $U'$ and $T'$ are finite). Therefore, the $(S',c^{U' \lor T'},P')$-reachable part of any Web of Linked Data is finite. As discussed before, this fact shows that $c^{U' \lor T'}$ ensures finiteness. However, we may also use this fact, together with Proposition~\ref{Proposition:Reach:InfiniteneWebFindings}, case~\ref{Proposition:Reach:InfiniteneWebFindings:Case5}, and Fact~\ref{Fact:ConstReachCriteria}, cases~\ref{Fact:ConstReachCriteria:Case1} and~\ref{Fact:ConstReachCriteria:Case2}, to show that $c^{U'}\!$, $c^{T'}\!$, and $c^{U' \land T'}$ ensure finiteness, respectively.
\end{myproof}

\end{document}